\pdfoutput=1
\documentclass[12pt]{article}
\usepackage{physics,amssymb,mathtools,bm,multirow,hyperref,color,bbm}
\allowdisplaybreaks[1]
\usepackage[top=24truemm,bottom=24truemm,left=19truemm,right=19truemm]{geometry}
\def \lrangle#1 {\langle #1 \rangle}

\def \Im {\mathrm{Im}\,}
\def \cA {\mathcal{A}}
\def \cN {\mathcal{N}}
\def \tQ {\tilde{Q}}
\def \vv {\vec{v}}

\usepackage{soul}

\begin{document}

\title{
  \begin{flushright}
    \begin{minipage}{0.2\linewidth}
      \normalsize
      WU-HEP-26-03 \\*[50pt]
    \end{minipage}
  \end{flushright}
  {\Large \bf
    MSSM flavors from 7-brane configurations of magnetized SYM on $R^{1,3} \times (T^2)^3/(Z_2 \times Z'_2)$\\*[20pt]}}

\author{Hiroyuki~Abe\footnote{
    E-mail address: abe@waseda.jp}, \
  Molin~Chen\footnote{
    E-mail address: molin.13@moegi.waseda.jp},
  Itsuki~Miyane\footnote{
    E-mail address: ocojo.iki@akane.waseda.jp} \ and
  Naoki~Shimada\footnote{
    E-mail address: snaoki223223@ruri.waseda.jp
  }\\*[20pt]
  {\it \normalsize
    Department of Physics, Waseda University,
    Tokyo 169-8555, Japan} \\*[50pt]}

\date{
  \centerline{\small \bf Abstract}
  \begin{minipage}{0.9\linewidth}
    \medskip
    \medskip
    \small
    We show that chiral matters in the minimal supersymmetric standard model (MSSM) with semi-realistic flavor structures can be obtained from 7-brane configurations of magnetized super Yang-Mills (SYM) theory on a toroidal orbifold $R^{1,3} \times (T^2)^3/(Z_2 \times Z'_2)$,
    where background magnetic fluxes and Wilson-lines are turned on preserving four-dimensional ${\cal N}=1$ supersymmetry.
    The zero-mode spectrum of chiral multiplets in total is just MSSM ones, except the existence of those for three generations of right-handed neutrino and extra generations of MSSM Higgs pairs.
    Hierarchical Yukawa couplings can be obtained from the overlap integrals of wavefunctions localized in extra dimensions, allowing semi-realistic patterns of flavor structures for quarks and charged leptons.
    We also develop a systematic way to embed additional 7-branes into the configuration,
    those are sequestered from the visible sector
    toward a hidden sector model building.
  \end{minipage}
}

\begin{titlepage}
  \maketitle
  \thispagestyle{empty}
  \clearpage
  \tableofcontents
  \thispagestyle{empty}
\end{titlepage}

\renewcommand{\thefootnote}{\arabic{footnote}}
\setcounter{footnote}{0}

\section{Introduction}
The standard model (SM) of elementary particles is the extremely successful model to describe high-energy phenomena, but has so many free parameters like gauge and Yukawa coupling constants as well as Higgs mass and quartic coupling constant. Due to the freedom for Yukawa couplings, for instance, masses and mixing angles of the \rm observed three generations of quarks and leptons cannot be predicted by the model, while their experimental values are quite nontrivial.

In quantum field theory, symmetries play an essential role to reduce the number of free parameters.
In such a sense, higher symmetries mean higher predictability (regardless of whether it matches experiments). For example, if we consider grand unified models, where the three SM gauge groups are unified in a lager one, like SU(5)~\cite{Georgi:1974sy}, the number of gauge (as well as some of Yukawa) coupling constants can be reduced. In a similar way, flavor symmetries,\footnote{See, for a review, Ref.~\cite{Ishimori:2010au}.} whose group space is identified as the flavor (generation) one, can reduce the number of Yukawa coupling constants.
They are classified as internal symmetries, where the spacetime coordinates are unchanged under the symmetry transformation.

Another possibility of symmetries is the spacetime ones.
The Lorentz symmetry, which is necessary to describe high-energy physics, restrict the possible form of the Lagrangian density, and of course SM adopts it. If we consider a higher-dimensional spacetime, the (nontrivial) irreducible representations of the Lorentz algebra have lager degrees of freedom than the corresponding ones in lower dimensions. Especially, spinor fields in higher-dimensions become vector-like from a four-dimensional (4D) perspective. However, the free parameters in the Lagrangian like Yukawa coupling constants are not reduced compared with the corresponding ones in lower-dimensions, since the flavor space is untouched by the Lorentz transformation. Another point is that all the fermions are vector-like as mentioned above, while those in SM are chiral. Therefore we need a certain source for the chiral structure, if we consider models beyond SM in higher-dimensional spacetime.

The supersymmetry, where fermions and bosons are interchanged under the symmetry transformation, is one of the spacetime symmetries that can also restrict the possible form of the Lagrangian. However, since the former (quark or lepton) and the latter (gauge or Higgs boson) in SM cannot be a supersymmetric partner to each other, we need to duplicate the field contents introducing squarks, sleptons, gauginos and higgsinos, even in the minimal supersymmetric extension of SM called the minimal supersymmetric standard model (MSSM).\footnote{See, for a review, Ref.~\cite{Martin:1997ns}.} Due to such increase, the number of free parameters, especially Yukawa coupling constants, are not reduced.\footnote{Phenomenologically, soft supersymmetry breaking terms are required, and rather the number of free parameters becomes larger than the one in the original SM.}

It is notable that the higher supersymmetry such as ${\cal N}=2$ or $4$ dramatically restricts the form of the Lagrangian. Especially, with ${\cal N}=4$ supersymmetry, a single free parameter is allowed for a matter-coupled Yang-Mills system, that is the gauge coupling constant, and the Lagrangian is uniquely determined for each gauge group~\cite{Brink:1976bc}. It seems too much strong to construct phenomenological models, especially to explain the \rm observed hierarchical structure of quark/lepton masses and mixing angles, aside from the fact that all the fermions are vector-like with the higher supersymmetry ${\cal N}>1$.

It is known that the higher supersymmetry is related to a higher-dimensional spacetime~\cite{Brink:1976bc}, which is the reason why fermions are vector-like. The lowest supersymmetry in ten-dimensional (10D) spacetime corresponds to ${\cal N}=4$ supersymmetry in 4D spacetime. The 4D ${\cal N}=4$ supersymmetric field theory is obtained by a dimensional reduction from the minimal one in 10D, that is, the super Yang-Mills (SYM) theory in 10D spacetime~\cite{Brink:1976bc}, where the matter fields in 4D originate from the zero-modes of extra-dimensional components in 10D vector field and their supersymmetric partners. In 10D, only gauge bosons and gauginos but no matter fields are allowed by supersymmetry, and the above-mentioned single free parameter in 4D ${\cal N}=4$ supersymmetry corresponds to the gauge coupling constant of 10D SYM. Nevertheless, 4D Yukawa couplings arise from the 10D gauge coupling, since the origin of 4D matter fields are as mentioned above.

Although 10D SYM is a unique theory allowing a single free parameter for a given gauge group, it is quite interesting to notice that chiral generations of matter fermions in 4D can be realized in such a framework, if the extra-dimensional components of 10D vector (and/or metric) field have a nontrivial background.
For instance, if 10D SYM has magnetic fluxes in extra dimensions, the zero-modes of charged fermions become chiral and degenerated~\cite{Bachas:1995ik,Cremades:2004wa} (even if the background geometry is flat) due to a nontrivial index~\cite{Atiyah:1963zz} of the Dirac operator with the gauge field, and the degeneracy is given by the number of fluxes. Moreover, their wavefunctions have a nontrivial profile~\cite{Cremades:2004wa}, and there is a possibility that the 4D Yukawa coupling constants become hierarchical to reproduce the \rm observed quark/lepton masses and mixing angles, since the coupling constants include the overlap integral of matter wavefunctions in extra-dimensions~\cite{Arkani-Hamed:1999ylh}, providing a geometrical origin of flavor symmetries.\footnote{On magnetized tori, certain discrete flavor symmetries appear~\cite{Abe:2009vi}, those originate from the modular symmetry of the torus~\cite{Kobayashi:2016ovu,Kobayashi:2017dyu}.} It is also argued that, by combining magnetic flux and orbifold geometry, the relationship between the flux number and the zero-mode degeneracy number becomes diverse~\cite{Abe:2008fi}, increasing variety in flavor model building~\cite{Abe:2008sx} (also with Wilson-lines~\cite{Abe:2015yva}). Due to the chiral projection by magnetic fluxes, the supersymmetry is reduced like ${\cal N}=4$ to $1$, depending on the flux configuration. The flux can reduce the gauge symmetry too, playing a role to realize SM-like gauge groups from larger ones.

From these observations, MSSM-like models were constructed based on 10D magnetized $U(8)$ SYM on $R^{1,3} \times (T^2)^3$~\cite{Abe:2012ya},
where a Pati-Salam~\cite{Pati:1974yy} like gauge symmetry is realized by magnetic fluxes, which is broken down to SM like ones by turning on Wilson-lines. In this model, it was shown that semi-realistic patterns of quark and charged lepton masses as well as mixing angles can be obtained due to the effect of background magnetic fluxes mentioned above. The model was extended to the ones on orbifolds, first, to eliminate some of exotic matter fields (those are not included in the matter representations of MSSM) on $R^{1,3} \times (T^2)^3/Z_2$~\cite{Abe:2012fj,Abe:2014vza}, then, to construct the other types of flavor models on $R^{1,3} \times (T^2/Z_2)^3$~\cite{Abe:2016jsb}. Through the series of studies, despite the remarkable successes in flavor structures, it was argued that some exotic matter and non-Abelian gauge fields cannot be eliminated in the zero-mode spectra of 10D magnetized SYM on these orbifolds.

On the other hand, certain classes of SYM can be considered as low-energy effective theories of superstrings which are expected to provide a consistent quantum description of gravitational interactions among elementary particles.  Especially, $U(N)$ SYM in $(p+1)$-dimensional spacetime appears as a low-energy effective theory of $N$ coincident D$p$-branes~\cite{Polchinski:1995mt},
whose action for $p<9$ can be obtained by a dimensional reduction from $p=9$.
In such a framework, chiral matters arise from D-brane intersections, and the intersection number can be interpreted as a number of magnetic fluxes in the T-dual picture~\cite{Berkooz:1996km}, that is, magnetized D-branes whose low-energy effective theory is a magnetized SYM. In D-brane models, those with different dimensionality are mixed usually (like D9-D5 or D7-D3 system) and put on nontrivial background spacetime such as orbifolds (orientifolds)~\cite{Blumenhagen:2000fp,Aldazabal:2000sa}. It was argued that bi-fundamental chiral matters on (two stacks of) lower-dimensional D-branes can be interpreted as the ones on (a single stack of) higher-dimensional D-branes with infinite magnetic fluxes~\cite{Cremades:2004wa}. Those provide a perspective also for bottom-up model buildings and motivate us to consider SYM in lower-dimensions to overcome the problems in the previous model building based on 10D magnetized SYM~\cite{Abe:2015jqa}.

Therefore, in this paper, we consider 7-brane configurations of magnetized SYM on $R^{1,3} \times (T^2)^3/(Z_2 \times Z'_2)$ to construct MSSM like models (with right-handed neutrinos) those allow realistic flavor structures without exotic modes (except extra generations of MSSM Higgs pairs, which can play a role for realizing the flavor structures~\cite{Abe:2014vza}). The 7-brane configurations considered in this paper are such ones that SYM fields on them have a delta-function type localized wavefunction on one of three tori in 10D spacetime. We also consider matter fields in bi-fundamental representations under two SYM gauge groups on 7-branes localized on different torus, in addition to the  eight-dimensional (8D) SYM fields on each brane obtained by a naive dimensional reduction of 10D SYM ones. We interpret the wavefunctions of such bi-fundamental fields as one of the (infinite) eigenfunctions of the covariant derivative on the torus with infinite magnetic fluxes~\cite{Abe:2015jqa}, from the perspective that 7-brane configurations where 8D SYM theories reside are obtained as low energy effective theories of some D7-brane systems. In the previous flavor models based on 10D magnetized SYM, the background spacetime was (unintentionally) chosen as $R^{1,3} \times (T^2/Z_2)^3$ (to simply assure three generations of quarks and leptons) which could break 4D ${\cal N}=1$ supersymmetry in the gravitational sector. In this paper, instead, we adopt $R^{1,3} \times (T^2)^3/(Z_2 \times Z'_2)$ which can preserve 4D ${\cal N}=1$ supersymmetry even if the original model is embedded into the one with the local supersymmetry, namely, into the 10D supergravity~\cite{Bergshoeff:1981um}.

The following sections are organized as follows. In Section~\ref{sec:review}, the basics of 10D magnetized SYM on $R^{1,3} \times (T^2)^3$ and those on $R^{1,3} \times (T^2)^3/(Z_2 \times Z'_2)$ are reviewed. How to build 7-brane configurations based on them is also shown. Using such configurations, in Section~\ref{sec:visible}, MSSM-like models with semi-realistic flavor structures are constructed by identifying the orbifold projections, magnetic fluxes as well as (discrete) Wilson-lines. In Section~\ref{sec:hidden}, a systematic way is proposed for embedding additional 7-branes into the configurations, those are sequestered from the visible sector for a hidden sector model building. The conclusions are given in Section~\ref{sec:conclusion}. All the obtained three generation models and the numerically evaluated flavor structures are listed in Appendix~\ref{sec:appendix}.

\section{10D SYM with background magnetic fluxes}
\label{sec:review}
In this section, we first review 10D SYM theory compactified on magnetized tori, where background magnetic fluxes are introduced in the ten-dimensional action written in terms of 4D $\mathcal{N}=1$ superfields~\cite{Abe:2012ya}. For fluxes (as well as orbifold projections) preserving the $\mathcal{N}=1$ supersymmetry,
the zero-mode effective action can be also written in terms of the $\mathcal{N}=1$ superfields.

\subsection{10D magnetized SYM action on \texorpdfstring{$R^{1,3} \times (T^2)^3$}{R13*T23}}
The 10D SYM action is given by
\begin{equation}
  S
  =
  \int \dd^{10} X
  \sqrt{-G}\frac{1}{g^2}
  \Tr
  \left[
    -\frac{1}{4}F^{MN}F_{MN}
    +\frac{i}{2}\bar{\lambda}\Gamma^M D_M \lambda
    \right],
  \label{eq:10Daction}
\end{equation}
where $G$ is the determinant of the 10D metric $G_{MN}$,
$g$ is the 10D gauge coupling constant,
$F_{MN}=\partial_M A_N - \partial_N A_M - i[A_M,A_N]$ is the field strength of the gauge field $A_M$,
and $\lambda$ is a 10D Majorana--Weyl spinor in the adjoint representation of the gauge group.
The covariant derivative is defined as $D_M \lambda = \partial_M \lambda - i[A_M,\lambda]$.

We consider a compactification on the factorizable 6D torus $(T^2)^3$, where the 10D spacetime coordinates are decomposed as $X^{M}=(x^\mu,y^m)$ for $\mu=0,1,2,3$ and $m=4,\cdots,9$. The torus coordinates $y^m$ satisfy the periodic boundary condition $y^{m}\sim y^{m}+2$.The line element of the 10D spacetime is described as
\begin{equation*}
  \dd s^2
  =
  g_{\mu\nu}\dd x^\mu \dd x^\nu
  +
  g_{mn}\dd y^m \dd y^n,
  \quad
  g_{mn}
  =
  \begin{pmatrix}
    g^{(1)} & 0       & 0       \\
    0       & g^{(2)} & 0       \\
    0       & 0       & g^{(3)}
  \end{pmatrix}
  ,\quad
  g^{(i)}
  =
  (\pi R_i)^2
  \begin{pmatrix}
    1          & \Re \tau_i \\
    \Re \tau_i & |\tau_i|^2
  \end{pmatrix},
\end{equation*}
where $R_i$ and $\tau_i$ are the radius and complex structure of the $i$-th torus ($i=1,2,3$), respectively.
In the following, we use the complex coordinates $z^{i}\equiv(y^{2+2i}+\tau_i y^{3+2i})/2$ and their conjugates $\bar{z}^{\bar{i}}\equiv(y^{2+2i}+\bar{\tau}_{\bar{i}} y^{3+2i})/2$ with the metric $h_{i\bar{j}}=2(\pi R_i)^2 \delta_{i\bar{j}}$ satisfying $g_{mn}\dd y^m\dd y^n=2h_{i\bar{j}}\dd z^i \dd\bar{z}^{\bar{j}}$.
The boundary conditions can be reinterpreted as $z^i \sim z^i + 1, z^i \sim z^i + \tau_i$ in the complex basis, and the area of the $i$-th torus is denoted as $\cA^{(i)}=\int \dd z^{i}\dd \bar{z}^{\bar{j}}\ h_{i\bar{j}}=(2\pi R_i)^2 \Im \tau_i$.

The 10D SYM action \eqref{eq:10Daction} can be written in terms of 4D $\mathcal{N}=1$ superfields~\cite{Arkani-Hamed:2001vvu}, where the 10D vector multiplet is decomposed into a 4D vector multiplet and three chiral multiplets.
The 10D gauge field is decomposed as $A_M=(A_\mu,A_m)$ like $X^M=(x^\mu,y^m)$ described above, and the extra-dimensional components $A_m$ ($m=4,\cdots,9$) are combined into three complex fields
\begin{equation}
  a_i
  =
  -\frac{1}{\Im \tau_i}
  (\tau_i^{*}A_{2+2i} - A_{3+2i}),
\end{equation}
those are identified as the scalar field components of three chiral superfields $\Phi_i$ ($i=1,2,3$), respectively, while the 4D components $A_\mu$ are carried by the vector field component of a vector superfield $V$.
The 10D Majorana--Weyl spinor $\lambda$ is decompose into four 4D Weyl spinors $\lambda_0$ and $\lambda_i$, those are eigenstates of the chirality operators $\Gamma^{(i)}$ on each $T^2$ as
\begin{equation}
  \Gamma^{(i)}\lambda_0 = +\lambda_0,
  \quad
  \Gamma^{(i)}\lambda_j =
  \begin{cases}
    +\lambda_j & (i=j)     \\
    -\lambda_j & (i\neq j)
  \end{cases}.
\end{equation}
Therefore, the corresponding 4D $\mathcal{N}=1$ superfields are expressed like
\begin{equation}
  \Phi_i
  =
  \frac{1}{\sqrt{2}}a_i
  + \sqrt{2}\theta\lambda_i
  + \theta\theta F_i,
  \quad
  V
  =
  -\theta\sigma^\mu\bar{\theta}A_\mu
  + i\bar{\theta}\bar{\theta}\theta\lambda_0
  - i\theta\theta\bar{\theta}\bar{\lambda}_0
  + \frac{1}{2}\theta\theta\bar{\theta}\bar{\theta}D,
\end{equation}
where $F_i$ and $D$ are auxiliary fields. The Wess-Zumino gauge is adapted for $V$.
In terms of these superfields, the 10D SYM action~\eqref{eq:10Daction} can be rewritten as~\cite{Arkani-Hamed:2001vvu}
\begin{equation}
  S
  =
  \int \dd^{10} X
  \sqrt{-G}
  \left[
    \int \dd^4 \theta\
    \mathcal{K}
    +
    \left\{
    \int \dd^2 \theta\
    \left(
    \frac{1}{g^2}\mathcal{W}^{\alpha}
    \mathcal{W}_{\alpha}
    +
    \mathcal{W}
    \right)
    +
    \textrm{h.c.}
    \right\}
    \right],
  \label{eq:10Dssaction}
\end{equation}
where
\begin{eqnarray}
  \mathcal{K}
  & = &
  \frac{2}{g^2}
  h^{\bar{i} j}
  \Tr
  \left[
    \left\{
    (\sqrt{2}\bar{\partial}_{\bar{i}} + \bar{\Phi}_{\bar{i}})e^{-V}
    \right\}
    \left\{
    (-\sqrt{2}\partial_j + \Phi_j)e^{V}
    \right\}
    +
    (\bar{\partial}_{\bar{i}}e^{-V})
    (\partial_j e^{V})
    \right],
  \label{eq:K}
  \\
  \mathcal{W}
  & = &
  \frac{1}{2\sqrt{2} \pi^3 g^2 R_1 R_2 R_3}
  \epsilon^{ijk}
  \Tr
  \left[
    \sqrt{2}\Phi_{i}
    \left(
    \partial_{j}\Phi_{k}
    -\frac{1}{3\sqrt{2}}[\Phi_{j},\Phi_{k}]
    \right)
    \right]
  \label{eq:W}
\end{eqnarray}
and $\mathcal{W}^{\alpha}$ is the gauge field-strength superfield.
Note that $\epsilon^{ijk}$ is the totally antisymmetric tensor with $\varepsilon^{123}=1$.

In the following, we consider 10D $U(N)$ SYM and introduce the background magnetic fluxes and Wilson-lines
\begin{equation}
  \lrangle{a_i}
  =
  \frac{\pi}{\Im \tau_i}
  (
  M^{(i)}\bar{z}^{\bar{i}}
  +
  \bar{\zeta}^{(i)}
  ),
  \label{eq:background}
\end{equation}
so that they satisfy $\ev*{D}=\ev*{F_i}=0$ as well as $\ev*{A_{\mu}}=0$ to preserve 4D $\mathcal{N}=1$ supersymmetry and Lorentz invariance of $R^{1,3}$.
Here, $M^{(i)}$ and $\zeta^{(i)}$ are $N\times N$ diagonal matrices whose elements $M_a^{(i)}$ and $\zeta_a^{(i)}$ ($a=1,\ldots,n$) correspond to the magnetic fluxes and Wilson-lines on the $i$-th torus, respectively,
\begin{equation}
  M^{(i)}
  =
  \begin{pmatrix}
    M_1^{(i)}\mathbf{1}_{N_1} & 0                         & \cdots & 0                         \\
    0                         & M_2^{(i)}\mathbf{1}_{N_2} & \cdots & 0                         \\
    \vdots                    & \vdots                    & \ddots & \vdots                    \\
    0                         & 0                         & \cdots & M_n^{(i)}\mathbf{1}_{N_n}
  \end{pmatrix},
  \quad
  \zeta_{(i)}
  =
  \begin{pmatrix}
    \zeta^{(i)}_{1}\mathbf{1}_{N_1} & 0                               & \cdots & 0                               \\
    0                               & \zeta^{(i)}_{2}\mathbf{1}_{N_2} & \cdots & 0                               \\
    \vdots                          & \vdots                          & \ddots & \vdots                          \\
    0                               & 0                               & \cdots & \zeta^{(i)}_{n}\mathbf{1}_{N_n}
  \end{pmatrix},
\end{equation}
where $M_a^{(i)}\in Z$ due to the Dirac's quantization condition, $\mathbf{1}_{N_a}$ is the $N_a \times N_a$ unit matrix, and $\sum_{a=1}^nN_a=N$.
In this case, the background~\eqref{eq:background} breaks $U(N)$ gauge symmetry down to $U(N_1)\times U(N_2)\times \cdots \times U(N_n)$.
We also find that the F-flat condition, $\langle F_i\rangle=0$, is trivially satisfied, while the D-flat one,
\begin{equation}
  \lrangle{D}
  =
  -h^{\bar{i} j}
  \left(
  \bar{\partial}_{\bar{i}}\lrangle{a_j}
  +
  \partial_j \lrangle{\bar{a}_{\bar{i}}}
  +
  \frac{1}{2}\left[ \lrangle{\bar{a}_{\bar{i}}} , \lrangle{a_j} \right]
  \right)
  =
  -\sum_{i=1,2,3}\frac{2\pi M^{(i)}}{\cA^{(i)}}
  =
  0, \label{eq:D-flat}
\end{equation}
gives a constraint on the magnetic fluxes $M^{(i)}$.

Substituting $\Phi_i \rightarrow \langle \Phi_{i} \rangle + \Phi_i$, Eqs.~\eqref{eq:K} and \eqref{eq:W} respectively become
\begin{align}
  \mathcal{K}
   & =
  \frac{2}{g^2} h^{\bar{i} j}
  \Tr
  \left[
    \bar{\Phi}_{\bar{i}} \Phi_j
    +
    \sqrt{2}
    \left(
    \bar{\partial}_{\bar{i}}\langle\Phi_j\rangle
    +
    \partial_j \langle\bar{\Phi}_{\bar{i}}\rangle
    \right)V
    \right.
    \nonumber
    \\
    & \qquad
    +
    \sqrt{2}
    \left\{
    \left( \bar{\partial}_{\bar{i}}\Phi_j + \frac{1}{\sqrt{2}}[\lrangle{\bar{\Phi}_{\bar{i}}} ,\Phi_j] \right)
    +
    \left( \partial_j\bar{\Phi}_{\bar{i}}+\frac{1}{\sqrt{2}}[\lrangle{\Phi_j} ,\bar{\Phi}_{\bar{i}}] \right)
    \right\}
    V
      +
      [\bar{\Phi}_{\bar{i}},\Phi_j]V
    \nonumber
    \\
    & \qquad
    \left.
    +
    \left(
    \bar{\partial}_{\bar{i}}V
    +
    \frac{1}{\sqrt{2}}
    [\lrangle{\bar{\Phi}_{\bar{i}}} ,V]
    +
    \frac{1}{\sqrt{2}}
    [\bar{\Phi}_{\bar{i}},V]
    \right)
    \left(
    \partial_j V
    -
    \frac{1}{\sqrt{2}}
    [\lrangle{\Phi_j} ,V]
    -
    \frac{1}{\sqrt{2}}
    [\Phi_j,V]
    \right)
    \right],
  \\
  \mathcal{W}
   & =
  \frac{1}{2\pi^3 g^2 R_1 R_2 R_3}
  \epsilon^{ijk}
  \Tr
  \left[
    \left(
    \partial_i \Phi_j
    -
    \frac{1}{\sqrt{2}}
    [\lrangle{\Phi_i} ,\Phi_j]
    \right)
    \Phi_k
    -
    \frac{\sqrt{2}}{3}
    \Phi_i \Phi_j \Phi_k
    \right].
  \label{eq:fluctuation-superpotential}
\end{align}

\subsection{Zero-mode profiles on \texorpdfstring{$R^{1,3} \times (T^2)^3$}{R13*T23}}
To derive coupling constants, especially Yukawa couplings, in 4D effective theory below the compactification scale, we perform the dimensional reduction neglecting massive modes on $(T^2)^3$ and performing integrals over the coordinates $z^i$ and $\bar{z}^{\bar{i}}$ in the 10D action~\eqref{eq:10Dssaction}.
For such a purpose, we derive zero-mode wavefunctions for the fluctuations $\Phi_i$ around the background $\langle \Phi_i \rangle=\langle a_i \rangle/\sqrt{2}$, since quarks, leptons and Higgs bosons are carried by the chiral superfields $\Phi_i$.

The mode expansion is described as
\begin{equation}
  \Phi_i(x^{\mu},z^j,\bar{z}^{\bar{j}})
  =
  \sum_{\bm{n}}
  \left(
  \prod_{j=1}^{3}
  \phi_{i}^{(j),n_j}(z^j,\bar{z}^{\bar{j}})
  \right)
  \psi_{i}^{\bm{n}}(x^\mu),
  \label{eq:KKdecomp}
\end{equation}
where $\phi_{i}^{(j),n_j}(z_j,\bar{z}^{\bar{j}})$ and $\psi_{i}^{\bm{n}}(x^\mu)$ denote the wavefunctions on the $j$-th torus and the corresponding 4D chiral superfields on $R^{1,3}$, respectively, and $\bm{n}=(n_1,n_2,n_3)$ is a set of the Kaluza-Klein mode numbers $n_j = 0,1,\cdots$ on each torus.
Since only zero modes $\bm{n}=(0,0,0)$ are relevant to the 4D effective theory,
we will focus on them and omit the mode index $\bm{n}$ in the following.
Substituting \eqref{eq:KKdecomp} into \eqref{eq:10Dssaction} and \rm observing the bilinear terms, we obtain the zero-mode equations on each torus as
\begin{align}
  \left[
    \bar{\partial}_{\bar{i}}
    +
    \frac{\pi}{2\Im \tau_i}
    (M_{ab}^{(i)}z^i+\zeta_{ab}^{(i)})
    \right]\phi_{j,ab}^{(i)}
   & = 0
  \quad \textrm{for } i=j, \label{eq:zero-mode eq. i = j}
  \\
  \left[
    \partial_{i}
    -
    \frac{\pi}{2\Im \tau_i}
    (M_{ab}^{(i)}\bar{z}^{\bar{i}}+\bar{\zeta}_{ab}^{(i)})
    \right]\phi_{j,ab}^{(i)}
   & = 0
  \quad \textrm{for } i\neq j, \label{eq:zero-mode eq. i neq j}
\end{align}
where $M_{ab}^{(i)}\equiv M_a^{(i)}-M_b^{(i)}$ and $\zeta_{ab}^{(i)}\equiv \zeta_a^{(i)}-\zeta_b^{(i)}$.
Note that the indices $a,b$ label the gauge components of the bi-fundamental representation $(N_a,\bar{N}_b)$ under $U(N_a)\times U(N_b)$ for $a \ne b$ and those of the adjoint representations under $U(N_a)$ for $a=b$.

The solutions of Eqs.~\eqref{eq:zero-mode eq. i = j} and \eqref{eq:zero-mode eq. i neq j} are as follows~\cite{Cremades:2004wa}.
For $i=j$, the differential equation~\eqref{eq:zero-mode eq. i = j} has $M_{ab}^{(i)}$ normalizable solutions for $M_{ab}^{(i)}>0$
while no normalizable solution for $M_{ab}^{(i)}<0$, i.e.,
\begin{equation}
  \phi_{i,ab}^{(i)}
  =
  \begin{cases}
    \Theta^{I_{i,ab}^{(i)},M_{ab}^{(i)}}(\tilde{z}^{i}) & (M_{ab}^{(i)}>0) \\
    \pi R_i (\cA^{(i)})^{-\frac{1}{2}} g^{\frac{1}{3}}  & (M_{ab}^{(i)}=0) \\
    0                                                   & (M_{ab}^{(i)}<0)
  \end{cases},
  \label{eq:wf_phi_i}
\end{equation}
where $\tilde{z}^{i} \equiv z^i + \frac{\zeta_{ab}^{(i)}}{|M_{ab}^{(i)}|}$ and the index $I_{i,ab}^{(i)}=0,1,\cdots,M_{ab}^{(i)}-1$ labels the $M_{ab}^{(i)}$ independent solutions.
Here $\Theta^{I,M}(z)$ is defined by Jacobi theta function $\vartheta$
\begin{equation}
  \vartheta
  \begin{bmatrix}
    a \\
    b
  \end{bmatrix}
  (\nu,\tau)
  =
  \sum_{l\in Z}
  e^{\pi i(a+l)^2 \tau}
  e^{2\pi i(a+l)(\nu+b)}
\end{equation}
as
\begin{align}
  \Theta^{I,M}(z)
   & =
  \cN_{M}
  e^{\pi iMz\Im z/\Im\tau}
  \vartheta
  \begin{bmatrix}
    I/M \\
    0
  \end{bmatrix}
  (Mz,M\tau)
  ,\quad
  I=0, 1, \cdots, |M|-1.
  \label{eq:wavefunc-solution}
\end{align}
To have the 4D effective K\"ahler metrics in the canonical form, the normalization factor $\cN_M$ is determined by the condition
\begin{equation}
  2g^{-\frac{2}{3}}
  \int \dd z^{i}\dd \bar{z}^{\bar{i}}\ \sqrt{\det g^{(i)}} h^{i\bar{i}} \Theta^{I,M}(z^i)(\Theta^{J,M}(z^i))^{*}= \delta^{IJ}.
  \label{eq:normalization i eq j}
\end{equation}
for $i=j$. Solution for $M_{ab}^{(i)} = 0$ satisfies the same normalization condition.

On the other hand, for $i\neq j$, the differential equation~\eqref{eq:zero-mode eq. i neq j} has no normalizable solution for $M_{ab}^{(i)}>0$:
\begin{equation}
  \phi_{j,ab}^{(i)}
  =
  \begin{cases}
    0                                                            & (M_{ab}^{(i)}>0) \\
    (\cA^{(i)})^{-\frac{1}{2}} g^{\frac{1}{3}}                   & (M_{ab}^{(i)}=0) \\
    (\Theta^{I_{j,ab}^{(i)},|M_{ab}^{(i)}|}(\tilde{z}^{i}))^\ast & (M_{ab}^{(i)}<0)
  \end{cases},
  \label{eq:wf_phi_j}
\end{equation}
where the index $I_{j,ab}^{(i)}=0,1,\cdots,|M_{ab}^{(i)}|-1$ labels the $|M_{ab}^{(i)}|$ independent solutions. To make the 4D effective K\"ahler metrics canonical, the normalization for $i \neq j$ is determined by the condition
\begin{equation}
  g^{-\frac{2}{3}}
  \int \dd z^{i}\dd \bar{z}^{\bar{i}}\ \sqrt{\det g^{(i)}} \Theta^{I,M}(z^i)(\Theta^{J,M}(z^i))^{*}= \delta^{IJ},
  \label{eq:normalization i neq j}
\end{equation}
which is different from the one~(\ref{eq:normalization i eq j}) for $i = j$.

The important point here for a particle model building is that the normalizable solutions exist for either $(\phi_{j,ab}^{(i)})$ or $(\phi_{j,ba}^{(i)})$ depending on the sign of $M_{ab}^{(i)}=M_a^{(i)}-M_b^{(i)}$ and the degeneracy (determining the generation number in the model building) is given by the absolute value $|M_{ab}^{(i)}|$.

\subsection{Zero-mode profiles on \texorpdfstring{$R^{1,3} \times (T^2)^3/(Z_2 \times Z'_2)$}{R13*T23/Z2*Z2'}}
As we saw just above, the number of degenerate zero-modes on $R^{1,3} \times (T^2)^3$ is determined by the difference of magnetic fluxes they feel on each torus.
The situation changes on orbifolds~\cite{Abe:2016jsb,Abe:2017gye} and, in this paper, we consider $R^{1,3} \times (T^2)^3/(Z_2 \times Z'_2)$ where two kinds of orbifold projections are introduced.
The first $Z_2$ acts on the first and second tori as $(z^1,z^2,z^3)\to(-z^1,-z^2,z^3)$, while the second $Z'_2$ acts on the second and third tori as $(z^1,z^2,z^3)\to(z^1,-z^2,-z^3)$.
In the following, we will see how the zero mode profiles are modified from those on $R^{1,3} \times (T^2)^3$.

The transformation of $\Phi_i$ under $Z_2$ is given by
\begin{align}
  \Phi_1(-z^1,-z^2,z^3)
   & =
  -P \Phi_1(z^1,z^2,z^3) P^{-1},
  \nonumber
  \\
  \Phi_2(-z^1,-z^2,z^3)
   & =
  -P \Phi_2(z^1,z^2,z^3) P^{-1},
  \label{eq:PhiZ2}
  \\
  \Phi_3(-z^1,-z^2,z^3)
   & =
  +P \Phi_3(z^1,z^2,z^3) P^{-1}
  \nonumber
\end{align}
where $P$ is a projection operator satisfying $P^2=\mathbf{1}_N$. Note that the coordinates $x^\mu$ of $R^{1,3}$ are omitted in the arguments.
The overall sign in the right-hand side of each equation are determined by the fact that $\Phi_i$ includes $A_i$ whose transformation is the same as $z_i$ (up to $P$ acting on the gauge space).
In this paper, for simplicity, we only consider the case that $P$ is a diagonal matrix,
\begin{equation}
  P=
  \begin{pmatrix}
    \eta_1 \mathbf{1}_{N_1} & 0                       & \cdots & 0                       \\
    0                       & \eta_2 \mathbf{1}_{N_2} & \cdots & 0                       \\
    \vdots                  & \vdots                  & \ddots & \vdots                  \\
    0                       & 0                       & \cdots & \eta_n \mathbf{1}_{N_n}
  \end{pmatrix}
\end{equation}
with $\eta_a=\pm 1$.
Then the transformation of the $ab$-component $(\Phi_i)_{ab} \equiv \Phi_{i,ab}$  becomes as
\begin{align}
  \Phi_{1,ab}(-z^1,-z^2,z^3)
   & =
  -\eta_a \eta_b \Phi_{1,ab}(z^1,z^2,z^3),
  \nonumber
  \\
  \Phi_{2,ab}(-z^1,-z^2,z^3)
   & =
  -\eta_a \eta_b \Phi_{2,ab}(z^1,z^2,z^3),
  \nonumber
  \\
  \Phi_{3,ab}(-z^1,-z^2,z^3)
   & =
  +\eta_a \eta_b \Phi_{3,ab}(z^1,z^2,z^3).
  \nonumber
\end{align}
Since $\Phi_{i}$ is expanded like \eqref{eq:KKdecomp}, we can extract the transformations of zero-mode wavefunctions $(\phi_{i}^{(j)}(z^j))_{ab} \equiv \phi_{i,ab}^{(j)}(z^j)$ as
\begin{equation}
  \phi_{i,ab}^{(1)}(-z^1) \phi_{i,ab}^{(2)}(-z^2)
  =
  -\eta_a \eta_b \phi_{i,ab}^{(1)}(z^1)\phi_{i,ab}^{(2)}(z^2)
  =
  (-1)^{p_{i,ab}^{(1)}+p_{i,ab}^{(2)}} \phi_{i,ab}^{(1)}(z^1)\phi_{i,ab}^{(2)}(z^2)
  \label{eq:phi12Z2}
\end{equation}
for $i=1,2$, while
\begin{equation}
  \phi_{3,ab}^{(1)}(-z^1) \phi_{3,ab}^{(2)}(-z^2)
  =
  +\eta_a \eta_b \phi_{3,ab}^{(1)}(z^1)\phi_{3,ab}^{(2)}(z^2)
  =
  (-1)^{p_{3,ab}^{(1)}+p_{3,ab}^{(2)}} \phi_{3,ab}^{(1)}(z^1)\phi_{3,ab}^{(2)}(z^2)
  \label{eq:phi3Z2}
\end{equation}
for $i=3$, where $p_{i,ab}^{(j)}=0,1$ ($j=1,2$) parametrizes the parity $(-1)^{p_{i,ab}^{(j)}}$ of the wavefunction $\phi_{i,ab}^{(j)}(z^j)$ under $Z_2$.
From Eqs.~\eqref{eq:phi12Z2} and \eqref{eq:phi3Z2}, we find that possible values of $(p_{i,ab}^{(1)},p_{i,ab}^{(2)})$ are limited by the sign of $\eta_a \eta_b$, while the wavefunctions $\phi_{i,ab}^{(3)}(z^3)$ are not restricted by $Z_2$ projection.

Similarly, the transformation of $\Phi_i$ under $Z'_2$ is given by
\begin{align}
  \Phi_1(z^1,-z^2,-z^3)
   & =
  +P^{\prime} \Phi_1(z^1,z^2,z^3) P^{\prime -1},
  \nonumber
  \\
  \Phi_2(z^1,-z^2,-z^3)
   & =
  -P^{\prime} \Phi_2(z^1,z^2,z^3) P^{\prime -1},
  \label{eq:PhiZ2p}
  \\
  \Phi_3(z^1,-z^2,-z^3)
   & =
  -P^{\prime} \Phi_3(z^1,z^2,z^3) P^{\prime -1}.
  \nonumber
\end{align}
We only consider the case that $P^{\prime}$ is a diagonal matrix of the same structure as $P$,
\begin{equation}
  P^{\prime}=
  \begin{pmatrix}
    \eta_1^{\prime} \mathbf{1}_{N_1} & 0                                & \cdots & 0                                \\
    0                                & \eta_2^{\prime} \mathbf{1}_{N_2} & \cdots & 0                                \\
    \vdots                           & \vdots                           & \ddots & \vdots                           \\
    0                                & 0                                & \cdots & \eta_n^{\prime} \mathbf{1}_{N_n}
  \end{pmatrix}
\end{equation}
with $\eta_a^{\prime}=\pm 1$, resulting
\begin{align}
  \Phi_{1,ab}(z^1,-z^2,-z^3)
   & =
  +\eta_a^{\prime} \eta_b^{\prime} \Phi_{1,ab}(z^1,z^2,z^3),
  \nonumber
  \\
  \Phi_{2,ab}(z^1,-z^2,-z^3)
   & =
  -\eta_a^{\prime} \eta_b^{\prime} \Phi_{2,ab}(z^1,z^2,z^3),
  \nonumber
  \\
  \Phi_{3,ab}(z^1,-z^2,-z^3)
   & =
  -\eta_a^{\prime} \eta_b^{\prime} \Phi_{3,ab}(z^1,z^2,z^3),
  \nonumber
\end{align}
and then,
\begin{equation}
  \phi_{1,ab}^{(2)}(-z^2) \phi_{1,ab}^{(3)}(-z^3)
  =
  +\eta_a^{\prime} \eta_b^{\prime} \phi_{1,ab}^{(2)}(z^2) \phi_{1,ab}^{(3)}(z^3)
  =
  (-1)^{p_{1,ab}^{\prime(2)}+p_{1,ab}^{\prime(3)}} \phi_{1,ab}^{(2)}(z^2)\phi_{1,ab}^{(3)}(z^3)
  \label{eq:phi1Z2p}
\end{equation}
for $i=1$, while
\begin{equation}
  \phi_{i,ab}^{(2)}(-z^2) \phi_{i,ab}^{(3)}(-z^3)
  =
  -\eta_a^{\prime} \eta_b^{\prime} \phi_{i,ab}^{(2)}(z^2) \phi_{i,ab}^{(3)}(z^3)
  =
  (-1)^{p_{i,ab}^{\prime(2)}+p_{i,ab}^{\prime(3)}} \phi_{i,ab}^{(2)}(z^2)\phi_{i,ab}^{(3)}(z^3)
  \label{eq:phi23Z2p}
\end{equation}
for $i=2,3$, where $p_{i,ab}^{\prime(j)}=0,1$ ($j=2,3$) parametrizes the parity $(-1)^{p_{i,ab}^{\prime(j)}}$ of the wavefunction $\phi_{i,ab}^{(j)}(z^j)$ under $Z'_2$.
From Eqs.~\eqref{eq:phi23Z2p} and \eqref{eq:phi1Z2p}, we find that possible values of $(p_{i,ab}^{\prime(2)},p_{i,ab}^{\prime(3)})$ are limited by the sign of $\eta_a^{\prime} \eta_b^{\prime}$, while the wavefunctions $\phi_{i,ab}^{(1)}(z^1)$ are not restricted by $Z_2$ projection.

Finally, the transformation of $\Phi_i$ under $Z_2 \times Z'_2$ is given by combining Eq.~\eqref{eq:PhiZ2} and \eqref{eq:PhiZ2p}.
Gathering the above separate arguments on $Z_2$ and $Z'_2$, we find that the parity of the wavefunction $\phi_{i,ab}^{(j)}(z^j)$ under $Z_2 \times Z'_2$ can be parameterized as $(-1)^{p_{i,ab}^{(j)}}$ ($j=1,2,3$) with $p_{i,ab}^{(3)} \equiv p_{i,ab}^{\prime(3)}$.
It is important to notice that, since the second torus is involved in both $Z_2$ and $Z'_2$, the parity of $\phi_{i,ab}^{(2)}(z^2)$ should be consistent, i.e., the condition $p_{i,ab}^{(2)}=p_{i,ab}^{\prime(2)}$ must be imposed.
Taking all into account, Table~\ref{tab:parity} summarizes the possible values of $(p_{i,ab}^{(1)},p_{i,ab}^{(2)},p_{i,ab}^{(3)})$ for each value of $\eta_a \eta_b$ and $\eta_a' \eta_b'$.

\begin{table}[t]
  \centering
  \begin{tabular}{cccc}\hline
    $\eta_a\eta_b$ & $\eta_a'\eta_b'$ & $\Phi_{i,ab}$ & $(p_{i,ab}^{(1)},p_{i,ab}^{(2)},p_{i,ab}^{(3)})$ \\ \hline
                   &                  & $\Phi_{1,ab}$ & $(1,0,0), (0,1,1)$                               \\
    $+1$           & $+1$             & $\Phi_{2,ab}$ & $(0,1,0), (1,0,1)$                               \\
                   &                  & $\Phi_{3,ab}$ & $(0,0,1), (1,1,0)$                               \\ \hline
                   &                  & $\Phi_{1,ab}$ & $(1,0,1), (0,1,0)$                               \\
    $+1$           & $-1$             & $\Phi_{2,ab}$ & $(0,1,1), (1,0,0)$                               \\
                   &                  & $\Phi_{3,ab}$ & $(0,0,0), (1,1,1)$                               \\ \hline
                   &                  & $\Phi_{1,ab}$ & $(1,1,1), (0,0,0)$                               \\
    $-1$           & $+1$             & $\Phi_{2,ab}$ & $(0,0,1), (1,1,0)$                               \\
                   &                  & $\Phi_{3,ab}$ & $(0,1,0), (1,0,1)$                               \\ \hline
                   &                  & $\Phi_{1,ab}$ & $(1,1,0), (0,0,1)$                               \\
    $-1$           & $-1$             & $\Phi_{2,ab}$ & $(0,0,0), (1,1,1)$                               \\
                   &                  & $\Phi_{3,ab}$ & $(0,1,1), (1,0,0)$                               \\ \hline
  \end{tabular}
  \caption{Possible values of $(p_{i,ab}^{(1)},p_{i,ab}^{(2)},p_{i,ab}^{(3)})$ for each values of $\eta_a \eta_b$ and $\eta_a' \eta_b'$.}
  \label{tab:parity}
\end{table}

Now we consider the zero-mode wavefunctions on each torus with the $Z_2 \times Z'_2$ orbifold projection.
On such an orbifold, (differences of) Wilson-line parameters $\zeta_{ab}^{(i)}$ are discretized for each torus, allowing only four values $\zeta_{ab}^{(i)}=0,\frac{1}{2},\frac{\tau_i}{2}$ and $\frac{1+\tau_i}{2}$. For the vanishing one, $\zeta_{ab}^{(i)}=0$, from Eqs.~\eqref{eq:wf_phi_i} and \eqref{eq:wf_phi_j}, we find that normalizable zero-mode wavefunctions on the $i$-th torus $(\phi_{j}^{(i)}(z^i))_{ab}$ for $j=i$ and $j \ne i$ are given by $\Theta^{I_{j,ab}^{(i)},|M_{ab}^{(i)}|}(z^i)$ defined in Eq.~\eqref{eq:wavefunc-solution} and its complex conjugate for $M_{ab}^{(i)}>0$ and $M_{ab}^{(i)}<0$, respectively.
Using a shorthand notation $\phi^I(z) \equiv \Theta^{I,|M|}(z)$, we find $\phi^I(-z)=\phi^{|M|-I}(z)$ from the property of Jacobi theta function $\vartheta$. Among the linear combinations of $\phi^I(z)$ ($I=0,1,\ldots,|M|-1$), the even and odd functions $\phi_{\pm}^{\tilde{I}}(z)=\pm \phi_{\pm}^{\tilde{I}}(-z)$ are formally expressed as $\phi_{+}^{\tilde{I}}(z)=\frac{1}{\sqrt{2}}(\phi^{\tilde{I}}(z)+\phi^{|M|-\tilde{I}}(z))$ and $\phi_{-}^{\tilde{I}}(z)=\frac{1}{\sqrt{2}}(\phi^{\tilde{I}}(z)-\phi^{|M|-\tilde{I}}(z))$, where $\tilde{I}=0,1,\ldots,N_+-1$ for $\phi_+^{\tilde{I}}$ and $\tilde{I}=0,1,\ldots,N_--1$ for $\phi_-^{\tilde{I}}$. The numbers of even and odd zero-modes $N_+$ and $N_-$, respectively, are given as $N_+=\frac{|M|}{2}+1$ and $N_-=\frac{|M|}{2}-1$ if $|M|$ is a positive even number, while $N_+=\frac{|M|+1}{2}$ and $N_-=\frac{|M|-1}{2}$ if $|M|$
is a positive odd number.
Using the floor function $[x]$, they are expressed as $(N_+,N_-)=([\frac{|M|}{2}]+1,[\frac{|M|-1}{2}])$ for a positive integer $|M|$~\cite{Abe:2013bca,Abe:2015yva}.

In a similar way, we can find the even and odd zero-mode functions for all values of $\zeta_{ab}^{(i)}$ by rearranging the wavefunctions $\Theta^{I_{i,ab}^{(i)},M_{ab}^{(i)}}(z^{i+})$ or $(\Theta^{I_{j,ab}^{(i)},|M_{ab}^{(i)}|}(z^{i-}))^\ast$ in Eq.~\eqref{eq:wf_phi_i} or \eqref{eq:wf_phi_j} to the linear combinations
$\tilde{\Theta}_{M_{ab}^{(i)};(-1)^{p_{i,ab}^{(i)}}}^{(\tilde{I}_{i,ab}^{(i)}+\alpha_{ab}^{(i)},\beta_{ab}^{(i)})}(z^i,\tau_i)$
or
$\Big(\tilde{\Theta}_{M_{ab}^{(i)};(-1)^{p_{j,ab}^{(i)}}}^{(\tilde{I}_{j,ab}^{(i)}-\alpha_{ab}^{(i)},-\beta_{ab}^{(i)})}(z^i,\tau_i)\Big)^\ast$, respectively,
where $\tilde{I}_{j,ab}^{(i)}=0,1,\ldots,N_{j,ab;\pm}^{(i)}-1$ for $(-1)^{p_{j,ab}^{(i)}}=\pm 1$, and
\begin{eqnarray}
  \tilde{\Theta}_{M;(-1)^p}^{(\tilde{I}+\alpha,\beta)}(z,\tau)
  &=&
  \sum_{I=0}^{|M|-1}
  \Delta_{\tilde{I},I}^{M,\alpha,\beta;(-1)^p}
  \Theta_{M}^{(I+\alpha,\beta)}(z,\tau)
  \label{eq:orbifold-wavefunc}
\end{eqnarray}
expresses the linear combination of
\begin{align}
  \Theta_{M}^{(I+\alpha,\beta)}(z,\tau)
   & =
  \cN_M
  e^{\pi i M z \Im z/\Im \tau}
  \vartheta
  \begin{bmatrix}
    \frac{I+\alpha}{M} \\
    -\beta
  \end{bmatrix}
  (Mz,M\tau)
\end{align}
with the coefficients
\begin{align}
  \Delta_{\tilde{I},I}^{M,\alpha,\beta;(-1)^p}
   & =
  \begin{cases}
    \delta_{\tilde{I}+\delta_{\alpha,0}\delta_{p,1},I}
     & \hspace{-220pt} (\tilde{I}+\delta_{\alpha,0}\delta_{p,1}+\alpha=\frac{M}{2},2\beta=p) \\
    \delta_{\delta_{\alpha,0}\delta_{p,1},I}
     & \hspace{-220pt} (\tilde{I}+\delta_{\alpha,0}\delta_{p,1}+\alpha=0, p=0)               \\
    \frac{1}{\sqrt{2}}
    \left(
    \delta_{\tilde{I}+\delta_{\alpha,0}\delta_{p,1},I}
    + (-1)^p
    e^{-2\pi i\frac{2\beta}{M}(\tilde{I}+\alpha)}
    \delta_{M-2\alpha-\tilde{I}-\delta_{\alpha,0}\delta_{p,1},I}
    \right)
     & (\text{others})
  \end{cases}
\end{align}
determined by the normalization condition~\eqref{eq:normalization i eq j} and \eqref{eq:normalization i neq j}.
Here, the Wilson-line degrees of freedom $\zeta_{ab}^{(i)}=0,\frac{1}{2},\frac{\tau_i}{2}$ and $\frac{1+\tau_i}{2}$ are reparameterized by $(\alpha_{ab}^{(i)},\beta_{ab}^{(i)})=(0,0),(0,\frac{1}{2}),(\frac{1}{2},0)$ and $(\frac{1}{2},\frac{1}{2})$, respectively.

The numbers of even and odd zero-modes of $\Phi_{j,ab}$ on the $i$-th torus, $N_{j,ab;+}^{(i)}$ and $N_{j,ab;-}^{(i)}$, are identified as those of the above independent linear combinations for $p_{j,ab}^{(i)}=0$ and $1$, respectively.
In the case $i=j$ with $M_{ab}^{(i)}>0$ and $i \ne j$ with $M_{ab}^{(i)}<0$, they are expressed as~\cite{Abe:2015yva}
$(N_{j,ab;+}^{(i)},N_{j,ab;-}^{(i)})
  =([\frac{|M_{ab}^{(i)}|}{2}]+1,[\frac{|M_{ab}^{(i)}|-1}{2}])$,
$([\frac{|M_{ab}^{(i)}|}{2}+1],[\frac{|M_{ab}^{(i)}|}{2}])$,
$([\frac{|M_{ab}^{(i)}|}{2}+1],[\frac{|M_{ab}^{(i)}|}{2}])$ and
$([\frac{|M_{ab}^{(i)}|}{2}],[\frac{|M_{ab}^{(i)}|+1}{2}])$ for
$\zeta_{ab}^{(i)}=0,\frac{1}{2},\frac{\tau_i}{2}$ and $\frac{1+\tau_i}{2}$, respectively,
otherwise $(N_{j,ab;+}^{(i)},N_{j,ab;-}^{(i)})=(0,0)$, Those are summarized in Table~\ref{tab:num-of-zero-modes}.
Under the $Z_2 \times Z'_2$ orbifold projection, either even or odd modes in $\Phi_{j,ab}$ are projected out on the $i$-th torus depending on $p_{j,ab}^{(i)}=1$ or $0$, respectively, and the survived ones on all the three tori participate in the field contents of the 4D effective theory.

\begin{table}[t]
  \centering
  \begin{tabular}{c||c|ccccccccccccccc}\hline
    $\zeta_{ab}^{(i)}$                              & $|M_{ab}^{(i)}|$ & 0 & 1 & 2 & 3 & 4 & 5 & 6 & 7 & 8 & 9 & 10 & 11 & 12 & 13 & 14 \\ \hline \hline
    \multirow{2}{*}{0}                              & even             & 1 & 1 & 2 & 2 & 3 & 3 & 4 & 4 & 5 & 5 & 6  & 6  & 7  & 7  & 8  \\
                                                    & odd              & 0 & 0 & 0 & 1 & 1 & 2 & 2 & 3 & 3 & 4 & 4  & 5  & 5  & 6  & 6  \\ \hline
    \multirow{2}{*}{$\frac{1}{2},\frac{\tau_i}{2}$} & even             & 0 & 1 & 1 & 2 & 2 & 3 & 3 & 4 & 4 & 5 & 5  & 6  & 6  & 7  & 7  \\
                                                    & odd              & 0 & 0 & 1 & 1 & 2 & 2 & 3 & 3 & 4 & 4 & 5  & 5  & 6  & 6  & 7  \\ \hline
    \multirow{2}{*}{$\frac{\tau_i+1}{2}$}           & even             & 0 & 0 & 1 & 1 & 2 & 2 & 3 & 3 & 4 & 4 & 5  & 5  & 6  & 6  & 7  \\
                                                    & odd              & 0 & 1 & 1 & 2 & 2 & 3 & 3 & 4 & 4 & 5 & 5  & 6  & 6  & 7  & 7  \\ \hline
  \end{tabular}
  \caption{The numbers of even and odd zero-modes on the $i$-th torus.}
  \label{tab:num-of-zero-modes}
\end{table}

We would like to find the explicit forms of Yukawa coupling constants in the 4D effective theory,
those originate from $\epsilon^{ijk} \Tr [\Phi_i \Phi_j \Phi_k]$-terms
in the superpotential~\eqref{eq:fluctuation-superpotential}. Substituting the zero-mode wavefunctions and integrating it over the extra-dimensional coordinates, the effective (holomorphic) Yukawa coupling constants
$y_{\mathcal{I}_i\mathcal{J}_j\mathcal{K}_k} \propto \prod_{l=1}^3
  y^{(l)}_{\tilde{I}_{i,ab}^{(l)}\tilde{J}_{j,bc}^{(l)}\tilde{K}_{k,ca}^{(l)}}$ among (normalizable) zero-modes symbolically labeled by
$\mathcal{I}_i=(\tilde{I}_{i,ab}^{(1)},\tilde{I}_{i,ab}^{(2)},\tilde{I}_{i,ab}^{(3)})$,
$\mathcal{J}_j=(\tilde{J}_{j,bc}^{(1)},\tilde{J}_{j,bc}^{(2)},\tilde{J}_{j,bc}^{(3)})$ and
$\mathcal{K}_k=(\tilde{K}_{k,ca}^{(1)},\tilde{K}_{k,ca}^{(2)},\tilde{K}_{k,ca}^{(3)})$ for $i \ne j$, $j \ne k$ and $k \ne i$
are given by the product of overlap integrals,
\begin{align}
  y^{(l)}_{\tilde{I}_{i,ab}^{(l)}\tilde{J}_{j,bc}^{(l)}\tilde{K}_{k,ca}^{(l)}}
  =
  \int \dd^2 z_{l}\
  \Big( \tilde{\Theta}_{|M_{ab}^{(l)}|;(-1)^{p_{i,ab}^{(l)}}}^{(\tilde{I}_{i,ab}^{(l)}+\alpha_{ab}^{(l)},\beta_{ab}^{(l)})}(z^l,\tau_l) \Big)^\ast
  \Big( \tilde{\Theta}_{|M_{bc}^{(l)}|;(-1)^{p_{j,bc}^{(l)}}}^{(\tilde{J}_{j,bc}^{(l)}+\alpha_{bc}^{(l)},\beta_{bc}^{(l)})}(z^l,\tau_l) \Big)^\ast
  \tilde{\Theta}_{M_{ca}^{(l)};(-1)^{p_{k,ca}^{(l)}}}^{(\tilde{K}_{k,ca}^{(l)}+\alpha_{ca}^{(l)},\beta_{ca}^{(l)})}(z^l,\tau_l),
  \label{eq:overlap_integral}
\end{align}
for $l=1,2$ and $3$, where $M_{ab}^{(l)},M_{bc}^{(l)}<0$ and $M_{ca}^{(l)}>0$ for $l=k$ are assumed for the normalizability of wavefunctions in the integrand. Note that the normalization constants for the last $\tilde{\Theta}$
without complex conjugate are different from the other two $\tilde{\Theta}$ because of the difference between \eqref{eq:normalization i eq j} and \eqref{eq:normalization i neq j}. Here we denote them as $\cN_{M+}$ for $\tilde{\Theta}$ with the normalization \eqref{eq:normalization i eq j}, and $\cN_{M-}$ for $\tilde{\Theta}$ with the normalization \eqref{eq:normalization i neq j}.

To calculate this, we consider its complex conjugate $\lambda_{\tilde{I}\tilde{J}\tilde{K}}^{(i)} \equiv (y_{\tilde{I}\tilde{J}\tilde{K}}^{(i)})^*$. From a product rule of theta function, we find
\begin{eqnarray}
  \lefteqn{
    \tilde{\Theta}_{M_1;(-1)^{p_1}}^{(\tilde{I}+\alpha_1,\beta_1)}
    \tilde{\Theta}_{M_2;(-1)^{p_2}}^{(\tilde{J}+\alpha_2,\beta_2)}
  } \nonumber \\ &=&
  \left(
  \sum_{I=0}^{|M_1|-1}
  \Delta_{\tilde{I},I}^{M_1,\alpha_1,\beta_1;(-1)^{p_1}}
  \Theta_{M_1}^{(i+\alpha_1,\beta_1)}
  \right)
  \left(
  \sum_{J=0}^{|M_2|-1}
  \Delta_{\tilde{J},J}^{M_2,\alpha_2,\beta_2;(-1)^{p_2}}
  \Theta_{M_2}^{(J+\alpha_2,\beta_2)}
  \right)
  \nonumber
  \\
  &=&
  \sum_{I=0}^{|M_1|-1}
  \sum_{J=0}^{|M_2|-1}
  \Delta_{\tilde{I},I}^{M_1,\alpha_1,\beta_1;(-1)^{p_1}}
  \Delta_{\tilde{J},J}^{M_2,\alpha_2,\beta_2;(-1)^{p_2}}
  \frac{\cN_{M_1-}\cN_{M_2-}}{\cN_{M_3+}}
  \sum_{m\in Z_{M_3}}
  \Theta_{M_3}^{(\tilde{I}+\tilde{J}+M_1 m+\alpha_3,\beta_3)}
  \nonumber \\
  && \qquad
  \times
  \vartheta
  \begin{bmatrix}
    \frac{M_2(I+\alpha_1)-M_1(J+\alpha_2)+M_1 M_2 m}{M_1 M_2 M_3} \\
    0
  \end{bmatrix}
  \left(M_1\beta_2-M_2\beta_1,M_1M_2M_3\tau\right)
\end{eqnarray}
for $M_1+M_2-M_3=0$, and obtain the expression~\cite{Cremades:2004wa,Abe:2015yva}
\begin{align}
  \lambda_{\tilde{I}\tilde{J}\tilde{K}}^{(i)}
   & \equiv
  \int \dd^2 z\
  \tilde{\Theta}_{M_1;(-1)^{p_1}}^{(\tilde{I}+\alpha_1,\beta_1)}(z,\tau)
  \tilde{\Theta}_{M_2;(-1)^{p_2}}^{(\tilde{J}+\alpha_2,\beta_2)}(z,\tau)
  (\tilde{\Theta}_{M_3;(-1)^{p_3}}^{(\tilde{K}+\alpha_3,\beta_3)}(z,\tau))^{*}
  \nonumber
  \\
   & =
  \frac{\cN_{M_1-}\cN_{M_2-}}{\cN_{M_3+}}
  \sum_{I=0}^{|M_1|-1}\sum_{J=0}^{|M_2|-1}
  \Delta_{\tilde{I},I}^{M_1,\alpha_1,\beta_1;(-1)^{p_1}}
  \Delta_{\tilde{J},J}^{M_2,\alpha_2,\beta_2;(-1)^{p_2}}
  \nonumber
  \\
   & \qquad
  \times
  \sum_{m\in Z_{M_3}}
  \vartheta
  \begin{bmatrix}
    \frac{M_2(I+\alpha_1)-M_1(J+\alpha_2)+M_1 M_2 m}{M_1 M_2 M_3} \\
    0
  \end{bmatrix}
  \left(M_1\beta_2-M_2\beta_1,M_1M_2M_3\tau\right)
  \nonumber
  \\
   & \qquad
  \times
  \int_{T^2} \dd z^2\
  \Theta_{M_3}^{I+J+M_1 m+\alpha_3,\beta_3}(z,\tau)
  (\tilde{\Theta}_{M_3;(-1)^{p_3}}^{\tilde{K}+\alpha_3,\beta_3}(z,\tau))^{*}
  \nonumber
  \\
   & =
  \frac{\cN_{M_1-}\cN_{M_2-}}{\cN_{M_3+}} \frac{g^{\frac{2}{3}}}{4 \Im \tau_i}
  \sum_{I=0}^{|M_1|-1}\sum_{J=0}^{|M_2|-1}\sum_{K=0}^{|M_3|-1}
  \Delta_{\tilde{I},I}^{M_1,\alpha_1,\beta_1;(-1)^{p_1}}
  \Delta_{\tilde{J},J}^{M_2,\alpha_2,\beta_2;(-1)^{p_2}}
  (\Delta_{\tilde{K},K}^{M_3,\alpha_3,\beta_3;(-1)^{p_3}})^{*}
  \nonumber
  \\
   & \quad
  \times
  \sum_{m\in Z_{M_3}}
  \vartheta
  \begin{bmatrix}
    \frac{M_2(I+\alpha_1)-M_1(J+\alpha_2)+M_1 M_2 m}{M_1 M_2 M_3} \\
    0
  \end{bmatrix}
  \left(M_1\beta_2-M_2\beta_1,M_1M_2M_3\tau\right)
  \times
  \delta_{I+J+M_1 m,K+M_3 \ell},
  \label{eq:yukawa-orbifold}
\end{align}
where $\ell$ is an arbitrary integer, which is useful to evaluate (the complex conjugate of) the overlap integral~\eqref{eq:overlap_integral}. To obtain this, we used the normalization condition \eqref{eq:normalization i eq j} in the last equal sign.

\section{MSSM-like models from 7-brane configurations}
\label{sec:visible}
As stated in the introduction, MSSM-like models with semi-realistic flavor structures were built upon 10D magnetized SYM theory outlined in the previous section~\cite{Abe:2012fj}. However, it was argued that some exotic matters and non-Abelian gauge fields cannot be eliminated in the zero-mode spectra, with the magnetic fluxes satisfying 4D ${\cal N}=1$ supersymmetry conditions~\cite{Abe:2014vza}. If the conditions are not satisfied, some charged scalars become tachyonic indicating that the system is unstable.\footnote{Although the tachyonic field can be identified as the supersymmetric partner of neutrino in certain special cases~\cite{Abe:2016jsb}, such an instability usually causes phenomenologically/theoretically unwanted features like charge and/or color breaking.}
Therefore, it is desirable if the exotic matters can be somehow eliminated without breaking the supersymmetry conditions on the fluxes. For such a purpose, in this paper, we consider to realize similar MSSM-like models using lower-dimensional SYM theories embedded in 10D spacetime with magnetic fluxes on their world volumes. Since the non-vanishing flux always breaks the supersymmetry conditions in six-dimensions with a single torus, we consider 8D SYM theories located on 7-brane configurations spreading over two of three tori in 10D spacetime.

In the following, first, the 7-brane configuration is introduced as the important ingredient in our model building. It is remarkable that the wavefunction of charged field on one of three tori where 8D SYM localizes can be interpreted as one of the (infinite) eigenfunctions of the covariant derivative in 10D SYM with infinite magnetic fluxes on the torus~\cite{Cremades:2004wa,Abe:2015jqa}. Then, we will find that the chiral multiplets in MSSM with semi-realistic flavor structures can be obtained from such configurations, without any exotic ones except those for three generations of right-handed neutrino and extra generations of MSSM Higgs pairs in the zero-mode spectrum. The existence of such multiple pairs of Higgs doublets will play a role to make the flavor structure semi-realistic as in the original 10D models~\cite{Abe:2012fj,Abe:2014vza}.

\subsection{7-brane configurations derived from 10D magnetized SYM}
Referring phenomenological configurations of 7-branes in Ref.~\cite{Abe:2017gye}, we consider two kinds of 7-branes,  7$_A$ and 7$_B$, on which 8D SYM is located in each as shown in Table~\ref{tab:D7-brane-embedding}.
\begin{table}[t]
  \centering
  \begin{tabular}{cccc}\hline
          & $T^2$        & $T^2$        & $T^2$        \\ \hline
    7$_A$ & $\checkmark$ & $\times$     & $\checkmark$ \\
    7$_B$ & $\checkmark$ & $\checkmark$ & $\times$     \\ \hline
  \end{tabular}
  \caption{Embedding of two 7-branes, 7$_A$ and 7$_B$, into three tori. The check mark $\checkmark$ shows that the 7-brane extends in the corresponding $T^2$ while the cross mark $\times$ indicates that it does not.}
  \label{tab:D7-brane-embedding}
\end{table}
Each 8D SYM can be obtained from 10D SYM through the dimensional reduction in a usual way.
However, since some chiral matters in bi-fundamental representations in the original 10D models~\cite{Abe:2012fj,Abe:2014vza} will be charged under both the gauge groups on 7$_A$- and 7$_B$-branes and localized at the intersection points in the 7-brane framework, there are ambiguities to introduce them. If we just consider the embedding of multiple 8D SYM theories into a certain 7-brane configuration in 10D spacetime, we can not uniquely determine the numbers of such bi-fundamental representations charged under any two of them at the intersection points.
Therefore, in this paper, we don't put such bi-fundamental fields by hand, but derive them by introducing certain infinite magnetic fluxes~\cite{Cremades:2004wa,Abe:2015jqa} on one of three tori in the original 10D SYM, which makes the wavefunctions of target fields localize on the torus.

In fact, by evaluating the integral of $\Re z$ in the normalization conditions, we find~\cite{Cremades:2004wa}
\begin{align}
  \int_{T^2} \dd(\Re z) \Theta^{I,M} (\Theta^{J,M})^* =
   & \cN_M^2 \sum_n e^{-2\pi |M| \Im \tau (n+\frac{I}{M}+\frac{\Im z}{\Im \tau})} \delta^{IJ} \nonumber                   \\
  \xrightarrow{M \rightarrow \infty}
   & \frac{\cN_M^2}{\sqrt{2 \Im \tau |M|}} \sum_n \delta \left( n+\frac{I}{M}+\frac{\Im z}{\Im \tau} \right) \delta^{IJ},
  \label{aln:infinite-wavefunctions}
\end{align}
which means that the infinite numbers of localized zero-modes exist with the width of $1/|M|$ in the $\Im z$-direction.
If we pick up one $I$th mode and drop the others, it corresponds to the wavefunction that localizes on the torus. We identify this localized zero mode as the target bi-fundamental field. Note however that $\Theta^{I,M} (\Theta^{J,M})^*$ is rather a constant than the delta-function in $\Re z$-direction as also pointed out in Ref.~\cite{Cremades:2004wa}. We just replace this constant by the delta-function $\delta(\Re z)$ to obtain the bi-fundamental fields charged under 8D SYM theories on the 7-brane configurations, those localize at the intersection points. Such a replacement will be interpreted as the T-duality transformation in $\Re z$-direction in the D-brane picture. Certain D7-brane system can be formally obtained from D9-branes with infinite magnetic fluxes on one of three tori,\footnote{If we introduce infinite fluxes on two of three tori, we obtain D5-branes as shown in Ref.~\cite{Cremades:2004wa}.} by taking T-duality transformations twice in a single direction of the torus. Since the first transformation yields infinite D8-branes, we need to pick up one of them before the second transformation. Infinite numbers of localized wavefunctions~(\ref{aln:infinite-wavefunctions}) in our case would correspond to these D8-branes. Finally, we remark that the localized points will be determined by Wilson-lines associated with the infinite fluxes, although vanishing ones have been assumed in the above arguments.

Then, to realize the 7-brane configuration in Table~\ref{tab:D7-brane-embedding},
we introduce the positive infinite magnetic flux $H$ as
\begin{equation}
  M^{(1)}
  =
  \left(
  \begin{array}{c|c}
    * & 0 \\ \hline
    0 & *
  \end{array}
  \right)
  ,\quad
  M^{(2)}
  =
  \left(
  \begin{array}{c|c}
    H\mathbf{1}_{N_A} & 0 \\ \hline
    0                 & *
  \end{array}
  \right)
  ,\quad
  M^{(3)}
  =
  \left(
  \begin{array}{c|c}
    * & 0                 \\ \hline
    0 & H\mathbf{1}_{N_B}
  \end{array}
  \right).
  \label{eq:infiniteflux}
\end{equation}
The top-left and the bottom-right blocks of each matrix correspond to the elements of SYM gauge groups $U(N_A)$ and $U(N_B)$ on 7$_A$- and 7$_B$-branes, respectively.
The stars $*$ represent the finite magnetic fluxes in the diagonal components.
The remaining zero-mode fields under the chiral projections (\ref{eq:wf_phi_i}) and (\ref{eq:wf_phi_j}) triggered by the above fluxes are
\begin{equation}
  \Phi_1
  =
  \left(
  \begin{array}{c|c}
    \Phi_1^A & 0        \\ \hline
    0        & \Phi_1^B
  \end{array}
  \right)
  ,\quad
  \Phi_2
  =
  \left(
  \begin{array}{c|c}
    \tilde{\Phi}_2^A & \Phi_2^{AB} \\ \hline
    0                & \Phi_2^B
  \end{array}
  \right)
  ,\quad
  \Phi_3
  =
  \left(
  \begin{array}{c|c}
    \Phi_3^A    & 0                \\ \hline
    \Phi_3^{BA} & \tilde{\Phi}_3^B
  \end{array}
  \right),
  \label{eq:phiAB}
\end{equation}
where the single indices $A$ and $B$ show that the fields are in the adjoint representation of gauge groups on 7$_A$- and 7$_B$-brane, respectively, while doubled indices $AB$ ($BA$) indicate that the fields are in the bi-fundamental, namely, fundamental and anti-fundamental representation of SYM gauge groups on 7$_A$- and 7$_B$-branes (7$_B$- and 7$_A$-branes), respectively. The diagonal components of the tilded fields are so-called position moduli in the D-brane picture, whose vacuum expectation value (VEV) determines the position of the localization points~\cite{Abe:2015jqa}.

In the following model building, chiral superfields for quarks and leptons will be assigned to $\Phi_2^{AB}$ and $\Phi_3^{BA}$, while those for Higgs bosons to $\Phi_1^B$. The Yukawa couplings among them originate from the three-point coupling $\Phi_2^{AB} \Phi_1^B \Phi_3^{BA}$.  Since the squared-wavefunctions of the bi-fundamental fields $\Phi_2^{AB}$ and $\Phi_3^{BA}$ are both localized as delta-functions on the second and the third tori due to the above argument, the products of these two wavefunctions in the overlap integrals in the expressions of 4D effective Yukawa coupling constants are treated as the delta-function on these two tori.\footnote{It is shown in Ref.~\cite{Cremades:2004wa} that such a treatment in the magnetized brane picture reproduces the Yukawa couplings obtained in the T-dual intersecting brane picture.} Then, hierarchical structures of the Yukawa couplings will be realized by the overlap integrals of quasi-localized wavefunctions on the first torus, where both 7$_A$- and 7$_B$-branes with finite fluxes on their world volumes spread.

Now we consider the orbifolding. Since wavefunctions along the localized directions are given by the delta functions from the above argument, we put them on the orbifold fixed points. Then, their parities must be even on the $i$-th torus they localize, forcing $p_i=0$ in Table~\ref{tab:parity}. The modified tables reflecting such limitations are shown in Table~\ref{tab:parity-D7}.

\begin{table}[t]
  \centering
  \begin{tabular}{cccc}\hline
    $\eta_a^A\eta_b^A$ & ${\eta_a^{A}}'{\eta_b^{A}}'$ &                 & $(p_1,p_2,p_3)$ \\ \hline
    \multirow{3}{*}{$+1$}
                       & \multirow{3}{*}{$+1$}
                       & $\Phi^A_{1,ab}$              & (1,0,0)                           \\
                       &                              & $\Phi^A_{2,ab}$ & (1,0,1)         \\
                       &                              & $\Phi^A_{3,ab}$ & (0,0,1)         \\ \hline
    \multirow{3}{*}{$+1$}
                       & \multirow{3}{*}{$-1$}
                       & $\Phi^A_{1,ab}$              & (1,0,1)                           \\
                       &                              & $\Phi^A_{2,ab}$ & (1,0,0)         \\
                       &                              & $\Phi^A_{3,ab}$ & (0,0,0)         \\ \hline
    \multirow{3}{*}{$-1$}
                       & \multirow{3}{*}{$+1$}
                       & $\Phi^A_{1,ab}$              & (0,0,0)                           \\
                       &                              & $\Phi^A_{2,ab}$ & (0,0,1)         \\
                       &                              & $\Phi^A_{3,ab}$ & (1,0,1)         \\ \hline
    \multirow{3}{*}{$-1$}
                       & \multirow{3}{*}{$-1$}
                       & $\Phi^A_{1,ab}$              & (0,0,1)                           \\
                       &                              & $\Phi^A_{2,ab}$ & (0,0,0)         \\
                       &                              & $\Phi^A_{3,ab}$ & (1,0,0)         \\ \hline
  \end{tabular}
  \hspace{1cm}
  \begin{tabular}{cccc}\hline
    $\eta_a^B\eta_b^B$ & ${\eta_a^{B}}'{\eta_b^{B}}'$ &                 & $(p_1,p_2,p_3)$ \\ \hline
    \multirow{3}{*}{$+1$}
                       & \multirow{3}{*}{$+1$}
                       & $\Phi^B_{1,ab}$              & (1,0,0)                           \\
                       &                              & $\Phi^B_{2,ab}$ & (0,1,0)         \\
                       &                              & $\Phi^B_{3,ab}$ & (1,1,0)         \\ \hline
    \multirow{3}{*}{$+1$}
                       & \multirow{3}{*}{$-1$}
                       & $\Phi^B_{1,ab}$              & (0,1,0)                           \\
                       &                              & $\Phi^B_{2,ab}$ & (1,0,0)         \\
                       &                              & $\Phi^B_{3,ab}$ & (0,0,0)         \\ \hline
    \multirow{3}{*}{$-1$}
                       & \multirow{3}{*}{$+1$}
                       & $\Phi^B_{1,ab}$              & (0,0,0)                           \\
                       &                              & $\Phi^B_{2,ab}$ & (1,1,0)         \\
                       &                              & $\Phi^B_{3,ab}$ & (0,1,0)         \\ \hline
    \multirow{3}{*}{$-1$}
                       & \multirow{3}{*}{$-1$}
                       & $\Phi^B_{1,ab}$              & (1,1,0)                           \\
                       &                              & $\Phi^B_{2,ab}$ & (0,0,0)         \\
                       &                              & $\Phi^B_{3,ab}$ & (1,0,0)         \\ \hline
  \end{tabular} \\*[1cm]
  \begin{tabular}{cccc} \hline
    $\eta_a^A\eta_b^B$ & ${\eta_a^{A}}'{\eta_b^{B}}'$ &                    & $(p_1,p_2,p_3)$ \\ \hline
    \multirow{2}{*}{$+1$}
                       & \multirow{2}{*}{$+1$}
                       & $\Phi^{AB}_{2,ab}$           & $-$                                  \\
                       &                              & $\Phi^{BA}_{3,ab}$ & $-$
    \\ \hline
    \multirow{2}{*}{$+1$}
                       & \multirow{2}{*}{$-1$}        & $\Phi^{AB}_{2,ab}$ & (1,0,0)         \\
                       &                              & $\Phi^{BA}_{3,ab}$ & (0,0,0)         \\ \hline
    \multirow{2}{*}{$-1$}
                       & \multirow{2}{*}{$+1$}        & $\Phi^{AB}_{2,ab}$ & $-$             \\
                       &                              & $\Phi^{BA}_{3,ab}$ & $-$             \\ \hline
    \multirow{2}{*}{$-1$}
                       & \multirow{2}{*}{$-1$}        & $\Phi^{AB}_{2,ab}$ & (0,0,0)         \\
                       &                              & $\Phi^{BA}_{3,ab}$ & (1,0,0)         \\ \hline
  \end{tabular} \\*[0.5cm]
  \caption{Possible boundary conditions for 7-brane configurations on the $Z_2 \times Z'_2$ orbifold.}
  \label{tab:parity-D7}
\end{table}

Finally, we remark that no new fundamental parameter will arise in our construction of 7-brane configurations at least at the classical level, since they are derived from 10D SYM by suitably selecting background fluxes, Wilson-lines and orbifold parities. Note also that only discrete numbers are allowed for these background values.

\subsection{SM gauge symmetry}

Referring again the phenomenological configurations of 7-branes in Ref.~\cite{Abe:2017gye}, we consider the case $N_A=N_B=4$ in the following model building and express the corresponding gauge symmetries $U(N_{A,B})$ as $U(4)_{A,B}$, respectively. To realize the SM gauge symmetry $SU(3)_C \times SU(2)_L \times U(1)_Y$ up to $U(1)$ factors, the original $U(4)_A \times U(4)_B$ gauge symmetry of 8D SYM theories on 7$_A$- and 7$_B$-branes has to be broken down to $U(3)_C \times U(2)_L \times U(1)^3$. We first introduce the following background magnetic fluxes $M^{(i)}$ in the extra-dimensional components of 10D vector field~(\ref{eq:background}),
\begin{eqnarray}
  M^{(1)} & = &
  \left(
  \begin{array}{cc|ccc}
    m_C^{(1)}\times\bm{1}_3 &           &                         &           &           \\
                            & m_l^{(1)} &                         &           &           \\ \hline
                            &           & m_L^{(1)}\times\bm{1}_2 &           &           \\
                            &           &                         & m_R^{(1)} &           \\
                            &           &                         &           & m_R^{(1)}
  \end{array}
  \right), \nonumber \\
  M^{(2)} & = &
  \left(
  \begin{array}{cc|ccc}
    H \times\bm{1}_3 &   &                         &           &           \\
                     & H &                         &           &           \\ \hline
                     &   & m_L^{(2)}\times\bm{1}_2 &           &           \\
                     &   &                         & m_R^{(2)} &           \\
                     &   &                         &           & m_R^{(2)}
  \end{array}
  \right), \nonumber \\
  M^{(3)} & = &
  \left(
  \begin{array}{cc|ccc}
    m_C^{(3)}\times\bm{1}_3 &           &                 &   &   \\
                            & m_l^{(3)} &                 &   &   \\ \hline
                            &           & H\times\bm{1}_2 &   &   \\
                            &           &                 & H &   \\
                            &           &                 &   & H
  \end{array}
  \right),
  \label{eq:magnetic flux}
\end{eqnarray}
that break the gauge symmetry as
\begin{equation}
  U(4)_A \times U(4)_B \rightarrow
  U(3)_C \times U(1)_l \times U(2)_L \times U(2)_{R},
  \label{eq:Wilson line}
\end{equation}
where $H$ represents the positive infinite fluxes taken as $H \rightarrow \infty$ to obtain the profile of the 7-brane configuration shown in Table~\ref{tab:D7-brane-embedding}.

To eliminate the exotic modes later while keeping the gauge symmetries preserved by the above fluxes, we parameterize the $Z_2 \times Z_2'$ orbifold projections $P$ and $P'$ in the following forms:
\begin{equation}
  P =
  \left(
  \begin{array}{c|c}
    P_A &     \\ \hline
        & P_B
  \end{array}
  \right), \quad
  P' =
  \left(
  \begin{array}{c|c}
    P'_A &      \\ \hline
         & P'_B
  \end{array}
  \right)
\end{equation}
\begin{equation}
  P_A^{(')} =
  \left(
  \begin{matrix}
    \eta^{(')}_c\times\bm{1}_3 &              \\
                               & \eta^{(')}_l \\
  \end{matrix}
  \right) , \quad
  P_B^{(')} =
  \left(
  \begin{matrix}
    \eta^{(')}_L\times\bm{1}_2 &              &              \\
                               & \eta^{(')}_R &              \\
                               &              & \eta^{(')}_R
  \end{matrix}
  \right).
\end{equation}
The parameters $\eta^{(')}_a=\pm 1$ ($a=c,l,L,R$) will be suitably fixed so that the exotic zero-modes are eliminated in each model.

The remaining $U(2)_R$ is broken down to $U(1)_{R'} \times U(1)_{R''}$ by introducing the following Wilson-lines:
\begin{equation}
  \zeta^{(1)} =
  \left(
  \begin{array}{cc|ccc}
    0\times\bm{1}_3 &   &                 &                  &                   \\
                    & 0 &                 &                  &                   \\ \hline
                    &   & 0\times\bm{1}_2 &                  &                   \\
                    &   &                 & \zeta_{R'}^{(1)} &                   \\
                    &   &                 &                  & \zeta_{R''}^{(1)}
  \end{array}
  \right) , \quad
  \zeta^{(2)} = \zeta^{(3)} = \bm{0}_8,
\end{equation}
where $\zeta_{R'} \neq \zeta_{R''}$.
Note that the numerical values are restricted as $\zeta_{R',R''}=0,1/2,\tau/2,(1+\tau)/2$ on the $Z_2 \times Z_2'$ orbifold.

With the above breaking pattern of the gauge symmetry, the matrix components of $\Phi_i$ are expressed as
\begin{equation}
  \Phi_i =
  \left(
  \begin{array}{cc|ccc}
    \Omega_C & \Xi_{Cl} & Q'       & U           & D            \\
    \Xi_{lC} & \Omega_l & L'       & N           & E            \\ \hline
    Q        & L        & \Omega_L & H_u'        & H_d'         \\
    U'       & N'       & H_u      & \Omega_{R'} & \Xi_{R'R''}  \\
    D'       & E'       & H_d      & \Xi_{R''R'} & \Omega_{R''}
  \end{array}
  \right), \label{eq:chiral superfield}
\end{equation}
where $Q^{(')}$ and $L^{(')}$ are identified as the chiral superfields for left-handed quarks and leptons, $U^{(')}$ and $D^{(')}$ for the right-handed up- and down-type quarks, $N^{(')}$ and $E^{(')}$ for the right-handed neutrinos and charged leptons, and $H_{u,d}^{(')}$ for the MSSM Higgs pairs, respectively, by their representations under the SM gauge group. On the other hand, $\Omega_a$ and $\Xi_{ab}$ are the exotic modes to be eliminated by suitably selecting the concrete values of the above magnetic fluxes, Wilson-lines and orbifold projections.

Finally, we remark that the above assignment of symmetries and their breaking pattern is not unique for the present 7-brane configuration shown in Table~\ref{tab:D7-brane-embedding}. We adopt one of the simple cases. A systematic search for the configurations realizing $U(3)_C \times U(2)_L \times U(1)^3$ symmetry from $U(N_A) \times U(N_B)$ will be possible, which remains as a future work.

\subsection{Constraints on configurations for eliminating exotic modes}

Since the (block-)off-diagonal components of $\Phi_i$ are bi-fundamental representations under the unbroken gauge groups, each of them feels the difference of the fluxes~\eqref{eq:magnetic flux} as can be seen in the zero-mode equation~(\ref{eq:zero-mode eq. i = j}) and (\ref{eq:zero-mode eq. i neq j}). So, there remains a freedom to shift the fluxes~\eqref{eq:magnetic flux} by an overall constant, and we fix it so that $m_L^{(1)}=0$. Similarly, we can take $\eta_C=\eta_C'=1$ using a freedom for the overall sign flip of each orbifold projection. On top of them, we will search the magnetic fluxes, Wilson-lines and orbifold projections satisfying the conditions for D-flatness~(\ref{eq:D-flat}) and projecting out exotic modes, as well as those for realizing three generations of quarks and leptons.

First, we consider the D-flat conditions~(\ref{eq:D-flat}). For 8D SYM theories on 7$_A$- and 7$_B$-branes, they are described as\footnote{Note that the infinite fluxes in Eq.~(\ref{eq:magnetic flux}) do not contribute here, those are formally introduced to systematically extract 7-brane configurations from 10D SYM.}
\begin{equation}
  0 = \frac{m_C^{(1)}}{\mathcal{A}^{(1)}} + \frac{m_C^{(3)}}{\mathcal{A}^{(3)}}, \quad
  0 = \frac{m_l^{(1)}}{\mathcal{A}^{(1)}} + \frac{m_l^{(3)}}{\mathcal{A}^{(3)}}, \quad
  0 = \frac{m_L^{(1)}}{\mathcal{A}^{(1)}} + \frac{m_L^{(2)}}{\mathcal{A}^{(2)}}, \quad
  0 = \frac{m_R^{(1)}}{\mathcal{A}^{(1)}} + \frac{m_R^{(2)}}{\mathcal{A}^{(2)}}.
  \label{eq:D-flat_condition_visible}
\end{equation}
The third equation determines $m_L^{(2)}=0$ since $m_L^{(1)}=0$ from the starting arguments. Then, we find that $\mathcal{A}^{(2)}/\mathcal{A}^{(1)}$ is determined by $m_R^{(1,2)}$ through the fourth equation. With $m_L^{(2)}=0$ and $\zeta^{(2)}=0\times\bm{1}_8$, we find $|m_R^{(2)}|=0,1$ and $|m_R^{(2)}|=3,4$ are allowed for $Z_2$-even and -odd non-degenerate Higgs wavefunction on the second torus, respectively.  In this paper, we consider the $Z_2$-even case, and select $|m_R^{(2)}|=1$ to break $U(4)_B$. On the other hand, the first and the second equations lead to
\begin{equation}
  \frac{m_C^{(3)}}{m_C^{(1)}} = \frac{m_l^{(3)}}{m_l^{(1)}} = - \frac{\mathcal{A}^{(3)}}{\mathcal{A}^{(1)}},
\end{equation}
and we consider the case with $\mathcal{A}^{(3)}/\mathcal{A}^{(1)}=1$ for simplicity, that is,  the flux numbers satisfying
\begin{equation}
  m_C^{(3)} = - m_C^{(1)} , \quad m_l^{(3)} = - m_l^{(1)},
  \label{eq:1st-3rd ratio}
\end{equation}
are searched in the following model building.

Next, we consider the restrictions related to zero-mode degeneracies identified as generations of corresponding particles, also relevant for eliminating exotic modes. Since the bi-fundamental fields between $U(4)_A$ and $U(4)_B$ should include quarks and leptons, $\eta_a'\eta_b'=-1$ is required for the zero-modes to remain, that leads to
\begin{equation}
  P' =
  \left(
  \begin{array}{c|c}
    P_A' &      \\ \hline
         & P_B'
  \end{array}
  \right) =
  \left(
  \begin{array}{cc|ccc}
    \bm{1}_3 &   &           &         \\
             & 1 &           &         \\ \hline
             &   & -\bm{1}_2 &    &    \\
             &   &           & -1 &    \\
             &   &           &    & -1
  \end{array}
  \right). \label{eq:P'}
\end{equation}
As for the exotic zero-modes, the diagonal components $\Omega_a$ in all $\Phi_i$s are eliminated by the orbifold projections. Some of the off-diagonal ones $\Xi_{R'R'',R''R'}$ disappear due to the conditions $m_{R'}=m_{R''}$ and $\zeta_{R'}\neq\zeta_{R''}$ assumed in this paper, while the disappearance of $\Xi_{Cl,lC}$ depends on fluxes and orbifold parities they feel. Note that $\Phi_2$ does not have $\Xi_{Cl,lC}$ due to the sign of flux difference they feel satisfying the D-flat condition~(\ref{eq:1st-3rd ratio}). The degeneracies of $\Xi_{Cl,lC}$ can be determined based on the first row of Table~\ref{tab:num-of-zero-modes}, since they feel vanishing Wilson-lines~(\ref{eq:Wilson line}) on each torus. Therefore, they can be eliminated if their parity is odd and $|m_C^{(i)}-m_l^{(i)}|\le2$ is satisfied on at least one of three tori. Since $P'$ is fixed as Eq.~(\ref{eq:P'}), taking the 7-brane configuration in consideration where wavefunctions on the second torus is the delta-function (parity even), it is found from Table~\ref{tab:num-of-zero-modes} that  $\Xi_{Cl,lC}$ in $\Phi_3$ is eliminated if $|m_C^{(3)}-m_l^{(3)}|\le2$ is satisfied. Then, $|m_C^{(1)}-m_l^{(1)}|=|m_C^{(3)}-m_l^{(3)}|\le2$ follows from Eq.~(\ref{eq:1st-3rd ratio}). Similar arguments suggest that not only the conditions on fluxes but also $\eta_C\eta_l=1$ is required for eliminating $\Xi_{Cl,lC}$ in $\Phi_1$, that determines $P_A$ as
\begin{equation}
  P_A =
  \left(
  \begin{array}{cc}
    \bm{1}_3 &   \\
             & 1
  \end{array}
  \right). \label{eq:PA}
\end{equation}

\subsection{Three generation models}

In the following, we construct MSSM-like models with three generations of quarks and leptons based on the parameter values or relations discussed previously. We adopt 7-brane configurations with the infinite fluxes~(\ref{eq:magnetic flux}) expressed by $H \to \infty$, where only the unprimed quarks, leptons and Higgs zero-modes in the chiral superfields~(\ref{eq:chiral superfield}),
\begin{equation}
  \Phi_1 =
  \left(
  \begin{array}{cc|ccc}
    0 & 0 & 0   & 0 & 0 \\
    0 & 0 & 0   & 0 & 0 \\ \hline
    0 & 0 & 0   & 0 & 0 \\
    0 & 0 & H_u & 0 & 0 \\
    0 & 0 & H_d & 0 & 0
  \end{array}
  \right) ,
  \Phi_2 =
  \left(
  \begin{array}{cc|ccc}
    0 & 0 & 0 & U & D \\
    0 & 0 & 0 & N & E \\ \hline
    0 & 0 & 0 & 0 & 0 \\
    0 & 0 & 0 & 0 & 0 \\
    0 & 0 & 0 & 0 & 0
  \end{array}
  \right) ,
  \Phi_3 =
  \left(
  \begin{array}{cc|ccc}
    0 & 0 & 0 & 0 & 0 \\
    0 & 0 & 0 & 0 & 0 \\ \hline
    Q & L & 0 & 0 & 0 \\
    0 & 0 & 0 & 0 & 0 \\
    0 & 0 & 0 & 0 & 0
  \end{array}
  \right), \label{eq:model goal}
\end{equation}
remain with suitable choices of parameters.

To construct such models, we first search possible flux values yielding three generations of $Q$ and $L$ zero-modes. The Wilson-lines felt by them are both vanishing, and the parities of their wavefunctions on the first torus are equal to each other. Then, from Tables~\ref{tab:num-of-zero-modes} and \ref{tab:parity-D7}, the possible absolute values of fluxes in $M^{(1)}$ those yield three degeneracies for these zero-modes are found as
\begin{equation}
  ( |m_C^{(1)}|,|m_l^{(1)}| ) =
  \begin{cases}
    (4,5) , (5,4) & \qquad \text{for} \quad \eta_L = +1  \\
    (7,8) , (8,7) & \qquad \text{for} \quad \eta_L = -1.
  \end{cases}
\end{equation}
Since $Q$ and $L$ are carried by $\Phi_3$, the net flux they feel on the first torus are both negative valued, that determines the flux values as
\begin{equation}
  ( m_C^{(1)},m_l^{(1)} ) =
  \begin{cases}
    (4,5) , (5,4) & \qquad \text{for} \quad \eta_L = +1  \\
    (7,8) , (8,7) & \qquad \text{for} \quad \eta_L = -1.
  \end{cases}
\end{equation}

Next, we consider possible patterns of fluxes and Wilson-lines as well as orbifold parities to realize three generations of $U$, $D$, $N$ and $E$ zero-modes. In our setup, net fluxes felt by $U$ and $D$ are the same and as are $N$ and $E$, while those felt by $U$ and $N$ differ by 1. As for the Wilson-lines, net ones felt by $U$ and $N$ are the same and so are $D$ and $E$, while those felt by $U$ and $D$ are different from each other. Note also that $U$, $D$, $N$ and $E$ have common orbifold parities. Taking these \rm observations into account, the possible patterns so that all four have three generations can be identified by suitably choosing a pair of two flux values in Table~\ref{tab:num-of-zero-modes} for each field, those differ by 1 with different values of Wilson-lines but sharing a common orbifold parity.  We find such patterns as
\begin{equation}
  ( |m_C^{(1)}-m_R^{(1)}|,|m_l^{(1)}-m_R^{(1)}| ) =
  \begin{cases}
    (6,7) , (7,6) & \qquad \text{for} \quad \eta_R = +1 \\
    (5,6) , (6,5) & \qquad \text{for} \quad \eta_R = -1
  \end{cases},
\end{equation}
\begin{equation}
  (\zeta_{R'}^{(1)} , \zeta_{R''}^{(1)}) = (\frac{1}{2},\frac{\tau}{2}),(\frac{\tau}{2},\frac{1}{2}),
\end{equation}
those can be futher restricted to
\begin{equation}
  ( m_R^{(1)}-m_C^{(1)},m_R^{(1)}-m_l^{(1)} ) =
  \begin{cases}
    (6,7) , (7,6) & \qquad \text{for} \quad \eta_R = +1 \\
    (5,6) , (6,5) & \qquad \text{for} \quad \eta_R = -1
  \end{cases},
\end{equation}
since the net fluxes $U$, $D$, $N$ and $E$ feel are all negative valued meaning $m_C^{(1)},m_l^{(1)}<m_R^{(1)}$.

After all, the possible combinations of fluxes $M^{(1)}$, Wilson-lines $\zeta^{(1)}$ and orbifold projections $P$ on the first torus realizing three generations of quarks and leptons are classified into the following four cases.

\vspace{0.5cm}
\noindent
Case 1:
\begin{equation}
  \begin{aligned}
    M^{(1)}     & =
    \left(
    \begin{array}{cc|ccc}
      4\times\bm{1}_3 &   &          &    &    \\
                      & 5 &          &    &    \\ \hline
                      &   & \bm{0}_2 &    &    \\
                      &   &          & 10 &    \\
                      &   &          &    & 10
    \end{array}
    \right) \text{or}
    \left(
    \begin{array}{cc|ccc}
      5\times\bm{1}_3 &   &          &    &    \\
                      & 4 &          &    &    \\ \hline
                      &   & \bm{0}_2 &    &    \\
                      &   &          & 10 &    \\
                      &   &          &    & 10
    \end{array}
    \right) ,       \\
    \zeta^{(1)} & =
    \left(
    \begin{array}{cc|ccc}
      \bm{0}_3 &   &          &             &                \\
               & 0 &          &             &                \\ \hline
               &   & \bm{0}_2 &             &                \\
               &   &          & \frac{1}{2} &                \\
               &   &          &             & \frac{\tau}{2}
    \end{array}
    \right) \text{or}
    \left(
    \begin{array}{cc|ccc}
      \bm{0}_3 &   &          &                &             \\
               & 0 &          &                &             \\ \hline
               &   & \bm{0}_2 &                &             \\
               &   &          & \frac{\tau}{2} &             \\
               &   &          &                & \frac{1}{2}
    \end{array}
    \right), \quad
    P           =
    \left(
    \begin{array}{cc|ccc}
      \bm{1}_3 &   &          &    &    \\
               & 1 &          &    &    \\ \hline
               &   & \bm{1}_2 &    &    \\
               &   &          & -1 &    \\
               &   &          &    & -1
    \end{array}
    \right)
  \end{aligned}
\end{equation}

\vspace{0.5cm}
\noindent
Case 2:
\begin{equation}
  \begin{aligned}
    M^{(1)}     & =
    \left(
    \begin{array}{cc|ccc}
      4\times\bm{1}_3 &   &          &    &    \\
                      & 5 &          &    &    \\ \hline
                      &   & \bm{0}_2 &    &    \\
                      &   &          & 11 &    \\
                      &   &          &    & 11
    \end{array}
    \right) \text{or}
    \left(
    \begin{array}{cc|ccc}
      5\times\bm{1}_3 &   &          &    &    \\
                      & 4 &          &    &    \\ \hline
                      &   & \bm{0}_2 &    &    \\
                      &   &          & 11 &    \\
                      &   &          &    & 11
    \end{array}
    \right) ,       \\
    \zeta^{(1)} & =
    \left(
    \begin{array}{cc|ccc}
      \bm{0}_3 &   &          &             &                \\
               & 0 &          &             &                \\ \hline
               &   & \bm{0}_2 &             &                \\
               &   &          & \frac{1}{2} &                \\
               &   &          &             & \frac{\tau}{2}
    \end{array}
    \right) \text{or}
    \left(
    \begin{array}{cc|ccc}
      \bm{0}_3 &   &          &                &             \\
               & 0 &          &                &             \\ \hline
               &   & \bm{0}_2 &                &             \\
               &   &          & \frac{\tau}{2} &             \\
               &   &          &                & \frac{1}{2}
    \end{array}
    \right) ,\quad
    P           =
    \left(
    \begin{array}{cc|ccc}
      \bm{1}_3 &   &          &   &   \\
               & 1 &          &   &   \\ \hline
               &   & \bm{1}_2 &   &   \\
               &   &          & 1 &   \\
               &   &          &   & 1
    \end{array}
    \right)
  \end{aligned}
\end{equation}

\vspace{0.5cm}
\noindent
Case 3:
\begin{equation}
  \begin{aligned}
    M^{(1)}     & =
    \left(
    \begin{array}{cc|ccc}
      7\times\bm{1}_3 &   &          &    &    \\
                      & 8 &          &    &    \\ \hline
                      &   & \bm{0}_2 &    &    \\
                      &   &          & 13 &    \\
                      &   &          &    & 13
    \end{array}
    \right) \text{or}
    \left(
    \begin{array}{cc|ccc}
      8\times\bm{1}_3 &   &          &    &    \\
                      & 7 &          &    &    \\ \hline
                      &   & \bm{0}_2 &    &    \\
                      &   &          & 13 &    \\
                      &   &          &    & 13
    \end{array}
    \right) ,       \\
    \zeta^{(1)} & =
    \left(
    \begin{array}{cc|ccc}
      \bm{0}_3 &   &          &             &                \\
               & 0 &          &             &                \\ \hline
               &   & \bm{0}_2 &             &                \\
               &   &          & \frac{1}{2} &                \\
               &   &          &             & \frac{\tau}{2}
    \end{array}
    \right) \text{or}
    \left(
    \begin{array}{cc|ccc}
      \bm{0}_3 &   &          &                &             \\
               & 0 &          &                &             \\ \hline
               &   & \bm{0}_2 &                &             \\
               &   &          & \frac{\tau}{2} &             \\
               &   &          &                & \frac{1}{2}
    \end{array}
    \right) ,\quad
    P           =
    \left(
    \begin{array}{cc|ccc}
      \bm{1}_3 &   &           &    &    \\
               & 1 &           &    &    \\ \hline
               &   & -\bm{1}_2 &    &    \\
               &   &           & -1 &    \\
               &   &           &    & -1
    \end{array}
    \right)
  \end{aligned}
\end{equation}

\vspace{0.5cm}
\noindent
Case 4:
\begin{equation}
  \begin{aligned}
    M^{(1)}     & =
    \left(
    \begin{array}{cc|ccc}
      7\times\bm{1}_3 &   &          &    &    \\
                      & 8 &          &    &    \\ \hline
                      &   & \bm{0}_2 &    &    \\
                      &   &          & 14 &    \\
                      &   &          &    & 14
    \end{array}
    \right) \text{or}
    \left(
    \begin{array}{cc|ccc}
      8\times\bm{1}_3 &   &          &    &    \\
                      & 7 &          &    &    \\ \hline
                      &   & \bm{0}_2 &    &    \\
                      &   &          & 14 &    \\
                      &   &          &    & 14
    \end{array}
    \right) ,       \\
    \zeta^{(1)} & =
    \left(
    \begin{array}{cc|ccc}
      \bm{0}_3 &   &          &             &                \\
               & 0 &          &             &                \\ \hline
               &   & \bm{0}_2 &             &                \\
               &   &          & \frac{1}{2} &                \\
               &   &          &             & \frac{\tau}{2}
    \end{array}
    \right) \text{or}
    \left(
    \begin{array}{cc|ccc}
      \bm{0}_3 &   &          &                &             \\
               & 0 &          &                &             \\ \hline
               &   & \bm{0}_2 &                &             \\
               &   &          & \frac{\tau}{2} &             \\
               &   &          &                & \frac{1}{2}
    \end{array}
    \right) ,\quad
    P           =
    \left(
    \begin{array}{cc|ccc}
      \bm{1}_3 &   &           &   &   \\
               & 1 &           &   &   \\ \hline
               &   & -\bm{1}_2 &   &   \\
               &   &           & 1 &   \\
               &   &           &   & 1
    \end{array}
    \right)
  \end{aligned}
\end{equation}
The complete flux patterns realizing three generation models can be obtained by selecting $M^{(2)}$ and $M^{(3)}$ those satisfy the D-flat conditions with the above $M^{(1)}$ in each case. The concrete values of $M^{(i)}$, $\zeta^{(i)}$ and $P^{(\prime)}$ for obtained three generation models are shown in Appendix~\ref{sec:appendix}.

\subsection{Flavor structures}
\label{ssec:flavor_structure}

For each model in Appendix~\ref{sec:appendix}, we numerically checked whether the quark and charged lepton mass ratios and mixing angles can be realistic or not. In our setup, the parameters relevant to the flavor structures are $\rm{Im}\tau_1$, $\langle H_u \rangle$, $\langle H_d \rangle$ and $\tan \beta \equiv |\langle H_u \rangle| / |\langle H_d \rangle|$. We searched parameter values those reproduce the \rm observed values of
\begin{equation}
  (m_{\rm u} , m_{\rm c})/m_{\rm t} , \quad (m_{\rm d} , m_{\rm s})/m_{\rm b} , \quad (m_{\rm e} , m_{\rm \mu})/m_{\rm \tau}, \quad |V^{\rm{CKM}}_{\rm ij}| , \quad m_{\rm t}/m_{\rm b},
\end{equation}
where $m_{\rm i}$ describe the fermion masses and $|V^{\rm{CKM}}_{\rm ij}|$ is the absolute value of Cabibbo-Kobayashi-Maskawa (CKM) matrix element representing the quark mixings. The degree of fitting is evaluated by
\begin{eqnarray}
  \chi_{\rm u} &\equiv& \left( \log_{10} \left( \frac{m_{\rm u}^{\rm obs}}{m_{\rm t}^{\rm obs}} \right) - \log_{10} \left( \frac{m_{\rm u}}{m_{\rm t}} \right) \right)^2 + \left( \log_{10} \left( \frac{m_{\rm c}^{\rm obs}}{m_{\rm t}^{\rm obs}} \right) - \log_{10} \left( \frac{m_{\rm c}}{m_{\rm t}} \right) \right)^2, \nonumber \\
  \chi_{\rm d} &\equiv& \left( \log_{10} \left( \frac{m_{\rm d}^{\rm obs}}{m_{\rm b}^{\rm obs}} \right) - \log_{10} \left( \frac{m_{\rm d}}{m_{\rm b}} \right) \right)^2 + \left( \log_{10} \left( \frac{m_{\rm s}^{\rm obs}}{m_{\rm b}^{\rm obs}} \right) - \log_{10} \left( \frac{m_{\rm s}}{m_{\rm b}} \right) \right)^2, \nonumber \\
  \chi_{\rm e} &\equiv& \left( \log_{10} \left( \frac{m_{\rm e}^{\rm obs}}{m_{\rm \tau}^{\rm obs}} \right) - \log_{10} \left( \frac{m_{\rm e}}{m_{\rm \tau}} \right) \right)^2 + \left( \log_{10} \left( \frac{m_{\rm \mu}^{\rm obs}}{m_{\rm \tau}^{\rm obs}} \right) - \log_{10} \left( \frac{m_{\rm \mu}}{m_{\rm \tau}} \right) \right)^2, \nonumber \\
  \chi_V &\equiv& \sum_{\rm i,j} \left( \log_{10} |V^{\text{CKM},\rm obs}_{\rm ij}| - \log_{10} |V^{\text{CKM}}_{\rm ij}| \right)^2,                                                                                                                                                   \nonumber \\
  \chi_r &\equiv& \left( \frac{m_{\rm t}^{\rm obs}}{m_{\rm b}^{\rm obs}}-\frac{m_{\rm t}}{m_{\rm b}} \right)^2.
\end{eqnarray}
where variables with (without) the upper-script `$\rm obs$' represent the experimental (theoretical) values. For each model, first, the parameter values of $\rm{Im}\tau_1$, $\langle H_u \rangle$ and $\langle H_d \rangle$ those minimize
\begin{equation}
  \chi \equiv \chi_{\rm u} + \chi_{\rm d} + \chi_{\rm e} + \chi_V
\end{equation}
are searched, then, the value of  $\rm{tan}\beta$ minimizing $\chi_r$ is determined, since the latter one only affects $m_{\rm t}/m_{\rm b}$ while the other \rm observables are determined by the former three.

The numerical results for each model are shown in Appendix~\ref{sec:appendix}. Here, more details are explained based on the model giving the minimal value of $\chi$, that is Model 3-4 whose fluxes, Wilson-lines and orbifold projections are shown in Eq.~(\ref{eq:model3-4}). This model has six generations of MSSM Higgs pairs. The Yukawa coupling constants $Y_{IJK}^{U,D,N,E}$ between quarks or leptons and the $K$-th Higgs generation ($K=1,\ldots,6$) can be evaluated by Eq.~(\ref{eq:yukawa-orbifold}). For instance, using
\begin{equation}
  \eta_U(N) = \vartheta
  \left[
    \begin{matrix}
      \frac{N}{520} \\ 0
    \end{matrix}
    \right]
  (0, 520\tau)
\end{equation}
$Y_{IJK}^{U}$s are expressed as
\begin{eqnarray}
  (Y_{IJ1}^U)^* &\sim&
  \frac{1}{2\sqrt{2}} \left(
  \begin{matrix}
    (\eta_U(9) + \eta_U(-9))     &
    - (\eta_U(17) + \eta_U(-17)) &
    \sqrt{2} (\eta_U(95) + \eta_U(-95))  \\
    - (\eta_U(37) + \eta_U(-37)) &
    (\eta_U(22) + \eta_U(-22))   &
    -\sqrt{2} (\eta_U(30) + \eta_U(-30)) \\
    - (\eta_U(61) + \eta_U(-61)) &
    - (\eta_U(43) + \eta_U(-43)) &
    \sqrt{2} (\eta_U(35) + \eta_U(-35))
  \end{matrix}
  \right), \nonumber \\
  (Y_{IJ2}^U)^* &\sim&
  \frac{1}{2\sqrt{2}} \left(
  \begin{matrix}
    (\eta_U(1) + \eta_U(-1))     &
    (\eta_U(103) + \eta_U(-103)) &
    -\sqrt{2} (\eta_U(25) + \eta_U(-25)) \\
    (\eta_U(14) + \eta_U(-14))   &
    - (\eta_U(38) + \eta_U(-38)) &
    \sqrt{2} (\eta_U(90) + \eta_U(-90))  \\
    - (\eta_U(51) + \eta_U(-51)) &
    (\eta_U(27) + \eta_U(-27))   &
    -\sqrt{2} (\eta_U(155) + \eta_U(-155))
  \end{matrix}
  \right), \nonumber \\
  (Y_{IJ3}^U)^* &\sim&
  \frac{1}{2\sqrt{2}} \left(
  \begin{matrix}
    (\eta_U(71) + \eta_U(-71))   &
    (\eta_U(7) + \eta_U(-7))     &
    -\sqrt{2} (\eta_U(175) + \eta_U(-175)) \\
    (\eta_U(3) + \eta_U(-3))     &
    - (\eta_U(58) + \eta_U(-58)) &
    \sqrt{2} (\eta_U(110) + \eta_U(-110))  \\
    - (\eta_U(19) + \eta_U(-19)) &
    (\eta_U(123) + \eta_U(-123)) &
    -\sqrt{2} (\eta_U(45) + \eta_U(-45))
  \end{matrix}
  \right), \nonumber \\
  (Y_{IJ4}^U)^* &\sim&
  \frac{1}{2\sqrt{2}} \left(
  \begin{matrix}
    - (\eta_U(41) + \eta_U(-41)) &
    - (\eta_U(63) + \eta_U(-63)) &
    \sqrt{2} (\eta_U(15) + \eta_U(-15))  \\
    - (\eta_U(54) + \eta_U(-54)) &
    (\eta_U(2) + \eta_U(-2))     &
    -\sqrt{2} (\eta_U(50) + \eta_U(-50)) \\
    (\eta_U(11) + \eta_U(-11))   &
    - (\eta_U(37) + \eta_U(-37)) &
    \sqrt{2} (\eta_U(115) + \eta_U(-115))
  \end{matrix}
  \right), \nonumber \\
  (Y_{IJ5}^U)^* &\sim&
  \frac{1}{2\sqrt{2}} \left(
  \begin{matrix}
    - (\eta_U(49) + \eta_U(-49)) &
    (\eta_U(23) + \eta_U(-23))   &
    -\sqrt{2} (\eta_U(55) + \eta_U(-55)) \\
    (\eta_U(94) + \eta_U(-94))   &
    - (\eta_U(42) + \eta_U(-42)) &
    \sqrt{2} (\eta_U(10) + \eta_U(-10))  \\
    - (\eta_U(29) + \eta_U(-29)) &
    (\eta_U(3) + \eta_U(-3))     &
    -\sqrt{2} (\eta_U(75) + \eta_U(-75))
  \end{matrix}
  \right), \nonumber \\
  (Y_{IJ6}^U)^* &\sim&
  \frac{1}{2\sqrt{2}} \left(
  \begin{matrix}
    (\eta_U(31) + \eta_U(-31))   &
    - (\eta_U(47) + \eta_U(-47)) &
    \sqrt{2} (\eta_U(135) + \eta_U(-135)) \\
    - (\eta_U(34) + \eta_U(-34)) &
    (\eta_U(18) + \eta_U(-18))   &
    -\sqrt{2} (\eta_U(70) + \eta_U(-70))  \\
    - (\eta_U(21) + \eta_U(-21)) &
    - (\eta_U(83) + \eta_U(-83)) &
    \sqrt{2} (\eta_U(5) + \eta_U(-5))
  \end{matrix}
  \right).
\end{eqnarray}
The other ones can be similarly evaluated. Based on them, we performed the parameter search and found the parameter values~(\ref{eq:model3-4vev}) yield semi-realistic flavor structures shown in Table~\ref{tab:case 3-4}. Here $v_u=v\sin\beta$ and $v_d=v\cos\beta$ depend on $\tan\beta$ where $v$ is the \rm observed Higgs VEV, and $\mathcal{N}_{H_{u,d}}$ is the normalization factor for the ratio of multiple Higgs VEVs (e.g. $\mathcal{N}_{H_{u}} = 1/\sqrt{0.5^2+0.7^2}$ and $\mathcal{N}_{H_{d}} = 1/\sqrt{0.8^2+0.3^2}$ in Eq. \eqref{eq:model3-4vev}).

We remark that all the models in Appendix~\ref{sec:appendix} have multiple (such as five, six and seven) generations of the MSSM Higgs pairs ($H_u$,$H_d$). Though the $\chi$-values are different, all these models yield semi-realistic hierarchical values of mass ratios and mixing angles from non-hierarchical VEVs of multiple Higgs generations, thanks to the Gaussian structure of the wave function~(\ref{eq:wavefunc-solution}) caused by the magnetic flux.

\subsection{Hypercharge and other \texorpdfstring{$U(1)$}{U(1)}s}
\label{ssec:u1factors}

Finally in Section~\ref{sec:visible}, we describe the Abelian factors in the remaining gauge symmetries,
based on the similar ones in 10D SYM models of Ref.~\cite{Abe:2012fj}.
Since each $U(N)$ factor decomposes locally as $SU(N) \times U(1)$, the gauge groups in all the models in Appendix~\ref{sec:appendix} have five Abelian factors.
We denote these as $U(1)_I$ with $I=C,l,L,R',R''$ corresponding to the diagonal $U(1)$ generators of the blocks in Eq.~\eqref{eq:magnetic flux}.
The charge assignments of the MSSM fields under these $U(1)$'s are shown in Table~\ref{tab:matter-charge}.

\begin{table}[ht]
  \centering
  \begin{tabular}{c||c|ccccc}\hline
    Matter & $SU(3)_C \times SU(2)_L$ & $Q_C$ & $Q_l$ & $Q_L$ & $Q_{R'}$ & $Q_{R''}$ \\ \hline
    $Q$    & $(\bar{\bm{3}},\bm{2})$  & $-1$  & $0$   & $1$   & $0$      & $0$       \\
    $U$    & $(\bm{3},\bm{1})$        & $1$   & $0$   & $0$   & $-1$     & $0$       \\
    $D$    & $(\bm{3},\bm{1})$        & $1$   & $0$   & $0$   & $0$      & $-1$      \\
    $L$    & $(\bm{1},\bm{2})$        & $0$   & $-1$  & $1$   & $0$      & $0$       \\
    $N$    & $(\bm{1},\bm{1})$        & $0$   & $1$   & $0$   & $-1$     & $0$       \\
    $E$    & $(\bm{1},\bm{1})$        & $0$   & $1$   & $0$   & $0$      & $-1$      \\
    $H_u$  & $(\bm{1},\bm{2})$        & $0$   & $0$   & $-1$  & $1$      & $0$       \\
    $H_d$  & $(\bm{1},\bm{2})$        & $0$   & $0$   & $-1$  & $0$      & $1$       \\ \hline
  \end{tabular}
  \caption{MSSM matter fields and their gauge charges~\cite{Abe:2012fj}. Here, $Q_I$ describes the $U(1)_I$ charge for $I=C,l,L,R',R''$.}
  \label{tab:matter-charge}
\end{table}

There are two familiar $U(1)$ directions $U(1)_Y$ and $U(1)_{B-L}$ representing the hypercharge and the B-L symmetries, respectively, whose charges can be expressed as
\begin{align}
  Q_{Y}
   & =
  \alpha Q_C
  +
  \left( \frac{2}{3}+\alpha \right)
  Q_l
  +
  \left( \frac{1}{6}+\alpha \right)
  Q_L
  +
  \left( \frac{2}{3}+\alpha \right)
  Q_{R'}
  +
  \left( -\frac{1}{3}+\alpha \right)
  Q_{R''},
  \nonumber \\
  Q_{B-L}
   & =
  \left(-\frac{1}{3}+\beta\right) Q_C + \left( 1+\beta \right) Q_{l}
  +
  \beta Q_L
  +
  \beta Q_{R'}
  +
  \beta Q_{R''},
  \label{eq:u1charges}
\end{align}
where $\alpha$ and $\beta$ are arbitrary real constants.
From Table~\ref{tab:matter-charge}, we can check that Eq.~(\ref{eq:u1charges})
reproduces the actual values of hypercharge and B-L charge shown
in the first two columns of Table~\ref{tab:matter-charge-redefined}.
These $U(1)_Y$ and $U(1)_{B-L}$ are anomaly-free, since all the models in Appendix~\ref{sec:appendix}
have three generations of right-handed neutrinos.

\begin{table}[ht]
  \centering
  \begin{tabular}{c||ccccc}\hline
    Matter & $Q_Y$  & $Q_{B-L}$ & $Q_{A_1}$ & $Q_{A_2}$ & $Q_{A_3}$ \\ \hline
    $Q$    & $1/6$  & $1/3$     & $4/3$     & $1/3$     & $11/6$    \\
    $U$    & $-2/3$ & $-1/3$    & $-4/3$    & $-4/3$    & $-5/6$    \\
    $D$    & $1/3$  & $-1/3$    & $-7/3$    & $-1/3$    & $-5/6$    \\
    $L$    & $-1/2$ & $-1$      & $0$       & $0$       & $1$       \\
    $N$    & $0$    & $1$       & $0$       & $-1$      & $0$       \\
    $E$    & $1$    & $1$       & $-1$      & $0$       & $0$       \\
    $H_u$  & $1/2$  & $0$       & $0$       & $1$       & $-1$      \\
    $H_d$  & $-1/2$ & $0$       & $1$       & $0$       & $-1$      \\ \hline
  \end{tabular}
  \caption{Redefined charge assignments under the orthogonal $U(1)$ basis.}
  \label{tab:matter-charge-redefined}
\end{table}

We choose $\alpha$ and $\beta$ to make $U(1)_{Y}$ and $U(1)_{B-L}$ orthogonal to each other.
To achieve this,
$\vv_{Y} = (\alpha,\, 2/3 + \alpha,\, 1/6 + \alpha,\, 2/3 + \alpha,\, -1/3 + \alpha)$
and $\vv_{B-L} = (-1/3 + \beta,\, 1 + \beta,\, \beta,\, \beta,\, \beta)$ are orthogonal.
This condition yields $\beta = -4(\alpha + 1)/(30\alpha + 7)$.
For simplicity, we pick $(\alpha, \beta) = (-1, 0)$ in the subsequent discussion.

The possible other directions orthogonal to both $\vv_{Y}$ and $\vv_{B-L}$ are
\begin{equation}
  \vv_{A_1}
  =
  (-5/6,0,1,0,0)
  ,\quad
  \vv_{A_2}
  =
  (-1/3,0,0,1,0)
  ,\quad
  \vv_{A_3}
  =
  (-4/3,0,0,0,1),
\end{equation}
leading the orthogonal transformation
\begin{equation}
  \begin{pmatrix}
    Q_Y     \\
    Q_{B-L} \\
    Q_{A_1} \\
    Q_{A_2} \\
    Q_{A_3}
  \end{pmatrix}
  =
  \begin{pmatrix}
    -1   & -1/3 & -5/6 & -1/3 & -4/3 \\
    -1/3 & 1    & 0    & 0    & 0    \\
    -5/6 & 0    & 1    & 0    & 0    \\
    -1/3 & 0    & 0    & 1    & 0    \\
    -4/3 & 0    & 0    & 0    & 1
  \end{pmatrix}
  \begin{pmatrix}
    Q_C    \\
    Q_l    \\
    Q_L    \\
    Q_{R'} \\
    Q_{R''}
  \end{pmatrix}.
\end{equation}
The resulting charges are already shown in Table~\ref{tab:matter-charge-redefined}.
The last three columns of the table display the charge assignments of the matter fields under the three new $U(1)$'s,
denoted as $U(1)_{A_1}$, $U(1)_{A_2}$, and $U(1)_{A_3}$.
These last three $U(1)$'s are anomalous, in contrast to $U(1)_Y$ and $U(1)_{B-L}$,
and the corresponding gauge bosons may acquire masses via a Green-Schwarz (GS)~\cite{Green:1984sg}-like mechanism,
if the original 8D SYM theories couple to some external fields,
such as supergravity fields, in an appropriate way.\footnote{Anomalies on magnetized orbifolds were studied within the framework of 6D supergravity in Refs.~\cite{Buchmuller:2015eya}.}
Such couplings would be provided by the so-called Chern-Simons term~\cite{Deser:1981wh},
if our SYM action can somehow be embedded into the D-brane action~\cite{Green:1996dd}.
In that case, our 7-brane configurations could be regarded as low-energy effective theories of magnetized D7-brane systems.
In any case, if some GS-like mechanisms work,
these $U(1)$'s are no longer gauge symmetries at low energies,
but may instead impose certain coupling selection rules.

Finally, let us briefly remark on other physically meaningful $U(1)$ symmetries.
While we have manifestly realized the hypercharge and B-L symmetries,
it may be possible to realize other $U(1)$ factors simultaneously.
For example, in a previous work \cite{Abe:2012fj}, the authors found a transformation
\begin{equation}
  Q_{PQ} = \frac{1}{2}Q_C + \frac{1}{2}Q_l + Q_L
\end{equation}
that realizes a Peccei-Quinn (PQ)-like $U(1)$ symmetry.
However, this transformation cannot be implemented in our models
because it is not orthogonal to $U(1)_Y$ or $U(1)_{B-L}$.
(In fact, $\vec{v}_{PQ} = (1/2,\, 1/2,\, 1,\, 0,\, 0)$ is orthogonal to neither $\vec{v}_Y$ nor $\vec{v}_{B-L}$.)
This leads to kinetic mixing between $U(1)_{PQ}$ and $U(1)_Y$, which complicates the analysis of low-energy physics.

\section{Implementing hidden SYM sector}
\label{sec:hidden}

Even the models constructed so far originate from 10D SYM, its ${\cal N}=4$ supersymmetry is broken down to ${\cal N}=1$ by the magnetic fluxes satisfying the D-flat conditions~(\ref{eq:D-flat}) as well as the orbifold projections. Then, MSSM-like models are obtained in the 4D effective theory. Although the ${\cal N}=1$ supersymmetry would have played important roles for the stability of 7-brane configurations, we need a further supersymmetry breaking to obtain SM from MSSM. To still guarantee the stability, it is desirable that the ${\cal N}=1$ to $0$ breaking scale is smaller enough than the ${\cal N}=4$ to $1$ scale. Such a hierarchy would be generated by a dynamical supersymmetry breaking~\cite{Witten:1981nf} where nonperturbative effects caused by strong SYM dynamics trigger the breaking. Especially, if the ${\cal N}=1$ breaking scale is much lower than the 4D Planck scale, a radiatively stable intermediate (ultimately the electroweak) scale could arise in the 4D effective theory, which may play some roles in particle physics and cosmology. Furthermore, since there are no linear and bilinear terms of the O'Raifeartaigh~\cite{ORaifeartaigh:1975nky} type for a spontaneous supersymmetry breaking in the original superpotential~\eqref{eq:fluctuation-superpotential}, we need to consider such a dynamical scenario to break the remaining supersymmetry within our framework.

Following the previous idea~\cite{Abe:2016zgq}, we demonstrate how such a sector will be implemented within our framework. Since the appearance of Yukawa couplings among matter chiral multiplets in the 4D effective theory can be expected as in the visible sector, we consider a scenario based on the Affleck--Dine--Seiberg (ADS)~\cite{Affleck:1984xz} type nonperturbative effect in the hidden sector. If 4D $SU(N_c)$ SYM theory has $N_f$ flavors of matter chiral multiplets $Q$ ($\tilde{Q}$) in the (anti-)fundamental representation, this theory is asymptotically free for $N_f < N_c$ and a nonperturbative superpotential,
\begin{equation}
  W_{\rm ADS}
  =
  (N_c-N_f) \left( \frac{\Lambda^{3N_c-N_f}}{\det M} \right)^{1/(N_c-N_f)},
  \label{eq:ADSsuperpotential}
\end{equation}
will be generated, where $\Lambda$ is the dynamical scale of $SU(N_c)$ SYM and $M$ is the meson field $M=Q\tilde{Q}$ corresponding to the flat-direction of the D-term potential generated by integrating out the D-term auxiliary component of the $SU(N_c)$ vector superfield. Although sole $W_{\rm ADS}$ has a runaway property in the directions of meson fields, by combining it with a tree level Yukawa coupling in the superpotential like $SQ\tilde{Q}$ with the singlet chiral multiplet $S$, the meson can obtain non-vanishing vacuum expectation value $\langle M \rangle \sim \Lambda^2$, that yields an effective linear term of $S$ in the superpotential and causes a supersymmetry breaking (like the O'Raifeartaigh model~\cite{ORaifeartaigh:1975nky}) at low energies below $\Lambda$.

The above observations motivate us to build a systematic way to embed such a hidden SYM sector to the 7-brane configurations obtained in the previous section in a suitable way. In such a study, we should notice that couplings between visible and hidden sectors generically cause some instabilities and/or phenomenologically unwanted features. Therefore, we will develop a systematic way to implement a hidden sector so that there appear no zero-mode fields charged under both visible and hidden gauge groups (called charged messenger fields\footnote{If such fields satisfies certain conditions, they can work as messenger fields for the so called gauge-mediated supersymmetry breaking~\cite{Giudice:1998bp}. We don't consider such a case in this paper.}) in the following.

\subsection{7-brane configurations for hidden sectors}

The simplest way to eliminate the charged messenger fields is a spacial sequestering~\cite{Randall:1998uk} of hidden and visible sectors, under which such bi-fundamental fields can not be a local field. For such a purpose, we introduce the hidden 7-brane configurations such that they localize on the same tori as the visible ones do, and locate the former at different fixed points from those latter live. Then, we consider two 7-branes, 7$_C$ and 7$_D$, in the hidden sector, whose configuration is shown in Table \ref{tab:D7-brane-embedding-hidden} which is similar to that of the visible sector shown in Table \ref{tab:D7-brane-embedding}.

\begin{table}[t]
  \centering
  \begin{tabular}{cccc}\hline
           & $T^2$        & $T^2$        & $T^2$        \\ \hline
    D7$_C$ & $\checkmark$ & $\times$     & $\checkmark$ \\
    D7$_D$ & $\checkmark$ & $\checkmark$ & $\times$     \\ \hline
  \end{tabular}
  \caption{Embedding of two 7-branes, 7$_C$ and 7$_D$, into three tori. The notations are the same as those in Table~\ref{tab:D7-brane-embedding}.}
  \label{tab:D7-brane-embedding-hidden}
\end{table}

As a preliminary consideration, forget about the visible sector for a moment, and think that $U(N_C) \times U(N_D)$ is {\it not} broken by the orbifoldings. In this case, dynamical supersymmetry breaking does not occur, since the adjoint zero-modes in the diagonal blocks will be eliminated under the $Z_2 \times Z'_2$ projection as
\begin{equation}
  \Phi_1
  =
  \left(
  \begin{array}{c|c}
    0           & Q_1 \\ \hline
    \tilde{Q}_1 & 0
  \end{array}
  \right)
  ,\quad
  \Phi_2
  =
  \left(
  \begin{array}{c|c}
    0           & Q_2 \\ \hline
    \tilde{Q}_2 & 0
  \end{array}
  \right)
  ,\quad
  \Phi_3
  =
  \left(
  \begin{array}{c|c}
    0   & \tilde{Q}_3 \\ \hline
    Q_3 & 0
  \end{array}
  \right),
\end{equation}
where the top-left and the bottom-right blocks of each matrix correspond to the elements of SYM gauge groups $U(N_C)$ and $U(N_D)$ on 7$_C$- and 7$_D$-branes, respectively. The above matrices indicate that there are no tree level Yukawa couplings among zero-modes in the 4D effective superpotential~\eqref{eq:fluctuation-superpotential}, which is necessary for the ADS mechanism to work. Thus, at least one of $U(N_C)$ and $U(N_D)$ should be broken by the background, and we will consider the minimal breaking patterns. One is $U(N_C) \to U(N_C^{(1)}) \times U(N_C^{(2)})$ referred to as $C_2 D_1$ and the other is $U(N_D) \to U(N_D^{(1)}) \times U(N_D^{(2)})$ as $C_1 D_2$. Then, the desired zero-mode contents for the ADS mechanism are
\begin{equation}
  \Phi_i
  =
  \left(
  \begin{array}{ccc}
    0 & 0 & * \\
    * & 0 & 0 \\
    0 & * & 0
  \end{array}
  \right)
  \textrm{\ or\ }
  \left(
  \begin{array}{ccc}
    0 & * & 0 \\
    0 & 0 & * \\
    * & 0 & 0
  \end{array}
  \right),
\end{equation}
where the stars $*$ represent non-zero blocks for each $\Phi_i$.
If this is the case, the 4D effective superpotential~\eqref{eq:fluctuation-superpotential} has the Yukawa couplings.

First, we consider the pattern $C_2D_1$. The possible orbifold projections are
\begin{equation}
  (P_{\rm hid},P'_{\rm hid})
  =
  \left(
  \begin{array}{cc|c}
    (\pm1,\pm1) & 0           & 0           \\
    0           & (\mp1,\pm1) & 0           \\ \hline
    0           & 0           & (\pm1,\mp1)
  \end{array}
  \right)
  ,\quad
  \left(
  \begin{array}{cc|c}
    (\pm1,\pm1) & 0           & 0           \\
    0           & (\mp1,\pm1) & 0           \\ \hline
    0           & 0           & (\mp1,\mp1)
  \end{array}
  \right)
  \label{eq:BC-C2D1}
\end{equation}
where the upper left and the lower right blocks separated by the lines correspond to $U(N_C)$ and $U(N_D)$ gauge groups, respectively. The decoding order is the same for each $P$ (the left component in each block) and $P'$ (the right component in each block), since flipping the overall sign of one of $P$ and $P'$ does not change the spectrum\footnote{Thus, the first one of two matrices~(\ref{eq:BC-C2D1}), for instance, indicates the four patterns of the projection,
  \begin{align*}
    \left(
    \begin{array}{cc|c}
      (\pm1,\pm1) & 0           & 0           \\
      0           & (\mp1,\pm1) & 0           \\ \hline
      0           & 0           & (\pm1,\mp1)
    \end{array}
    \right)
     & =
    \left(
    \begin{array}{cc|c}
      (+1,+1) & 0       & 0       \\
      0       & (-1,+1) & 0       \\ \hline
      0       & 0       & (+1,-1)
    \end{array}
    \right),
    \left(
    \begin{array}{cc|c}
      (+1,-1) & 0       & 0       \\
      0       & (-1,-1) & 0       \\ \hline
      0       & 0       & (+1,+1)
    \end{array}
    \right),  \\
     & \qquad
    \left(
    \begin{array}{cc|c}
      (-1,+1) & 0       & 0       \\
      0       & (+1,+1) & 0       \\ \hline
      0       & 0       & (-1,-1)
    \end{array}
    \right),
    \left(
    \begin{array}{cc|c}
      (-1,-1) & 0       & 0       \\
      0       & (+1,-1) & 0       \\ \hline
      0       & 0       & (-1,+1)
    \end{array}
    \right).
  \end{align*}
}.
With this orbifolding, the zero-mode contents become
\begin{equation}
  \Phi_i
  =
  \left(
  \begin{array}{cc|c}
    0 & 0 & 0 \\
    X & 0 & 0 \\ \hline
    0 & 0 & 0
  \end{array}
  \right)
  ,\quad
  \Phi_j
  =
  \left(
  \begin{array}{cc|c}
    0 & 0 & Y \\
    0 & 0 & 0 \\ \hline
    0 & 0 & 0
  \end{array}
  \right)
  ,\quad
  \Phi_k
  =
  \left(
  \begin{array}{cc|c}
    0 & 0 & 0 \\
    0 & 0 & 0 \\ \hline
    0 & Z & 0
  \end{array}
  \right),
  \label{eq:hidden-sector-fields}
\end{equation}
where $(i,j,k)$ is a permutation of $(1,2,3)$, resulting from the choice of the projections\footnote{Note that the zero-mode contents are unique up to the inversion of the representation matrix. Namely, the following form,
  \begin{equation*}
    \Phi_{i}
    =
    \left(
    \begin{array}{cc|c}
      0 & 0  & 0 \\
      0 & 0  & 0 \\ \hline
      0 & X' & 0
    \end{array}
    \right)
    ,\quad  \Phi_{j}
    =
    \left(
    \begin{array}{cc|c}
      0 & Y' & 0 \\
      0 & 0  & 0 \\ \hline
      0 & 0  & 0
    \end{array}
    \right)
    ,\quad  \Phi_{k}
    =
    \left(
    \begin{array}{cc|c}
      0  & 0 & 0 \\
      Z' & 0 & 0 \\ \hline
      0  & 0 & 0
    \end{array}
    \right),
  \end{equation*}
  is also allowed. However, since the following discussions are the same, we only consider Eq.~(\ref{eq:hidden-sector-fields}).}. As expected, Yukawa couplings appear in the superpotential~\eqref{eq:fluctuation-superpotential} and the effective superpotential for the 4D fields $X^{(0)}$, $Y^{(0)}$ and $Z^{(0)}$ at the tree level will be expressed like
\begin{equation}
  W_{\mathrm{tree}}
  =
  \lambda \tr(X^{(0)}Y^{(0)}Z^{(0)})
  \label{eq:tree-level-superpotential}
\end{equation}
where $\lambda$ is the 4D holomorphic Yukawa coupling constant obtained by the overlap integral of the wave functions~(\ref{eq:yukawa-orbifold}). Note that there appear degenerate zero-modes for $X$, $Y$ and $Z$ depending on the magnetic fluxes they feel, which is implicit in the above expressions. The degeneracy will play a role of the flavor in the argument of ADS mechanism.

Next, we consider the pattern $C_1D_2$. The possible projections are
\begin{equation}
  (P_{\rm hid},P'_{\rm hid})
  =
  \left(
  \begin{array}{c|cc}
    (\pm1,\pm1) & 0           & 0           \\ \hline
    0           & (\mp1,\pm1) & 0           \\
    0           & 0           & (\pm1,\mp1)
  \end{array}
  \right)
  ,\quad
  \left(
  \begin{array}{c|cc}
    (\pm1,\pm1) & 0           & 0           \\ \hline
    0           & (\pm1,\mp1) & 0           \\
    0           & 0           & (\mp1,\mp1)
  \end{array}
  \right).
  \label{eq:BC-C1D2}
\end{equation}
The spectrum is similar to the previous case and the holomorphic Yukawa couplings required for the ADS mechanism can be obtained.

We have obtained totally 16 patterns of the orbifold projections~\eqref{eq:BC-C2D1} and \eqref{eq:BC-C1D2}. We can check whether the zero-modes of charged messenger fields $\Phi_i^{AC,CA,AD,DA,BC,CB,BD,DB}$ are eliminated or not on this orbifolds. The notations are similar to those in Eq.~(\ref{eq:phiAB}) for the visible sector, but here we started from $U(N_A) \times U(N_B) \times U(N_C) \times U(N_D)$ SYM in total. It is obvious from Tables~\ref{tab:D7-brane-embedding} and \ref{tab:D7-brane-embedding-hidden} that $\Phi_i^{AC,CA,BD,DB}$ can be eliminated if the localization points of the visible and hidden sectors are different from each other on the second and/or the third torus, as mentioned at the beginning of this section.

On the other hand, the zero-modes in $\Phi_i^{AD,DA,BC,CB}$ can be removed by suitably selecting the orbifold projections $P_{\rm hid}$ and $P'_{\rm hid}$ according to the ones in the visible sector $P$ and $P'$ shown in Appendix~\ref{sec:appendix} for each cases, those should be denoted here as $P_{\rm vis}$ and $P'_{\rm vis}$, respectively, and are summarized as follows.
\begin{eqnarray}
  {\rm Case\ 1} &:&
  (P_{\rm vis},P_{\rm vis}') =
  \left(
  \begin{array}{cc|ccc}
    (1,1) & 0     & 0      & 0       & 0       \\
    0     & (1,1) & 0      & 0       & 0       \\ \hline
    0     & 0     & (1,-1) & 0       & 0       \\
    0     & 0     & 0      & (-1,-1) & 0       \\
    0     & 0     & 0      & 0       & (-1,-1)
  \end{array}
  \right), \nonumber \\
  {\rm Case\ 2} &:&
  (P_{\rm vis},P_{\rm vis}') =
  \left(
  \begin{array}{cc|ccc}
    (1,1) & 0     & 0      & 0      & 0      \\
    0     & (1,1) & 0      & 0      & 0      \\ \hline
    0     & 0     & (1,-1) & 0      & 0      \\
    0     & 0     & 0      & (1,-1) & 0      \\
    0     & 0     & 0      & 0      & (1,-1)
  \end{array}
  \right), \nonumber \\
  {\rm Case\ 3} &:&
  (P_{\rm vis},P_{\rm vis}') =
  \left(
  \begin{array}{cc|ccc}
    (1,1) & 0     & 0       & 0       & 0       \\
    0     & (1,1) & 0       & 0       & 0       \\ \hline
    0     & 0     & (-1,-1) & 0       & 0       \\
    0     & 0     & 0       & (-1,-1) & 0       \\
    0     & 0     & 0       & 0       & (-1,-1)
  \end{array}
  \right), \nonumber \\
  {\rm Case\ 4} &:&
  (P_{\rm vis},P_{\rm vis}') =
  \left(
  \begin{array}{cc|ccc}
    (1,1) & 0     & 0       & 0      & 0      \\
    0     & (1,1) & 0       & 0      & 0      \\ \hline
    0     & 0     & (-1,-1) & 0      & 0      \\
    0     & 0     & 0       & (1,-1) & 0      \\
    0     & 0     & 0       & 0      & (1,-1)
  \end{array}
  \right),
\end{eqnarray}
Then, to eliminate the zero-modes in $\Phi_i^{AD,DA,BC,CB}$, we should choose the parities $(\eta_a^C\eta_b^D,{\eta_a^{C}}'{\eta_b^{D}}')=(\pm1,+1)$ in Table~\ref{tab:parity-D7} with the replacement $A \to C$ and $B \to D$. Due to this restriction, only the following patterns are available for each case:
\begin{align}
  (P_{\rm hid},P_{\rm hid}')
   & =
  \left(
  \begin{array}{cc|c}
    (\pm1,-1) &           &           \\
              & (\mp1,-1) &           \\ \hline
              &           & (\pm1,+1)
  \end{array}
  \right)
  ,\quad
  \left(
  \begin{array}{cc|c}
    (\pm1,-1) &           &           \\
              & (\mp1,-1) &           \\ \hline
              &           & (\mp1,+1)
  \end{array}
  \right),
  \nonumber
  \\
   & \qquad
  \left(
  \begin{array}{c|cc}
    (\pm1,-1) &           &           \\ \hline
              & (\mp1,+1) &           \\
              &           & (\pm1,+1)
  \end{array}
  \right)
  ,\quad
  \left(
  \begin{array}{c|cc}
    (\pm1,-1) &           &           \\ \hline
              & (\pm1,+1) &           \\
              &           & (\mp1,+1)
  \end{array}
  \right).
  \label{eq:available_parity_hidden}
\end{align}

We have found that the above eight patterns of the orbifold projections~(\ref{eq:available_parity_hidden}) lead to the tree-level Yukawa couplings among necessary fields $X^{(0)}$, $Y^{(0)}$ and $Z^{(0)}$ for the ADS mechanism, without yielding the charged messenger fields. Note that the same conclusions are derived for the zero-mode contents~\eqref{eq:hidden-sector-fields} up to the inversion of the representation.

As an example of Case 3, if we select
\begin{align}
  (P_{\rm hid},P_{\rm hid}')
   & =
  \left(
  \begin{array}{c|cc}
    (-1,-1) &         &         \\ \hline
            & (+1,+1) &         \\
            &         & (-1,+1)
  \end{array}
  \right),
\end{align}
for Model 3-4 of the visible sector, the total zero-mode contents are found as
\begin{eqnarray}
  \Phi_1 &=&
  \left(
  \begin{array}{cc|ccc||c|cc}
    0 & 0 & 0   & 0 & 0 & 0 & 0 & 0 \\
    0 & 0 & 0   & 0 & 0 & 0 & 0 & 0 \\ \hline
    0 & 0 & 0   & 0 & 0 & 0 & 0 & 0 \\
    0 & 0 & H_u & 0 & 0 & 0 & 0 & 0 \\
    0 & 0 & H_d & 0 & 0 & 0 & 0 & 0 \\ \hline \hline
    0 & 0 & 0   & 0 & 0 & 0 & 0 & 0 \\\hline
    0 & 0 & 0   & 0 & 0 & 0 & 0 & X \\
    0 & 0 & 0   & 0 & 0 & 0 & 0 & 0
  \end{array}
  \right),
  \nonumber \\
  \Phi_2 &=&
  \left(
  \begin{array}{cc|ccc||c|cc}
    0 & 0 & 0 & U & D & 0 & 0 & 0 \\
    0 & 0 & 0 & N & E & 0 & 0 & 0 \\ \hline
    0 & 0 & 0 & 0 & 0 & 0 & 0 & 0 \\
    0 & 0 & 0 & 0 & 0 & 0 & 0 & 0 \\
    0 & 0 & 0 & 0 & 0 & 0 & 0 & 0 \\ \hline \hline
    0 & 0 & 0 & 0 & 0 & 0 & Y & 0 \\\hline
    0 & 0 & 0 & 0 & 0 & 0 & 0 & 0 \\
    0 & 0 & 0 & 0 & 0 & 0 & 0 & 0
  \end{array}
  \right),
  \nonumber \\
  \Phi_3 &=&
  \left(
  \begin{array}{cc|ccc||c|cc}
    0 & 0 & 0 & 0 & 0 & 0 & 0 & 0 \\
    0 & 0 & 0 & 0 & 0 & 0 & 0 & 0 \\ \hline
    Q & L & 0 & 0 & 0 & 0 & 0 & 0 \\
    0 & 0 & 0 & 0 & 0 & 0 & 0 & 0 \\
    0 & 0 & 0 & 0 & 0 & 0 & 0 & 0 \\ \hline \hline
    0 & 0 & 0 & 0 & 0 & 0 & 0 & 0 \\\hline
    0 & 0 & 0 & 0 & 0 & 0 & 0 & 0 \\
    0 & 0 & 0 & 0 & 0 & Z & 0 & 0
  \end{array}
  \right),
\end{eqnarray}
where the four diagonal blocks separated by the lines correspond to $U(N_A)$, $U(N_B)$, $U(N_C)$ and $U(N_D)$ gauge groups from the upper left to the lower right, respectively. The double line separates visible and hidden sectors. Then, the cubic terms $\epsilon^{ijk} \Tr [\Phi_i \Phi_j \Phi_k]$ in the superpotential~(\ref{eq:fluctuation-superpotential}) yield Yukawa couplings among the chiral superfields $Q$, $U$, $D$, $L$, $N$ and $E$ in the visible sector and those among $X$, $Y$ and $Z$ in the hidden sector, separately. The former determined the flavor structures of quarks and charged leptons in Section~\ref{ssec:flavor_structure}, while the latter will support the dynamical supersymmetry breaking in the hidden sector as shown in the next subsection. Note also that all the charged messenger fields are successfully eliminated.

\subsection{Prospects for dynamical supersymmetry breaking}

For the breaking pattern $C_2D_1$ ($C_1D_2$ will give similar results) the 4D effective theory is $U(N_C^{(1)}) \times U(N_C^{(2)}) \times U(N_D)$ SYM with the tree-level superpotential~\eqref{eq:tree-level-superpotential}. Let us consider the case $N_C^{(2)}=N_D=1$ for simplicity and describe $U(N_C^{(1)})$,  $U(N_C^{(2)})$ and $U(N_D)$ as $U(N_c)$, $U(1)_1$ and $U(1)_2$, respectively, resulting in the $U(N_c) \times U(1)_1 \times U(1)_2$ SYM theory with three bi-fundamental chiral multiplets in the hidden sector. We denote the 4D fields of $U(N_c)$ singlet, fundamental and anti-fundamental representations by $S$, $Q$ and $\tQ$, respectively. Then, the tree-level superpotential after the dimensional reduction takes the form of
\begin{equation}
  W_{\rm tree}=\lambda S Q \tQ,
\end{equation}
where $\lambda$ is the 4D Yukawa coupling constant and the degeneracies of fields are implicit. Since the non-Abelian part in the 4D effective theory is $SU(N_c)$ SYM with $N_f$ flavors of $Q$ and $\tQ$ with $N_f$ determined by the background fluxes, the nonperturbative superpotential $W_{\rm ADS}$ shown in Eq.~(\ref{eq:ADSsuperpotential}) will be generated, whose runaway direction of meson fields $M=Q \tQ$ can be lifted by the above $W_{\rm tree}$ in the F-term scalar potential. On the other hand, since there is no mass term for the singlet $S$ in $W_{\rm tree}$, some higher-dimensional operators and/or higher-order corrections will be necessary to stabilize it.

Here, we comment about the Abelian factors in the remaining gauge symmetries in the hidden sector. They are anomalous in general, and the corresponding gauge bosons may get masses via GS like mechanism, as mentioned at the end of Section~\ref{ssec:u1factors} for those in the visible sector. Their D-term potential could affect the above argument of dynamical supersymmetry breaking, depending on the concrete GS couplings in the ultra-violet completion, that is beyond the scope of this paper.

These Abelian factors can also be spontaneously broken.
It is plausible that matter fields charged under these Abelian factors acquire non-vanishing VEVs
through the above SUSY-breaking scheme, thereby breaking the symmetries.
In this case, the gauge bosons of the broken Abelian factors no longer contribute to dark radiation,
and they do not necessarily need to be canceled by the GS mechanism.

\section{Conclusions and discussions}
\label{sec:conclusion}

10D SYM allows only a single free parameter, nevertheless, having a possibility to reproduce realistic patterns of flavor structure for quarks and leptons, if it possesses magnetic fluxes in extra-dimensions~\cite{Cremades:2004wa}. This is true also on orbifolds, despite the fact that Wilson-lines/fluxes are discretized/quantized~\cite{Abe:2008fi}. In this paper, we showed that this is still true for 7-brane configurations yielding MSSM flavors where all the exotic matter zero-modes can be eliminated. We constructed concrete 7-brane configurations where $SU(3)_C$ and $SU(2)_L$ gauge groups of SM originate from $U(N_A)$ and $U(N_B)$ on 7$_A$- and 7$_B$-branes in Table~\ref{tab:D7-brane-embedding}, respectively, and found that semi-realistic flavor structures of quarks and charged leptons can be obtained if a certain linear combination of multiple Higgs fields obtains a non-vanishing VEV in each configuration. Note that the assignment of symmetries and their breaking pattern adopted in this paper to the 7-brane configuration of Table~\ref{tab:D7-brane-embedding} is not unique. A comprehensive search for the assignments realizing $U(3)_C \times U(2)_L \times U(1)^3$ from $U(N_A) \times U(N_B)$ will be straightforward, that remains as a future work.

We also developed a systematic way to implement a hidden SYM sector into the 7-brane configurations so that there appear no zero-mode fields charged under both visible and hidden gauge groups at all. We remark that the way proposed here can be applied not only to the configurations considered in this paper but also to those with different assignments of symmetries in the visible sector. The hidden sectors obtained in this way allow the appearance of Yukawa couplings, which can play a role to prevent the meson fields in ADS superpotential from running away in the scalar potential. Analyzing the dynamics of supersymmetry breaking in phenomenologically promising configurations also remains as a future work.

It is remarkable that the MSSM-like models obtained so far from magnetized SYM including the present one,
those realize semi-realistic flavor structures, tend to have multiple generations of MSSM Higgs pairs.
Such models share a common problem to have supersymmetric mass terms for Higgs,
namely, the so-called $\mu$-terms.
Especially, the eigenvectors of Higgs mass matrix would determine the VEVs of Higgs generations fixing the flavor structure of quarks and leptons, although the VEVs are treated as parameters in this paper.
From a symmetry point of view, these mass terms can be put at the orbifold fixed points where the supersymmetry is reduced to 4D ${\cal N}=1$~\cite{Ishida:2017avx,Abe:2018qbp}.\footnote{The other terms like Yukawa coupling ones could also appear at the fixed points if the symmetries allow. If absolute values of some of them are lager than the bulk ones, some results in this paper could be modified.}
Note that the origin cannot be perturbative corrections, since they are superpotential terms.

All the results in this paper are at the tree-level, but the loop effects are definitely important to discuss the detailed phenomenological/cosmological aspects. Renormalization group (RG) running effects for Yukawa coupling constants should be included to determine the flavor structures more precisely. The RG running effects for gauge coupling constants would determine the supersymmetry breaking scale in the hidden sector, since all the 7-branes in each configuration share a single gauge coupling constant and, other than the geometric configuration, the RG effects cause those of unbroken gauge groups to differ from each other at low energies.

The loop effects are also important to confirm the stability of configuration itself on orbifolds. In general, anomalies localize on orbifold fixed points and the inflows from bulk to cancel them could affect bulk wavefunctions of chiral fields, then, the classical configurations can be modified~\cite{Arkani-Hamed:2001uol,Scrucca:2001eb}. The analogous phenomenon is that mixed gravitational anomalies induce Fayet-Iliopoulos terms at the orbifold fixed points~\cite{GrootNibbelink:2002qp,Lee:2003mc,Abe:2019nkv,Abe:2020vmv}, that can cause instabilities for bulk fields. In our framework, two kinds of chiral projections by orbifold and magnetic flux (through spin and gauge connections, respectively) are combined, that could cause nontrivial effects related to the anomalies. We would study the loop effects in our configurations elsewhere separately.

One of the other motivations to consider 7-brane configurations, than the phenomenological ones to eliminate exotic modes, is that such configurations would appear as low energy effective theories of magnetized D7-branes. In this case, SYM action is promoted to Dirac-Born-Infeld type non-linear one coupled with bulk supergravity~\cite{Green:1996dd} accompanied with Chern-Simons terms, where the possible magnetic fluxes will be further restricted by conditions to cancel various anomalies as well as tadpoles of supergravity fields. The anomaly cancellation via GS like mechanism may contribute to make, for instance, some of extra $U(1)$ (other than hypercharge) gauge fields massive~\cite{Deser:1981wh} in our 7-brane configurations. If some superstring embeddings are somehow possible,\footnote{Magnetized SYM models might be embedded into matrix models those are expected to provide nonperturbative formulations of superstrings~\cite{Aoki:2010gv}.} the quantum gravitational interactions missed in SM will be consistently described, allowing much more precise analyses for particle phenomenology/cosmology at higher energies.

In such a framework, the dynamical supersymmetry breaking in the hidden sector should be discussed with moduli stabilization issues, which would require some nonperturbative effects such as D-brane instantons~\cite{Ibanez:2006da,Cvetic:2007ku} as well as perturbative ones such as vacuum bubbles~\cite{Appelquist:1983vs}. If the hidden sector is geometrically sequestered from the visible sector like ours, the breaking effect will be mediated to the MSSM sector via moduli~\cite{Kaplunovsky:1993rd} and/or superconformal anomaly~\cite{Randall:1998uk,Giudice:1998xp} in a supergravity extension, that yields soft supersymmetry breaking terms~\cite{Dimopoulos:1981zb,Sakai:1981gr} reducing MSSM to SM. We also remark that the above mentioned nonperturbative effects can also contribute to generate supersymmetric masses for Higgs and right-handed neutrinos~\cite{Ibanez:2006da,Cvetic:2007ku} and are quite relevant for the phenomenology/cosmology. These directions also remain as future works.

\subsection*{Acknowledgement}
H.A. would like to thank Keigo~Sumita and Tatsuo~Kobayashi for early discussions which initiate this work.

\appendix

\section{Three generation models}
\label{sec:appendix}

In this appendix, numerical results of all three generation models obtained are listed.  In the following matrix expressions, the rank of sub-matrix proportional to the unit matrix is implicit and only the overall values are shown (e.g., $m_a\times\bm{1}_{n_a}$ is denoted just as $m_a$).

\subsection*{Case 1: Models with five-generation Higgs pairs (I)}

\subsubsection*{Model 1-1:}
\begin{equation}
  \begin{aligned}
    (M^{(1)},M^{(2)},M^{(3)})             & =
    \left(
\begin{array}{cc|ccc}
      (4,H,-4) &          &         &           &           \\
               & (5,H,-5) &         &           &           \\ \hline
               &          & (0,0,H) &           &           \\
               &          &         & (10,-1,H) &           \\
               &          &         &           & (10,-1,H)
    \end{array}
    \right) ,                                 \\
    (\zeta^{(1)},\zeta^{(2)},\zeta^{(3)}) & =
    \left(
    \begin{array}{cc|ccc}
      (0,0,0) &         &         &                   &                      \\
              & (0,0,0) &         &                   &                      \\ \hline
              &         & (0,0,0) &                   &                      \\
              &         &         & (\frac{1}{2},0,0) &                      \\
              &         &         &                   & (\frac{\tau}{2},0,0)
    \end{array}
    \right) ,                                 \\
    (P,P')                                & =
    \left(
    \begin{array}{cc|ccc}
      (1,1) &       &        &         &         \\
            & (1,1) &        &         &         \\ \hline
            &       & (1,-1) &         &         \\
            &       &        & (-1,-1) &         \\
            &       &        &         & (-1,-1)
    \end{array}
    \right)
  \end{aligned}
\end{equation}
For the above inputs, the flavor structures are obtained as Table~\ref{tab:case 1-1} with
\begin{equation}
  \begin{gathered}
    \text{Im}\tau_1 = 1.8 , \quad \tan\beta = 55, \\
    \langle H_u \rangle = (0,\ 0,\ 0,\ 0.2,\ 0.5)\times v_u \times \mathcal{N}_{H_u}, \quad \langle H_d \rangle = (0,\ 0,\ 0.1,\ 0.4,\ 0.9)\times v_d \times \mathcal{N}_{H_d}.
  \end{gathered}
\end{equation}

\begin{table}[t]
  \centering
  \begin{tabular}{c|cc} \hline
                                                     & Experimental value~\cite{ParticleDataGroup:2022pth} & Model 1-1     \\ \hline
    $(10^5m_{\rm u} , 10^3m_{\rm c})/m_{\rm t}$      & (1.25, 7.35)                                        & (0.535, 35.1) \\
    $(10^3m_{\rm d} , 10^2m_{\rm s})/m_{\rm b}$      & (1.12, 2.23)                                        & (0.529, 2.24) \\
    $(10^4m_{\rm e} , 10^2m_{\rm \mu})/m_{\rm \tau}$ & (2.88, 5.95)                                        & (5.05, 1.90)  \\
    $|V^{\rm{CKM}}|$                                 &
    $\left(
      \begin{matrix}
        0.974 & 0.225  & 0.00382 \\
        0.221 & 0.987  & 0.041   \\
        0.008 & 0.0388 & 1.01
      \end{matrix}
      \right)$                  &
    $\left(
      \begin{matrix}
        0.994  & 0.109 & 0.0179 \\
        0.107  & 0.988 & 0.112  \\
        0.0263 & 0.110 & 0.994
      \end{matrix}
      \right)$                                                                                            \\
    $m_{\rm t}/m_{\rm b}$                            & 41.3                                                & 41.2          \\
    $\chi$                                           & -                                                   & 2.32          \\ \hline
  \end{tabular}
  \caption{Flavor structures of Model 1-1}
  \label{tab:case 1-1}
\end{table}

\subsubsection*{Model 1-2:}
\begin{equation}
  \begin{aligned}
    (M^{(1)},M^{(2)},M^{(3)})             & =
    \left(
    \begin{array}{cc|ccc}
      (4,H,-4) &          &         &           &           \\
               & (5,H,-5) &         &           &           \\ \hline
               &          & (0,0,H) &           &           \\
               &          &         & (10,-1,H) &           \\
               &          &         &           & (10,-1,H)
    \end{array}
    \right) ,                                 \\
    (\zeta^{(1)},\zeta^{(2)},\zeta^{(3)}) & =
    \left(
    \begin{array}{cc|ccc}
      (0,0,0) &         &         &                      &                   \\
              & (0,0,0) &         &                      &                   \\ \hline
              &         & (0,0,0) &                      &                   \\
              &         &         & (\frac{\tau}{2},0,0) &                   \\
              &         &         &                      & (\frac{1}{2},0,0)
    \end{array}
    \right) ,                                 \\
    (P,P')                                & =
    \left(
    \begin{array}{cc|ccc}
      (1,1) &       &        &         &         \\
            & (1,1) &        &         &         \\ \hline
            &       & (1,-1) &         &         \\
            &       &        & (-1,-1) &         \\
            &       &        &         & (-1,-1)
    \end{array}
    \right)
  \end{aligned}
\end{equation}
For the above inputs, the flavor structures are obtained as Table~\ref{tab:case 1-2} with
\begin{equation}
  \begin{gathered}
    \text{Im}\tau_1 = 1.3 , \quad \tan\beta = 40, \\
    \langle H_u \rangle = (0,\ 0,\ 0,\ 0.2,\ 0.5)\times v_u \times \mathcal{N}_{H_u}, \quad \langle H_d \rangle = (0.8,\ 0.6,\ 0.1,\ 0,\ 0)\times v_d \times \mathcal{N}_{H_d}.
  \end{gathered}
\end{equation}

\begin{table}[t]
  \centering
  \begin{tabular}{c|cc} \hline
                                                     & Experimental value~\cite{ParticleDataGroup:2022pth} & Model 1-2     \\ \hline
    $(10^5m_{\rm u} , 10^3m_{\rm c})/m_{\rm t}$      & (1.25, 7.35)                                        & (1.13, 19.7)  \\
    $(10^3m_{\rm d} , 10^2m_{\rm s})/m_{\rm b}$      & (1.12, 2.23)                                        & (0.729, 6.96) \\
    $(10^4m_{\rm e} , 10^2m_{\rm \mu})/m_{\rm \tau}$ & (2.88, 5.95)                                        & (6.76, 7.38)  \\
    $|V^{\rm{CKM}}|$                                 &
    $\left(
      \begin{matrix}
        0.974 & 0.225  & 0.00382 \\
        0.221 & 0.987  & 0.041   \\
        0.008 & 0.0388 & 1.01
      \end{matrix}
      \right)$                  &
    $\left(
      \begin{matrix}
        0.994   & 0.107  & 0.00532 \\
        0.107   & 0.991  & 0.0754  \\
        0.00486 & 0.0754 & 0.997
      \end{matrix}
      \right)$                                                                                         \\
    $m_{\rm t}/m_{\rm b}$                            & 41.3                                                & 39.7          \\
    $\chi$                                           & -                                                   & 1.03          \\ \hline
  \end{tabular}
  \caption{Flavor structures of Model 1-2}
  \label{tab:case 1-2}
\end{table}

\subsubsection*{Model 1-3:}
\begin{equation}
  \begin{aligned}
    (M^{(1)},M^{(2)},M^{(3)})             & =
    \left(
    \begin{array}{cc|ccc}
      (5,H,-5) &          &         &           &           \\
               & (4,H,-4) &         &           &           \\ \hline
               &          & (0,0,H) &           &           \\
               &          &         & (10,-1,H) &           \\
               &          &         &           & (10,-1,H)
    \end{array}
    \right) ,                                 \\
    (\zeta^{(1)},\zeta^{(2)},\zeta^{(3)}) & =
    \left(
    \begin{array}{cc|ccc}
      (0,0,0) &         &         &                   &                      \\
              & (0,0,0) &         &                   &                      \\ \hline
              &         & (0,0,0) &                   &                      \\
              &         &         & (\frac{1}{2},0,0) &                      \\
              &         &         &                   & (\frac{\tau}{2},0,0)
    \end{array}
    \right) ,                                 \\
    (P,P')                                & =
    \left(
    \begin{array}{cc|ccc}
      (1,1) &       &        &         &         \\
            & (1,1) &        &         &         \\ \hline
            &       & (1,-1) &         &         \\
            &       &        & (-1,-1) &         \\
            &       &        &         & (-1,-1)
    \end{array}
    \right)
  \end{aligned}
\end{equation}
For the above inputs, the flavor structures are obtained as Table~\ref{tab:case 1-3} with
\begin{equation}
  \begin{gathered}
    \text{Im}\tau_1 = 1.4 , \quad \tan\beta = 55, \\
    \langle H_u \rangle = (0,\ 0,\ 0,\ 0.3,\ 1)\times v_u \times \mathcal{N}_{H_u}, \quad \langle H_d \rangle = (0,\ 0,\ 0,\ 0,\ 0.3)\times v_d \times \mathcal{N}_{H_d}.
  \end{gathered}
\end{equation}

\begin{table}[t]
  \centering
  \begin{tabular}{c|cc} \hline
                                                     & Experimental value~\cite{ParticleDataGroup:2022pth} & Model 1-3     \\ \hline
    $(10^5m_{\rm u} , 10^3m_{\rm c})/m_{\rm t}$      & (1.25, 7.35)                                        & (11.9, 19.5)  \\
    $(10^3m_{\rm d} , 10^2m_{\rm s})/m_{\rm b}$      & (1.12, 2.23)                                        & (0.147, 4.66) \\
    $(10^4m_{\rm e} , 10^2m_{\rm \mu})/m_{\rm \tau}$ & (2.88, 5.95)                                        & (1.47, 5.16)  \\
    $|V^{\rm{CKM}}|$                                 &
    $\left(
      \begin{matrix}
        0.974 & 0.225  & 0.00382 \\
        0.221 & 0.987  & 0.041   \\
        0.008 & 0.0388 & 1.01
      \end{matrix}
      \right)$                  &
    $\left(
      \begin{matrix}
        0.996   & 0.0864 & 0.00446 \\
        0.0862  & 0.995  & 0.0532  \\
        0.00751 & 0.0529 & 0.999
      \end{matrix}
      \right)$                                                                                         \\
    $m_{\rm t}/m_{\rm b}$                            & 41.3                                                & 41.8          \\
    $\chi$                                           & -                                                   & 2.48          \\ \hline
  \end{tabular}
  \caption{Flavor structures of Model 1-3}
  \label{tab:case 1-3}
\end{table}

\subsubsection*{Model 1-4:}
\begin{equation}
  \begin{aligned}
    (M^{(1)},M^{(2)},M^{(3)})             & =
    \left(
    \begin{array}{cc|ccc}
      (5,H,-5) &          &         &           &           \\
               & (4,H,-4) &         &           &           \\ \hline
               &          & (0,0,H) &           &           \\
               &          &         & (10,-1,H) &           \\
               &          &         &           & (10,-1,H)
    \end{array}
    \right) ,                                 \\
    (\zeta^{(1)},\zeta^{(2)},\zeta^{(3)}) & =
    \left(
    \begin{array}{cc|ccc}
      (0,0,0) &         &         &                      &                   \\
              & (0,0,0) &         &                      &                   \\ \hline
              &         & (0,0,0) &                      &                   \\
              &         &         & (\frac{\tau}{2},0,0) &                   \\
              &         &         &                      & (\frac{1}{2},0,0)
    \end{array}
    \right) ,                                 \\
    (P,P')                                & =
    \left(
    \begin{array}{cc|ccc}
      (1,1) &       &        &         &         \\
            & (1,1) &        &         &         \\ \hline
            &       & (1,-1) &         &         \\
            &       &        & (-1,-1) &         \\
            &       &        &         & (-1,-1)
    \end{array}
    \right)
  \end{aligned}
\end{equation}
For the above inputs, the flavor structures are obtained as Table~\ref{tab:case 1-4} with
\begin{equation}
  \begin{gathered}
    \text{Im}\tau_1 = 1.5 , \quad \tan\beta = 45, \\
    \langle H_u \rangle = (0.6,\ 0.1,\ 0,\ 0,\ 0)\times v_u \times \mathcal{N}_{H_u}, \quad \langle H_d \rangle = (0.8,\ 0.5,\ 0.1,\ 0,\ 0)\times v_d \times \mathcal{N}_{H_d}.
  \end{gathered}
\end{equation}

\begin{table}[t]
  \centering
  \begin{tabular}{c|cc} \hline
                                                     & Experimental value~\cite{ParticleDataGroup:2022pth} & Model 1-4     \\ \hline
    $(10^5m_{\rm u} , 10^3m_{\rm c})/m_{\rm t}$      & (1.25, 7.35)                                        & (0.795, 14.2) \\
    $(10^3m_{\rm d} , 10^2m_{\rm s})/m_{\rm b}$      & (1.12, 2.23)                                        & (0.450, 7.95) \\
    $(10^4m_{\rm e} , 10^2m_{\rm \mu})/m_{\rm \tau}$ & (2.88, 5.95)                                        & (4.65, 7.39)  \\
    $|V^{\rm{CKM}}|$                                 &
    $\left(
      \begin{matrix}
        0.974 & 0.225  & 0.00382 \\
        0.221 & 0.987  & 0.041   \\
        0.008 & 0.0388 & 1.01
      \end{matrix}
      \right)$                  &
    $\left(
      \begin{matrix}
        0.994   & 0.110  & 0.00731 \\
        0.110   & 0.990  & 0.0855  \\
        0.00824 & 0.0854 & 0.996
      \end{matrix}
      \right)$                                                                                         \\
    $m_{\rm t}/m_{\rm b}$                            & 41.3                                                & 41.9          \\
    $\chi$                                           & -                                                   & 1.12          \\ \hline
  \end{tabular}
  \caption{Flavor structures of Model 1-4}
  \label{tab:case 1-4}
\end{table}

\subsection*{Case 2: Models with five-generation Higgs pairs (II)}

\subsubsection*{Model 2-1:}
\begin{equation}
  \begin{aligned}
    (M^{(1)},M^{(2)},M^{(3)})             & =
    \left(
    \begin{array}{cc|ccc}
      (4,H,-4) &          &         &           &           \\
               & (5,H,-5) &         &           &           \\ \hline
               &          & (0,0,H) &           &           \\
               &          &         & (11,-1,H) &           \\
               &          &         &           & (11,-1,H)
    \end{array}
    \right) ,                                 \\
    (\zeta^{(1)},\zeta^{(2)},\zeta^{(3)}) & =
    \left(
    \begin{array}{cc|ccc}
      (0,0,0) &         &         &                   &                      \\
              & (0,0,0) &         &                   &                      \\ \hline
              &         & (0,0,0) &                   &                      \\
              &         &         & (\frac{1}{2},0,0) &                      \\
              &         &         &                   & (\frac{\tau}{2},0,0)
    \end{array}
    \right) ,                                 \\
    (P,P')                                & =
    \left(
    \begin{array}{cc|ccc}
      (1,1) &       &        &        &        \\
            & (1,1) &        &        &        \\ \hline
            &       & (1,-1) &        &        \\
            &       &        & (1,-1) &        \\
            &       &        &        & (1,-1)
    \end{array}
    \right)
  \end{aligned}
\end{equation}
For the above inputs, the flavor structures are obtained as Table~\ref{tab:case 2-1} with
\begin{equation}
  \begin{gathered}
    \text{Im}\tau_1 = 1.6 , \quad \tan\beta = 45, \\
    \langle H_u \rangle = (1,\ 0.2,\ 0,\ 0,\ 0)\times v_u \times \mathcal{N}_{H_u}, \quad \langle H_d \rangle = (0.9,\ 0.7,\ 0,\ 0,\ 0)\times v_d \times \mathcal{N}_{H_d}.
  \end{gathered}
\end{equation}

\begin{table}[t]
  \centering
  \begin{tabular}{c|cc} \hline
                                                     & Experimental value~\cite{ParticleDataGroup:2022pth} & Model 2-1     \\ \hline
    $(10^5m_{\rm u} , 10^3m_{\rm c})/m_{\rm t}$      & (1.25, 7.35)                                        & (2.37, 13.2)  \\
    $(10^3m_{\rm d} , 10^2m_{\rm s})/m_{\rm b}$      & (1.12, 2.23)                                        & (0.583, 3.20) \\
    $(10^4m_{\rm e} , 10^2m_{\rm \mu})/m_{\rm \tau}$ & (2.88, 5.95)                                        & (5.59, 3.20)  \\
    $|V^{\rm{CKM}}|$                                 &
    $\left(
      \begin{matrix}
        0.974 & 0.225  & 0.00382 \\
        0.221 & 0.987  & 0.041   \\
        0.008 & 0.0388 & 1.01
      \end{matrix}
      \right)$                  &
    $\left(
      \begin{matrix}
        0.990   & 0.139  & 0.00660 \\
        0.139   & 0.989  & 0.0471  \\
        0.00511 & 0.0473 & 0.999
      \end{matrix}
      \right)$                                                                                         \\
    $m_{\rm t}/m_{\rm b}$                            & 41.3                                                & 41.8          \\
    $\chi$                                           & -                                                   & 0.580         \\ \hline
  \end{tabular}
  \caption{Flavor structures of Model 2-1}
  \label{tab:case 2-1}
\end{table}

\subsubsection*{Model 2-2:}
\begin{equation}
  \begin{aligned}
    (M^{(1)},M^{(2)},M^{(3)})             & =
    \left(
    \begin{array}{cc|ccc}
      (4,H,-4) &          &         &           &           \\
               & (5,H,-5) &         &           &           \\ \hline
               &          & (0,0,H) &           &           \\
               &          &         & (11,-1,H) &           \\
               &          &         &           & (11,-1,H)
    \end{array}
    \right) ,                                 \\
    (\zeta^{(1)},\zeta^{(2)},\zeta^{(3)}) & =
    \left(
    \begin{array}{cc|ccc}
      (0,0,0) &         &         &                      &                   \\
              & (0,0,0) &         &                      &                   \\ \hline
              &         & (0,0,0) &                      &                   \\
              &         &         & (\frac{\tau}{2},0,0) &                   \\
              &         &         &                      & (\frac{1}{2},0,0)
    \end{array}
    \right) ,                                 \\
    (P,P')                                & =
    \left(
    \begin{array}{cc|ccc}
      (1,1) &       &        &        &        \\
            & (1,1) &        &        &        \\ \hline
            &       & (1,-1) &        &        \\
            &       &        & (1,-1) &        \\
            &       &        &        & (1,-1)
    \end{array}
    \right)
  \end{aligned}
\end{equation}
For the above inputs, the flavor structures are obtained as Table~\ref{tab:case 2-2} with
\begin{equation}
  \begin{gathered}
    \text{Im}\tau_1 = 1.2 , \quad \tan\beta = 50, \\
    \langle H_u \rangle = (0,\ 0,\ 0.1,\ 0.4\ 1)\times v_u \times \mathcal{N}_{H_u}, \quad \langle H_d \rangle = (0,\ 0,\ 0,\ 0.1,\ 0.6)\times v_d \times \mathcal{N}_{H_d}.
  \end{gathered}
\end{equation}

\begin{table}[t]
  \centering
  \begin{tabular}{c|cc} \hline
                                                     & Experimental value~\cite{ParticleDataGroup:2022pth} & Model 2-2     \\ \hline
    $(10^5m_{\rm u} , 10^3m_{\rm c})/m_{\rm t}$      & (1.25, 7.35)                                        & (5.46, 61.5)  \\
    $(10^3m_{\rm d} , 10^2m_{\rm s})/m_{\rm b}$      & (1.12, 2.23)                                        & (0.507, 5.03) \\
    $(10^4m_{\rm e} , 10^2m_{\rm \mu})/m_{\rm \tau}$ & (2.88, 5.95)                                        & (4.59, 4.17)  \\
    $|V^{\rm{CKM}}|$                                 &
    $\left(
      \begin{matrix}
        0.974 & 0.225  & 0.00382 \\
        0.221 & 0.987  & 0.041   \\
        0.008 & 0.0388 & 1.01
      \end{matrix}
      \right)$                  &
    $\left(
      \begin{matrix}
        0.985   & 0.174  & 0.00327 \\
        0.174   & 0.984  & 0.0450  \\
        0.00859 & 0.0443 & 0.999
      \end{matrix}
      \right)$                                                                                         \\
    $m_{\rm t}/m_{\rm b}$                            & 41.3                                                & 40.6          \\
    $\chi$                                           & -                                                   & 1.60          \\ \hline
  \end{tabular}
  \caption{Flavor structures of Model 2-2}
  \label{tab:case 2-2}
\end{table}

\subsubsection*{Model 2-3:}
\begin{equation}
  \begin{aligned}
    (M^{(1)},M^{(2)},M^{(3)})             & =
    \left(
    \begin{array}{cc|ccc}
      (5,H,-5) &          &         &           &           \\
               & (4,H,-4) &         &           &           \\ \hline
               &          & (0,0,H) &           &           \\
               &          &         & (11,-1,H) &           \\
               &          &         &           & (11,-1,H)
    \end{array}
    \right) ,                                 \\
    (\zeta^{(1)},\zeta^{(2)},\zeta^{(3)}) & =
    \left(
    \begin{array}{cc|ccc}
      (0,0,0) &         &         &                   &                      \\
              & (0,0,0) &         &                   &                      \\ \hline
              &         & (0,0,0) &                   &                      \\
              &         &         & (\frac{1}{2},0,0) &                      \\
              &         &         &                   & (\frac{\tau}{2},0,0)
    \end{array}
    \right) ,                                 \\
    (P,P')                                & =
    \left(
    \begin{array}{cc|ccc}
      (1,1) &       &        &        &        \\
            & (1,1) &        &        &        \\ \hline
            &       & (1,-1) &        &        \\
            &       &        & (1,-1) &        \\
            &       &        &        & (1,-1)
    \end{array}
    \right)
  \end{aligned}
\end{equation}
For the above inputs, the flavor structures are obtained as Table~\ref{tab:case 2-3} with
\begin{equation}
  \begin{gathered}
    \text{Im}\tau_1 = 1.7 , \quad \tan\beta = 45, \\
    \langle H_u \rangle = (1,\ 0.2,\ 0,\ 0,\ 0)\times v_u \times \mathcal{N}_{H_u}, \quad \langle H_d \rangle = (1,\ 0.7,\ 0,\ 0,\ 0)\times v_d \times \mathcal{N}_{H_d}.
  \end{gathered}
\end{equation}

\begin{table}[t]
  \centering
  \begin{tabular}{c|cc} \hline
                                                     & Experimental value~\cite{ParticleDataGroup:2022pth} & Model 2-3     \\ \hline
    $(10^5m_{\rm u} , 10^3m_{\rm c})/m_{\rm t}$      & (1.25, 7.35)                                        & (1.66, 8.87)  \\
    $(10^3m_{\rm d} , 10^2m_{\rm s})/m_{\rm b}$      & (1.12, 2.23)                                        & (0.383, 2.08) \\
    $(10^4m_{\rm e} , 10^2m_{\rm \mu})/m_{\rm \tau}$ & (2.88, 5.95)                                        & (3.93, 1.97)  \\
    $|V^{\rm{CKM}}|$                                 &
    $\left(
      \begin{matrix}
        0.974 & 0.225  & 0.00382 \\
        0.221 & 0.987  & 0.041   \\
        0.008 & 0.0388 & 1.01
      \end{matrix}
      \right)$                  &
    $\left(
      \begin{matrix}
        0.979  & 0.206  & 0.0104 \\
        0.205  & 0.977  & 0.0554 \\
        0.0119 & 0.0551 & 0.998
      \end{matrix}
      \right)$                                                                                           \\
    $m_{\rm t}/m_{\rm b}$                            & 41.3                                                & 41.0          \\
    $\chi$                                           & -                                                   & 0.750         \\ \hline
  \end{tabular}
  \caption{Flavor structures of Model 2-3}
  \label{tab:case 2-3}
\end{table}

\subsubsection*{Model 2-4:}
\begin{equation}
  \begin{aligned}
    (M^{(1)},M^{(2)},M^{(3)})             & =
    \left(
    \begin{array}{cc|ccc}
      (5,H,-5) &          &         &           &           \\
               & (4,H,-4) &         &           &           \\ \hline
               &          & (0,0,H) &           &           \\
               &          &         & (11,-1,H) &           \\
               &          &         &           & (11,-1,H)
    \end{array}
    \right) ,                                 \\
    (\zeta^{(1)},\zeta^{(2)},\zeta^{(3)}) & =
    \left(
    \begin{array}{cc|ccc}
      (0,0,0) &         &         &                      &                   \\
              & (0,0,0) &         &                      &                   \\ \hline
              &         & (0,0,0) &                      &                   \\
              &         &         & (\frac{\tau}{2},0,0) &                   \\
              &         &         &                      & (\frac{1}{2},0,0)
    \end{array}
    \right) ,                                 \\
    (P,P')                                & =
    \left(
    \begin{array}{cc|ccc}
      (1,1) &       &        &        &        \\
            & (1,1) &        &        &        \\ \hline
            &       & (1,-1) &        &        \\
            &       &        & (1,-1) &        \\
            &       &        &        & (1,-1)
    \end{array}
    \right)
  \end{aligned}
\end{equation}
For the above inputs, the flavor structures are obtained as Table~\ref{tab:case 2-4} with
\begin{equation}
  \begin{gathered}
    \text{Im}\tau_1 = 1.8 , \quad \tan\beta = 55, \\
    \langle H_u \rangle = (0,\ 0,\ 0,\ 0.2,\ 1)\times v_u \times \mathcal{N}_{H_u}, \quad \langle H_d \rangle = (0,\ 0,\ 0.1,\ 0.2,\ 0.7)\times v_d \times \mathcal{N}_{H_d}.
  \end{gathered}
\end{equation}

\begin{table}[t]
  \centering
  \begin{tabular}{c|cc} \hline
                                                     & Experimental value~\cite{ParticleDataGroup:2022pth} & Model 2-4     \\ \hline
    $(10^5m_{\rm u} , 10^3m_{\rm c})/m_{\rm t}$      & (1.25, 7.35)                                        & (3.11, 1.88)  \\
    $(10^3m_{\rm d} , 10^2m_{\rm s})/m_{\rm b}$      & (1.12, 2.23)                                        & (0.377, 5.91) \\
    $(10^4m_{\rm e} , 10^2m_{\rm \mu})/m_{\rm \tau}$ & (2.88, 5.95)                                        & (4.21, 7.82)  \\
    $|V^{\rm{CKM}}|$                                 &
    $\left(
      \begin{matrix}
        0.974 & 0.225  & 0.00382 \\
        0.221 & 0.987  & 0.041   \\
        0.008 & 0.0388 & 1.01
      \end{matrix}
      \right)$                  &
    $\left(
      \begin{matrix}
        0.993   & 0.122 & 0.00593 \\
        0.122   & 0.986 & 0.118   \\
        0.00871 & 0.118 & 0.993
      \end{matrix}
      \right)$                                                                                          \\
    $m_{\rm t}/m_{\rm b}$                            & 41.3                                                & 42.0          \\
    $\chi$                                           & -                                                   & 1.57          \\ \hline
  \end{tabular}
  \caption{Flavor structures of Model 2-4}
  \label{tab:case 2-4}
\end{table}

\subsection*{Case 3: Models with six-generation Higgs pairs}

\subsubsection*{Model 3-1:}
\begin{equation}
  \begin{aligned}
    (M^{(1)},M^{(2)},M^{(3)})             & =
    \left(
    \begin{array}{cc|ccc}
      (7,H,-7) &          &         &           &           \\
               & (8,H,-8) &         &           &           \\ \hline
               &          & (0,0,H) &           &           \\
               &          &         & (13,-1,H) &           \\
               &          &         &           & (13,-1,H)
    \end{array}
    \right) ,                                 \\
    (\zeta^{(1)},\zeta^{(2)},\zeta^{(3)}) & =
    \left(
    \begin{array}{cc|ccc}
      (0,0,0) &         &         &                   &                      \\
              & (0,0,0) &         &                   &                      \\ \hline
              &         & (0,0,0) &                   &                      \\
              &         &         & (\frac{1}{2},0,0) &                      \\
              &         &         &                   & (\frac{\tau}{2},0,0)
    \end{array}
    \right) ,                                 \\
    (P,P')                                & =
    \left(
    \begin{array}{cc|ccc}
      (1,1) &       &         &         &         \\
            & (1,1) &         &         &         \\ \hline
            &       & (-1,-1) &         &         \\
            &       &         & (-1,-1) &         \\
            &       &         &         & (-1,-1)
    \end{array}
    \right)
  \end{aligned}
\end{equation}
For the above inputs, the flavor structures are obtained as Table~\ref{tab:case 3-1} with
\begin{equation}
  \begin{gathered}
    \text{Im}\tau_1 = 1.5 , \quad \tan\beta = 50, \\
    \langle H_u \rangle = (0,\ 0,\ 0,\ 0.2,\ 0.9,\ 0)\times v_u \times \mathcal{N}_{H_u}, \quad \langle H_d \rangle = (0,\ 0,\ 0,\ 0,\ 0,\ 0.2)\times v_d \times \mathcal{N}_{H_d}.
  \end{gathered}
\end{equation}

\begin{table}[t]
  \centering
  \begin{tabular}{c|cc} \hline
                                                     & Experimental value~\cite{ParticleDataGroup:2022pth} & Model 3-1     \\ \hline
    $(10^5m_{\rm u} , 10^3m_{\rm c})/m_{\rm t}$      & (1.25, 7.35)                                        & (2.38, 5.31)  \\
    $(10^3m_{\rm d} , 10^2m_{\rm s})/m_{\rm b}$      & (1.12, 2.23)                                        & (0.149, 4.90) \\
    $(10^4m_{\rm e} , 10^2m_{\rm \mu})/m_{\rm \tau}$ & (2.88, 5.95)                                        & (1.46, 4.71)  \\
    $|V^{\rm{CKM}}|$                                 &
    $\left(
      \begin{matrix}
        0.974 & 0.225  & 0.00382 \\
        0.221 & 0.987  & 0.041   \\
        0.008 & 0.0388 & 1.01
      \end{matrix}
      \right)$                  &
    $\left(
      \begin{matrix}
        0.990  & 0.137  & 0.0166 \\
        0.137  & 0.989  & 0.0564 \\
        0.0161 & 0.0565 & 0.998
      \end{matrix}
      \right)$                                                                                           \\
    $m_{\rm t}/m_{\rm b}$                            & 41.3                                                & 41.7          \\
    $\chi$                                           & -                                                   & 1.71          \\ \hline
  \end{tabular}
  \caption{Flavor structures of Model 3-1}
  \label{tab:case 3-1}
\end{table}

\subsubsection*{Model 3-2:}
\begin{equation}
  \begin{aligned}
    (M^{(1)},M^{(2)},M^{(3)})             & =
    \left(
    \begin{array}{cc|ccc}
      (7,H,-7) &          &         &           &           \\
               & (8,H,-8) &         &           &           \\ \hline
               &          & (0,0,H) &           &           \\
               &          &         & (13,-1,H) &           \\
               &          &         &           & (13,-1,H)
    \end{array}
    \right) ,                                 \\
    (\zeta^{(1)},\zeta^{(2)},\zeta^{(3)}) & =
    \left(
    \begin{array}{cc|ccc}
      (0,0,0) &         &         &                      &                   \\
              & (0,0,0) &         &                      &                   \\ \hline
              &         & (0,0,0) &                      &                   \\
              &         &         & (\frac{\tau}{2},0,0) &                   \\
              &         &         &                      & (\frac{1}{2},0,0)
    \end{array}
    \right) ,                                 \\
    (P,P')                                & =
    \left(
    \begin{array}{cc|ccc}
      (1,1) &       &         &         &         \\
            & (1,1) &         &         &         \\ \hline
            &       & (-1,-1) &         &         \\
            &       &         & (-1,-1) &         \\
            &       &         &         & (-1,-1)
    \end{array}
    \right)
  \end{aligned}
\end{equation}
For the above inputs, the flavor structures are obtained as Table~\ref{tab:case 3-2} with
\begin{equation}
  \begin{gathered}
    \text{Im}\tau_1 = 1.4 , \quad \tan\beta = 25, \\
    \langle H_u \rangle = (0.4,\ 0.9,\ 0,\ 0\ 0,\ 0)\times v_u \times \mathcal{N}_{H_u}, \quad \langle H_d \rangle = (1,\ 0.3,\ 0,\ 0,\ 0,\ 0)\times v_d \times \mathcal{N}_{H_d}.
  \end{gathered}
\end{equation}

\begin{table}[t]
  \centering
  \begin{tabular}{c|cc} \hline
                                                     & Experimental value~\cite{ParticleDataGroup:2022pth} & Model 3-2     \\ \hline
    $(10^5m_{\rm u} , 10^3m_{\rm c})/m_{\rm t}$      & (1.25, 7.35)                                        & (1.55, 5.17)  \\
    $(10^3m_{\rm d} , 10^2m_{\rm s})/m_{\rm b}$      & (1.12, 2.23)                                        & (0.395, 2.62) \\
    $(10^4m_{\rm e} , 10^2m_{\rm \mu})/m_{\rm \tau}$ & (2.88, 5.95)                                        & (3.24, 2.66)  \\
    $|V^{\rm{CKM}}|$                                 &
    $\left(
      \begin{matrix}
        0.974 & 0.225  & 0.00382 \\
        0.221 & 0.987  & 0.041   \\
        0.008 & 0.0388 & 1.01
      \end{matrix}
      \right)$                  &
    $\left(
      \begin{matrix}
        0.987   & 0.163  & 0.00567 \\
        0.163   & 0.985  & 0.0485  \\
        0.00318 & 0.0487 & 0.999
      \end{matrix}
      \right)$                                                                                         \\
    $m_{\rm t}/m_{\rm b}$                            & 41.3                                                & 41.4          \\
    $\chi$                                           & -                                                   & 0.609         \\ \hline
  \end{tabular}
  \caption{Flavor structures of Model 3-2}
  \label{tab:case 3-2}
\end{table}

\subsubsection*{Model 3-3:}
\begin{equation}
  \begin{aligned}
    (M^{(1)},M^{(2)},M^{(3)})             & =
    \left(
    \begin{array}{cc|ccc}
      (8,H,-8) &          &         &           &           \\
               & (7,H,-7) &         &           &           \\ \hline
               &          & (0,0,H) &           &           \\
               &          &         & (13,-1,H) &           \\
               &          &         &           & (13,-1,H)
    \end{array}
    \right) ,                                 \\
    (\zeta^{(1)},\zeta^{(2)},\zeta^{(3)}) & =
    \left(
    \begin{array}{cc|ccc}
      (0,0,0) &         &         &                   &                      \\
              & (0,0,0) &         &                   &                      \\ \hline
              &         & (0,0,0) &                   &                      \\
              &         &         & (\frac{1}{2},0,0) &                      \\
              &         &         &                   & (\frac{\tau}{2},0,0)
    \end{array}
    \right) ,                                 \\
    (P,P')                                & =
    \left(
    \begin{array}{cc|ccc}
      (1,1) &       &         &         &         \\
            & (1,1) &         &         &         \\ \hline
            &       & (-1,-1) &         &         \\
            &       &         & (-1,-1) &         \\
            &       &         &         & (-1,-1)
    \end{array}
    \right)
  \end{aligned}
\end{equation}
For the above inputs, the flavor structures are obtained as Table~\ref{tab:case 3-3} with
\begin{equation}
  \begin{gathered}
    \text{Im}\tau_1 = 1.5 , \quad \tan\beta = 50, \\
    \langle H_u \rangle = (0,\ 0,\ 0,\ 0.2,\ 0.9,\ 0)\times v_u \times \mathcal{N}_{H_u}, \quad \langle H_d \rangle = (0,\ 0,\ 0,\ 0,\ 0,\ 0.2)\times v_d \times \mathcal{N}_{H_d}.
  \end{gathered}
\end{equation}

\begin{table}[t]
  \centering
  \begin{tabular}{c|cc} \hline
                                                     & Experimental value~\cite{ParticleDataGroup:2022pth} & Model 3-3     \\ \hline
    $(10^5m_{\rm u} , 10^3m_{\rm c})/m_{\rm t}$      & (1.25, 7.35)                                        & (2.31, 5.81)  \\
    $(10^3m_{\rm d} , 10^2m_{\rm s})/m_{\rm b}$      & (1.12, 2.23)                                        & (0.146, 4.71) \\
    $(10^4m_{\rm e} , 10^2m_{\rm \mu})/m_{\rm \tau}$ & (2.88, 5.95)                                        & (1.49, 4.90)  \\
    $|V^{\rm{CKM}}|$                                 &
    $\left(
      \begin{matrix}
        0.974 & 0.225  & 0.00382 \\
        0.221 & 0.987  & 0.041   \\
        0.008 & 0.0388 & 1.01
      \end{matrix}
      \right)$                  &
    $\left(
      \begin{matrix}
        0.994   & 0.107  & 0.00942 \\
        0.107   & 0.993  & 0.0400  \\
        0.00850 & 0.0402 & 0.999
      \end{matrix}
      \right)$                                                                                         \\
    $m_{\rm t}/m_{\rm b}$                            & 41.3                                                & 41.4          \\
    $\chi$                                           & -                                                   & 1.42          \\ \hline
  \end{tabular}
  \caption{Flavor structures of Model 3-3}
  \label{tab:case 3-3}
\end{table}

\subsubsection*{Model 3-4:} \label{subsec:model 3-4}
\begin{equation}
  \begin{aligned}
    (M^{(1)},M^{(2)},M^{(3)})             & =
    \left(
    \begin{array}{cc|ccc}
      (8,H,-8) &          &         &           &           \\
               & (7,H,-7) &         &           &           \\ \hline
               &          & (0,0,H) &           &           \\
               &          &         & (13,-1,H) &           \\
               &          &         &           & (13,-1,H)
    \end{array}
    \right) ,                                 \\
    (\zeta^{(1)},\zeta^{(2)},\zeta^{(3)}) & =
    \left(
    \begin{array}{cc|ccc}
      (0,0,0) &         &         &                      &                   \\
              & (0,0,0) &         &                      &                   \\ \hline
              &         & (0,0,0) &                      &                   \\
              &         &         & (\frac{\tau}{2},0,0) &                   \\
              &         &         &                      & (\frac{1}{2},0,0)
    \end{array}
    \right) ,                                 \\
    (P,P')                                & =
    \left(
    \begin{array}{cc|ccc}
      (1,1) &       &         &         &         \\
            & (1,1) &         &         &         \\ \hline
            &       & (-1,-1) &         &         \\
            &       &         & (-1,-1) &         \\
            &       &         &         & (-1,-1)
    \end{array}
    \right)
  \end{aligned}
  \label{eq:model3-4}
\end{equation}
For the above inputs, the flavor structures are obtained as Table~\ref{tab:case 3-4} with
\begin{equation}
  \begin{gathered}
    \text{Im}\tau_1 = 1.4 , \quad \tan\beta = 40, \\
    \langle H_u \rangle = (0.5,\ 0.7,\ 0,\ 0,\ 0,\ 0)\times v_u \times \mathcal{N}_{H_u}, \quad \langle H_d \rangle = (0.8,\ 0.3,\ 0,\ 0,\ 0,\ 0)\times v_d \times \mathcal{N}_{H_d}.
  \end{gathered}
  \label{eq:model3-4vev}
\end{equation}

\begin{table}[t]
  \centering
  \begin{tabular}{c|cc} \hline
                                                     & Experimental value~\cite{ParticleDataGroup:2022pth} & Model 3-4     \\ \hline
    $(10^5m_{\rm u} , 10^3m_{\rm c})/m_{\rm t}$      & (1.25, 7.35)                                        & (1.44, 15.0)  \\
    $(10^3m_{\rm d} , 10^2m_{\rm s})/m_{\rm b}$      & (1.12, 2.23)                                        & (0.633, 2.53) \\
    $(10^4m_{\rm e} , 10^2m_{\rm \mu})/m_{\rm \tau}$ & (2.88, 5.95)                                        & (3.51, 4.57)  \\
    $|V^{\rm{CKM}}|$                                 &
    $\left(
      \begin{matrix}
        0.974 & 0.225  & 0.00382 \\
        0.221 & 0.987  & 0.041   \\
        0.008 & 0.0388 & 1.01
      \end{matrix}
      \right)$                  &
    $\left(
      \begin{matrix}
        0.984   & 0.175  & 0.00566 \\
        0.175   & 0.983  & 0.0632  \\
        0.00905 & 0.0628 & 0.998
      \end{matrix}
      \right)$                                                                                         \\
    $m_{\rm t}/m_{\rm b}$                            & 41.3                                                & 39.0          \\
    $\chi$                                           & -                                                   & 0.319         \\ \hline
  \end{tabular}
  \caption{Flavor structures of Model 3-4}
  \label{tab:case 3-4}
\end{table}

\subsection*{Case 4: Models with seven-generation Higgs pairs}

\begin{table}[t]
  \centering
  \begin{tabular}{c|cc} \hline
                                                     & Experimental value~\cite{ParticleDataGroup:2022pth} & Model 4-1     \\ \hline
    $(10^5m_{\rm u} , 10^3m_{\rm c})/m_{\rm t}$      & (1.25, 7.35)                                        & (0.886, 5.07) \\
    $(10^3m_{\rm d} , 10^2m_{\rm s})/m_{\rm b}$      & (1.12, 2.23)                                        & (0.499, 6.11) \\
    $(10^4m_{\rm e} , 10^2m_{\rm \mu})/m_{\rm \tau}$ & (2.88, 5.95)                                        & (4.98, 6.11)  \\
    $|V^{\rm{CKM}}|$                                 &
    $\left(
      \begin{matrix}
        0.974 & 0.225  & 0.00382 \\
        0.221 & 0.987  & 0.041   \\
        0.008 & 0.0388 & 1.01
      \end{matrix}
      \right)$                  &
    $\left(
      \begin{matrix}
        0.994   & 0.113  & 0.00180 \\
        0.125   & 0.991  & 0.0498  \\
        0.00730 & 0.0467 & 0.999
      \end{matrix}
      \right)$                                                                                         \\
    $m_{\rm t}/m_{\rm b}$                            & 41.3                                                & 38.6          \\
    $\chi$                                           & -                                                   & 0.694         \\ \hline
  \end{tabular}
  \caption{Flavor structures of Model 4-1}
  \label{tab:case 4-1}
\end{table}

\subsubsection*{Model 4-1:}
\begin{equation}
  \begin{aligned}
    (M^{(1)},M^{(2)},M^{(3)})             & =
    \left(
    \begin{array}{cc|ccc}
      (7,H,-7) &          &         &           &           \\
               & (8,H,-8) &         &           &           \\ \hline
               &          & (0,0,H) &           &           \\
               &          &         & (14,-1,H) &           \\
               &          &         &           & (14,-1,H)
    \end{array}
    \right) ,                                 \\
    (\zeta^{(1)},\zeta^{(2)},\zeta^{(3)}) & =
    \left(
    \begin{array}{cc|ccc}
      (0,0,0) &         &         &                   &                      \\
              & (0,0,0) &         &                   &                      \\ \hline
              &         & (0,0,0) &                   &                      \\
              &         &         & (\frac{1}{2},0,0) &                      \\
              &         &         &                   & (\frac{\tau}{2},0,0)
    \end{array}
    \right) ,                                 \\
    (P,P')                                & =
    \left(
    \begin{array}{cc|ccc}
      (1,1) &       &         &        &        \\
            & (1,1) &         &        &        \\ \hline
            &       & (-1,-1) &        &        \\
            &       &         & (1,-1) &        \\
            &       &         &        & (1,-1)
    \end{array}
    \right)
  \end{aligned}
\end{equation}
For the above inputs, the flavor structures are obtained as Table~\ref{tab:case 4-1} with
\begin{equation}
  \begin{gathered}
    \text{Im}\tau_1 = 1.5 , \quad \tan\beta = 50, \\
    \langle H_u \rangle = (0.4,\ 0.6,\ 1,\ 0,\ 0,\ 0,\ 0)\times v_u \times \mathcal{N}_{H_u}, \quad \langle H_d \rangle = (0.2,\ 0,\ 0,\ 0,\ 0,\ 0,\ 0)\times v_d \times \mathcal{N}_{H_d}.
  \end{gathered}
\end{equation}

\subsubsection*{Model 4-2:}
\begin{equation}
  \begin{aligned}
    (M^{(1)},M^{(2)},M^{(3)})             & =
    \left(
    \begin{array}{cc|ccc}
      (7,H,-7) &          &         &           &           \\
               & (8,H,-8) &         &           &           \\ \hline
               &          & (0,0,H) &           &           \\
               &          &         & (14,-1,H) &           \\
               &          &         &           & (14,-1,H)
    \end{array}
    \right) ,                                 \\
    (\zeta^{(1)},\zeta^{(2)},\zeta^{(3)}) & =
    \left(
    \begin{array}{cc|ccc}
      (0,0,0) &         &         &                      &                   \\
              & (0,0,0) &         &                      &                   \\ \hline
              &         & (0,0,0) &                      &                   \\
              &         &         & (\frac{\tau}{2},0,0) &                   \\
              &         &         &                      & (\frac{1}{2},0,0)
    \end{array}
    \right) ,                                 \\
    (P,P')                                & =
    \left(
    \begin{array}{cc|ccc}
      (1,1) &       &         &        &        \\
            & (1,1) &         &        &        \\ \hline
            &       & (-1,-1) &        &        \\
            &       &         & (1,-1) &        \\
            &       &         &        & (1,-1)
    \end{array}
    \right)
  \end{aligned}
\end{equation}
For the above inputs, the flavor structures are obtained as Table~\ref{tab:case 4-2} with
\begin{equation}
  \begin{gathered}
    \text{Im}\tau_1 = 1.4 , \quad \tan\beta = 25, \\
    \langle H_u \rangle = (0,\ 0.8,\ 0.4,\ 0.1,\ 0\ 0,\ 0)\times v_u \times \mathcal{N}_{H_u}, \quad \langle H_d \rangle = (0.4,\ 0.2,\ 0,\ 0,\ 0,\ 0,\ 0)\times v_d \times \mathcal{N}_{H_d}.
  \end{gathered}
\end{equation}

\begin{table}[t]
  \centering
  \begin{tabular}{c|cc} \hline
                                                     & Experimental value~\cite{ParticleDataGroup:2022pth} & Model 4-2     \\ \hline
    $(10^5m_{\rm u} , 10^3m_{\rm c})/m_{\rm t}$      & (1.25, 7.35)                                        & (0.614, 76.2) \\
    $(10^3m_{\rm d} , 10^2m_{\rm s})/m_{\rm b}$      & (1.12, 2.23)                                        & (0.603, 5.47) \\
    $(10^4m_{\rm e} , 10^2m_{\rm \mu})/m_{\rm \tau}$ & (2.88, 5.95)                                        & (7.35, 5.18)  \\
    $|V^{\rm{CKM}}|$                                 &
    $\left(
      \begin{matrix}
        0.974 & 0.225  & 0.00382 \\
        0.221 & 0.987  & 0.041   \\
        0.008 & 0.0388 & 1.01
      \end{matrix}
      \right)$                  &
    $\left(
      \begin{matrix}
        0.986   & 0.166  & 0.00736 \\
        0.166   & 0.986  & 0.0306  \\
        0.00423 & 0.0312 & 1.00
      \end{matrix}
      \right)$                                                                                         \\
    $m_{\rm t}/m_{\rm b}$                            & 41.3                                                & 45.2          \\
    $\chi$                                           & -                                                   & 1.74          \\ \hline
  \end{tabular}
  \caption{Flavor structures of Model 4-2}
  \label{tab:case 4-2}
\end{table}

\begin{table}[t]
  \centering
  \begin{tabular}{c|cc} \hline
                                                     & Experimental value~\cite{ParticleDataGroup:2022pth} & Model 4-3     \\ \hline
    $(10^5m_{\rm u} , 10^3m_{\rm c})/m_{\rm t}$      & (1.25, 7.35)                                        & (2.00, 9.45)  \\
    $(10^3m_{\rm d} , 10^2m_{\rm s})/m_{\rm b}$      & (1.12, 2.23)                                        & (0.498, 6.42) \\
    $(10^4m_{\rm e} , 10^2m_{\rm \mu})/m_{\rm \tau}$ & (2.88, 5.95)                                        & (4.99, 6.11)  \\
    $|V^{\rm{CKM}}|$                                 &
    $\left(
      \begin{matrix}
        0.974 & 0.225  & 0.00382 \\
        0.221 & 0.987  & 0.041   \\
        0.008 & 0.0388 & 1.01
      \end{matrix}
      \right)$                  &
    $\left(
      \begin{matrix}
        0.990   & 0.138  & 0.00313 \\
        0.0888  & 0.993  & 0.0755  \\
        0.00541 & 0.0498 & 0.999
      \end{matrix}
      \right)$                                                                                         \\
    $m_{\rm t}/m_{\rm b}$                            & 41.3                                                & 40.5          \\
    $\chi$                                           & -                                                   & 0.767         \\ \hline
  \end{tabular}
  \caption{Flavor structures of Model 4-3}
  \label{tab:case 4-3}
\end{table}

\subsubsection*{Model 4-3:}
\begin{equation}
  \begin{aligned}
    (M^{(1)},M^{(2)},M^{(3)})             & =
    \left(
    \begin{array}{cc|ccc}
      (8,H,-8) &          &         &           &           \\
               & (7,H,-7) &         &           &           \\ \hline
               &          & (0,0,H) &           &           \\
               &          &         & (14,-1,H) &           \\
               &          &         &           & (14,-1,H)
    \end{array}
    \right) ,                                 \\
    (\zeta^{(1)},\zeta^{(2)},\zeta^{(3)}) & =
    \left(
    \begin{array}{cc|ccc}
      (0,0,0) &         &         &                   &                      \\
              & (0,0,0) &         &                   &                      \\ \hline
              &         & (0,0,0) &                   &                      \\
              &         &         & (\frac{1}{2},0,0) &                      \\
              &         &         &                   & (\frac{\tau}{2},0,0)
    \end{array}
    \right) ,                                 \\
    (P,P')                                & =
    \left(
    \begin{array}{cc|ccc}
      (1,1) &       &         &        &        \\
            & (1,1) &         &        &        \\ \hline
            &       & (-1,-1) &        &        \\
            &       &         & (1,-1) &        \\
            &       &         &        & (1,-1)
    \end{array}
    \right)
  \end{aligned}
\end{equation}
For the above inputs, the flavor structures are obtained as Table~\ref{tab:case 4-3} with
\begin{equation}
  \begin{gathered}
    \text{Im}\tau_1 = 1.2 , \quad \tan\beta = 25, \\
    \langle H_u \rangle = (0.7,\ 0.9,\ 1,\ 0,\ 0,\ 0,\ 0)\times v_u \times \mathcal{N}_{H_u}, \quad \langle H_d \rangle = (0.7,\ 0,\ 0,\ 0,\ 0,\ 0,\ 0)\times v_d \times \mathcal{N}_{H_d}.
  \end{gathered}
\end{equation}

\begin{table}[t]
  \centering
  \begin{tabular}{c|cc} \hline
                                                     & Experimental value~\cite{ParticleDataGroup:2022pth} & Model 4-4     \\ \hline
    $(10^5m_{\rm u} , 10^3m_{\rm c})/m_{\rm t}$      & (1.25, 7.35)                                        & (1.29, 6.65)  \\
    $(10^3m_{\rm d} , 10^2m_{\rm s})/m_{\rm b}$      & (1.12, 2.23)                                        & (0.659, 2.44) \\
    $(10^4m_{\rm e} , 10^2m_{\rm \mu})/m_{\rm \tau}$ & (2.88, 5.95)                                        & (6.79, 2.69)  \\
    $|V^{\rm{CKM}}|$                                 &
    $\left(
      \begin{matrix}
        0.974 & 0.225  & 0.00382 \\
        0.221 & 0.987  & 0.041   \\
        0.008 & 0.0388 & 1.01
      \end{matrix}
      \right)$                  &
    $\left(
      \begin{matrix}
        0.991  & 0.131 & 0.00509 \\
        0.130  & 0.984 & 0.118   \\
        0.0126 & 0.117 & 0.993
      \end{matrix}
      \right)$                                                                                           \\
    $m_{\rm t}/m_{\rm b}$                            & 41.3                                                & 41.9          \\
    $\chi$                                           & -                                                   & 0.919         \\ \hline
  \end{tabular}
  \caption{Flavor structures of Model 4-4}
  \label{tab:case 4-4}
\end{table}

\subsubsection*{Model 4-4:}
\begin{equation}
  \begin{aligned}
    (M^{(1)},M^{(2)},M^{(3)})             & =
    \left(
    \begin{array}{cc|ccc}
      (8,H,-8) &          &         &           &           \\
               & (7,H,-7) &         &           &           \\ \hline
               &          & (0,0,H) &           &           \\
               &          &         & (14,-1,H) &           \\
               &          &         &           & (14,-1,H)
    \end{array}
    \right) ,                                 \\
    (\zeta^{(1)},\zeta^{(2)},\zeta^{(3)}) & =
    \left(
    \begin{array}{cc|ccc}
      (0,0,0) &         &         &                      &                   \\
              & (0,0,0) &         &                      &                   \\ \hline
              &         & (0,0,0) &                      &                   \\
              &         &         & (\frac{\tau}{2},0,0) &                   \\
              &         &         &                      & (\frac{1}{2},0,0)
    \end{array}
    \right) ,                                 \\
    (P,P')                                & =
    \left(
    \begin{array}{cc|ccc}
      (1,1) &       &         &        &        \\
            & (1,1) &         &        &        \\ \hline
            &       & (-1,-1) &        &        \\
            &       &         & (1,-1) &        \\
            &       &         &        & (1,-1)
    \end{array}
    \right)
  \end{aligned}
\end{equation}
For the above inputs, the flavor structures are obtained as Table~\ref{tab:case 4-4} with
\begin{equation}
  \begin{gathered}
    \text{Im}\tau_1 = 2.0 , \quad \tan\beta = 50, \\
    \langle H_u \rangle = (0,\ 0,\ 0,\ 0,\ 0,\ 1,\ 0.5)\times v_u \times \mathcal{N}_{H_u}, \quad \langle H_d \rangle = (0,\ 0.5,\ 0,\ 0,\ 0,\ 1,\ 0.5)\times v_d \times \mathcal{N}_{H_d}.
  \end{gathered}
\end{equation}

\bibliography{ref}
\bibliographystyle{ytphys}

\end{document}